\documentclass[aps,reprint,floatfix,prx,nofootinbib]{revtex4-2}

\usepackage{graphicx} 

\usepackage{amsmath,amssymb,amsfonts,bm}
\usepackage{accents}

\usepackage{sansmath}
\usepackage{upgreek}

\usepackage{tikz}
\usetikzlibrary{arrows.meta}

\usepackage{booktabs}
\usepackage{tabularx}

\usepackage{xcolor}
\definecolor{linkColor}{RGB}{0,70,120}
\definecolor{darkgreen}{RGB}{0,128,0}
\definecolor{darkgray}{RGB}{90,90,90}
\definecolor{orange}{RGB}{180,50,0}

\usepackage[colorlinks=true, allcolors=linkColor,pdfborder={0 0 0},pdfencoding = auto]{hyperref}

\usepackage[normalem]{ulem}

\newcommand{\tvec}{{\bm{\uptau}}}
\newcommand{\tvecIt}{{\bm{\tau}}}

\newcommand{\xivecIt}{{\bm{\xi}}}

\makeatletter
\DeclareRobustCommand*\upell{\mathpalette\@upell\relax}
\newcommand*\@upell[2]{
  \setbox0=\hbox{$#1\ell$}
  \setbox1=\hbox{\rotatebox{10}{$#1\ell$}}
  \dimen0=\wd0 \advance\dimen0 by -\wd1 \divide\dimen0 by 2
  \mathord{\lower 0.1ex \hbox{\kern\dimen0\unhbox1\kern\dimen0}}
}

\newcommand{\rvec}{\mathbf{r}}
\newcommand{\nvec}{\mathbf{n}}
\newcommand{\rvecIt}{{\bm{r}}}

\newcommand{\upp}{\mathrm{p}}
\newcommand{\upa}{\mathrm{a}}
\newcommand{\uph}{\mathrm{h}}
\newcommand{\upv}{\mathrm{v}}

\newcommand{\Fiso}{\mathbf{F}}

\newcommand{\muT}{\mu_\mathrm{T}}
\newcommand{\muI}{\mu_\mathrm{I}}
\newcommand{\phiT}{\phi_\mathrm{T}}
\newcommand{\phiI}{\phi_\mathrm{I}}
\newcommand{\phiTI}{\phi_\mathrm{TI}}

\DeclareMathOperator{\tr}{Tr}
\let\div\relax
\DeclareMathOperator{\div}{div}

\begin{document}

\title{Tissue shape from cell-scale active tensions}

\author{Nikolas H.\ Claussen}
\email{nc1333@princeton.edu}
\affiliation{Princeton Center for Theoretical Science, Princeton University, Princeton, New Jersey 08542, USA}
\author{Fridtjof Brauns}
\email{fbrauns@pks.mpg.de}
\affiliation{Max Planck Institute for the Physics of Complex Systems, Nöthnitzer Straße 38, 01187 Dresden, Germany}
\affiliation{Max Planck Institute of Molecular Cell Biology and Genetics, Pfotenhauerstraße 108, 01307 Dresden, Germany}
\affiliation{Center for Systems Biology Dresden, 01307 Dresden, Germany}
\author{Boris I.\ Shraiman}
\email{shraiman@ucsb.edu}
\affiliation{Kavli Institute for Theoretical Physics, University of California, Santa Barbara, California 93106, USA}

\begin{abstract}
    Connecting cell and tissue mechanics is a fundamental challenge in the physics of morphogenesis.
    Conventional elasticity theory is generally inapplicable; cytoskeletal turnover precludes a fixed reference state, and the dominant forces are generated actively, independently of strain.   
    Here, we study epithelia in quasi-static force balance between intracellular pressure and active interfacial tensions.
    The theory starts from a set of prescribed active tensions, represented as a Maxwell--Cremona tension triangulation dual to the cell tiling. This geometric formulation brings to bear the mathematics of discrete conformal geometry. Using generalized Voronoi diagrams, we construct the mechanically balanced cell tiling resulting from given tensions and compute the tissue-scale stress.
    Conceptually, we treat cell interfaces as force dipoles whose embedding in physical space -- the physical cell tiling -- is uniquely determined by force balance.
    We find that the tensions define an effective reference state in which the macroscopic stress vanishes as pressures and tensions cancel locally.
    Deformations from this reference
    result in macroscopic stress, constituting an emergent form of elasticity.
    Adiabatic tension dynamics change the reference shape.
    Thus, rather than fluidizing, tissues can deform -- or ``morph'' --, while resisting external forces like a solid.
    Discrete conformal geometry also defines a geometric representation of cell network topology, plastically remodeled by cell rearrangements.
    We make several experimentally testable predictions, notably that macroscopic stress is independent of local tension anisotropy.
    Overall, the unified description of emergent elasticity and plastic flow in epithelia is a step towards a theory of morphogenesis that puts active forces first, and reflects the  differences between inert and biological matter.
    Our framework also applies to 2D fluid foams and to frictionless granular media via a mechanical duality.
\end{abstract}

\maketitle

\section*{Introduction}

Active materials locally generate active forces, independent of elastic or viscoelastic stresses~\cite{Marchetti.etal2013}. 
Living tissues are a paradigmatic example, generating active stress via motor molecules in the cytoskeleton~\cite{Heisenberg.Bellaiche2013}.
During morphogenesis, cells use these ``microscopic muscles'' to sculpt the form of the embryonic body~\cite{Gilbert.Barresi2016}.
Understanding how cell-scale active forces determine tissue-scale shape, and how tissues react to external forces (e.g., fluid- or solid-like), is an important question in biological physics. 
Physically, it amounts to determining the tissue's mechanical ground state and low-energy excitations.

In conventional (visco)elasticity, these two questions are answered from the outset by specifying a reference configuration (either via a stress-free rest state~\cite{Landau.Lifshitz1986} or a target metric~\cite{Efrati.etal2009}) and a constitutive relation between stress and strain, thereby determining the response to external forces. 
However, in a living tissue, rapid molecular turnover precludes a fixed reference, and active stress can be controlled \emph{independently of strain}. 
Despite this fundamental challenge, phenomenological continuum models have successfully reproduced observed morphogenetic dynamics (tissue flows) \cite{Streichan.etal2018,Caldarelli.etal2024,Serra.etal2023,Ioratim-Uba.etal2023}.
These models add active stress and feedback terms ad hoc on top of a viscous or viscoelastic background. However, they lack access to the cellular scale, where biological regulation takes place. Therefore, quantitatively linking specific cellular dynamics and regulation to tissue-scale outcomes remains an open problem.

\begin{figure}
    \centering
    \includegraphics{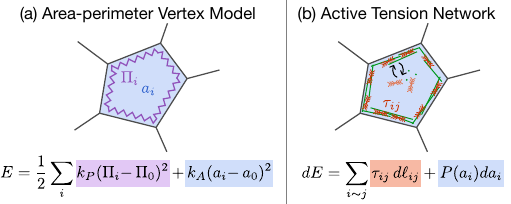}
    \caption{(a) The conventional vertex model posits that each cell has a spring-like constitutive relation relative to a preferred area $\upa_0$ and a preferred perimeter $\Pi_0$ (note that we reserve $p_i$ for the intracellular pressure).
    (b) Molecular turnover of the cytoskeleton (actin: green; myosin: red) rapidly relaxes passive stresses and thus precludes a fixed reference shape for the cell. Instead,  tensions $\tau_{ij}$ along cell interfaces are actively generated by contractile myosin. Cells can control these active tensions independently of strain, so there is no equation of state for the mechanical energy. We allow for a general equation of state for intracellular pressure $P(a_i)$.}
    \label{fig:vm-vs-atn}
\end{figure}

On the cell scale, the predominant modeling framework is the so-called vertex model \cite{Weliky.Oster1990,Hufnagel.etal2007,Farhadifar.etal2007,Alt.etal2017} which, in its most commonly used form, posits a target perimeter and area for each cell with an elastic energy penalty for deviations from these target values [Fig.~\ref{fig:vm-vs-atn}(a)].
How these cellular parameters control the model's rheology (fluid-to-solid transition) has been studied in great detail \cite{Farhadifar.etal2007,Bi.etal2015}. 
In the fluid regime, cell interfaces become ``slack'' as their interfacial tensions vanish, and the tissue's shear modulus goes to zero \cite{Yan.Bi2019}. 
However, in many systems, this phenomenology conflicts with experimental observations. Tissues can deform against external resistance~\cite{Collinet.etal2015}, and interfacial tensions are non-zero even in tissues changing shape (``flowing'')~\cite{Rauzi.etal2008,Fernandez-Gonzalez.etal2009}. Indeed, biological evidence suggests that it is precisely the dynamics of active tensions that drive tissue deformation~\cite{Bertet.etal2004,Saadaoui.etal2020,Brauns.etal2024}.

\subsubsection{Biology of active force generation}

The assumption of a fixed target shape made by many models for cell mechanics is contradicted by the highly dynamic nature of the cytoskeleton, which is essential for its function~\cite{Jodoin.etal2015,Thiyagarajan.etal2022}.
For instance, Ref.~\cite{Khalilgharibi.etal2019} shows that in MDCK epithelia, half of actomyosin turns over within two minutes, relaxing elastic stresses in under one minute. 
Instead, persistent cellular stress depends on myosin activity.
Thus, fast molecular turnover precludes the reference cellular configuration assumed by the models. In other words, passive stresses relax rapidly, so that on longer timescales, only actively generated stresses persist \cite{Noll.etal2017}.
Understanding active tissues requires an approach that fundamentally departs from passive elasticity. Here, we address these challenges starting from a type of vertex model in which all tensions are active -- an active tension network \cite{Noll.etal2017}. 

This type of model applies to tissues whose mechanics are dominated by active tensions, generated by actomyosin in the junctional cortex along cell interfaces.
There are, however, important effects that the model does not capture. 
First, relative to the junctional cytoskeleton, we neglect the activity of apical and basal myosin pools. These pools become important for cells whose shape changes out of the plane of the tissue, for example during invagination~\cite{Martin2020}.
Second, relative to tensile forces between cells, we neglect traction forces between cells and solid substrates. Traction forces are important for mesenchymal cells~\cite{Harris.etal1980}, but can be neglected in many embryonic tissues, which instead sit atop a fluid yolk~\cite{Saadaoui.etal2020,Streichan.etal2018}.
Third, we neglect the influence of intermediate filaments like keratin, which turn over on a much slower time scale than actomyosin. Keratin is important for the resistance of epithelial cells to extreme deformations~\cite{Harris.etal2012,Latorre.etal2018}.

\subsubsection{Effective elasticity}

The ATN model, as we will show, admits a systematic coarse-graining to the continuum level. Specifically, we develop a geometric theory for how cell-scale tensions determine macroscopic tissue shape and stress.
We find that elasticity and plasticity emerge on the tissue scale from purely active tensions on the cell scale.
More precisely, the intrinsic tensions along cell-cell junctions define a mechanically balanced, macroscopically stress-free reference configuration for the cell tiling.
Strain away from this reference generates stress, so that the tissue behaves like an elastic solid. Tension dynamics lead to plastic tissue deformation by changing the effective reference state. This explains how cells can control tissue shape using local tensions. Biologically, maintaining a certain active tension configuration only requires homeostasis of motor molecule concentrations, in contrast to a reference shape.

Our theory relies on the geometric constraints imposed by mechanical force balance. It treats the tissue as a \emph{distributed hydrostatic skeleton} in which cell pressures balance interfacial tensions. 
Through adiabatic changes of active tensions, the tissue can change shape while remaining in force balance. Importantly, this involves cell rearrangements (active T1 events) that drive plastic shape change \cite{Brauns.etal2024,Claussen.Brauns2025}.
Our results explain the success of phenomenological continuum theories and provide insights into the character of active stress across scales.
In a companion paper, Ref.~\cite{Claussen.etal2026}, we complement the present cell-scale theory by deriving the effective long-wavelength theory directly in the continuum.

\subsubsection{Introductory example: soap foams}

Before embarking on the technical analysis, let us prime our intuition with an everyday example: a soap foam.
The interfaces of the foam are fluid -- they can freely change length while under constant surface tension, in marked contrast to a Hookean spring. 
Despite the fluidity of its interface, a single soap bubble has a defined shape, a sphere, defined by the force balance of surface tension and internal pressure. 
Note that for mechanical stability, the internal pressure must depend on the bubble volume -- prescribing both tension and pressure independently of shape makes the bubble unstable to collapse or expansion\footnote{This can be demonstrated by connecting two foam bubbles by a thin pipe, equalizing their pressures. One foam bubble will collapse while the other one will expand.}

A foam, assembled from many bubbles, also takes on a shape defined by force balance: the interfaces meet at $120^\circ$ angles, defined by the balance of surface tensions, and internal pressure prevents the foam cells from collapsing \footnote{Gas exchange between the foam cells leads to a coarsening process where large cells grow at the expense of small ones, which eventually collapse. In living cells, water exchange due to hydrostatic pressure differences is balanced by osmotic pressure.}.
A soap foam can hold its shape against (sufficiently weak) external shear forces. It behaves like a solid, even though it is microscopically composed of a fluid. 

To understand this elastic response to external forces, consider each interface as a force dipole. The macroscopic tensile stress is their local average, made precise in the Batchelor formula (see Eq.~\eqref{eq:batchelor_general}).
Even though interface tensions are fixed, the contribution of each force dipole depends on interface length and orientation. Therefore, the macroscopic stress depends on the shape of the foam cells, which leads to the emergence of an effective constitutive relation on the tissue scale.
Thus, force balance determines both cell and tissue geometry.
In the remainder of this work, we will generalize the above intuition to a setting where each interface can have a different tension and make it mathematically precise using ideas from discrete conformal geometry.

\subsection{Background and motivation}
\label{sec:discrete_summary}

\subsubsection{Active tension networks in 2D}

Our starting point is a minimal model for a tissue where mechanics is dominated by active tensions generated along cell-cell interfaces (also referred to as ``junctions''). The dominance of active tensions is motivated by the rapid turnover of the junctional cytoskeleton, which relaxes passive stresses. We refer to this type of model as an active tension network (ATN)~\cite{Noll.etal2017,Claussen.etal2024}.
Intuitively, the ATN model describes a tissue as a generalized, active foam~\cite{Kim.etal2021}. In contrast to a conventional fluid foam, where the fluid's surface tension is a fixed material parameter, cells actively regulate their interfacial tensions.
More precisely, the cell-based ATN model~\cite{Noll.etal2017,Claussen.etal2024} models a confluent 2d epithelial tissue whose cells $i,j,\dots$ form a tiling.
We enumerate cell-cell interfaces by indices $ij$, and tricellular junctions by indices $ijk$, indicating the cells which meet there.
We assume the mechanics are dominated by interfacial tensions $\uptau_{ij}$ and intracellular pressure $\upp_i$. The tri-cellular vertices of the cell tiling are written $\rvec_{ijk}$. ATNs are thus a type of vertex model~\cite{Alt.etal2017}: they represent the tissue by a polygonal tiling with vertices $\rvec_{ijk}$. However, as we will see below, the effective degrees of freedom are geometric ``collective modes'' coupling multiple vertices.

The configuration of the tissue is determined by mechanical balance -- virtual work has to vanish
\begin{align}
    0 = d E = \sum_{i\sim j} \uptau_{ij} d\upell_{ij} - \sum_i \upp_i d\upa_i 
    \label{eq:dE}
\end{align}
where $\upell_{ij}$ and $\upa_i$ are interface lengths and cell areas and $i{\sim}j$ denotes all adjacent cell pairs.
A constitutive relation (cell compressibility) determines the pressures $\upp_i$:
\begin{align}\label{eq:constitutive}
    \upp_i = P(\upa_i)
\end{align}
As explained for a soap bubble, actively prescribing both tensions and pressures is mechanically unstable.

By contrast, the interfacial tensions $\uptau_{ij}$ are dynamical variables independent of interfacial lengths $\upell_{ij}$, and instead determined by the local activity of motor molecules.
This is the limit of a more detailed description~\cite{Noll.etal2017} of the actomyosin cortex as a spring with an adjustable rest length $\upell_{ij}^0$. 
Viscous relaxation of external pulling forces elongates $\upell_{ij}^0$, while internal motor molecule activity contracts it. On short timescales (relative to the viscous relaxation time), the interface behaves elastically, while on long timescales, contraction and elongation of $\upell_{ij}^0$ balance, so the interface tension is given by the intrinsic motor molecule activity. The latter setting will be considered here. Due to these internal dynamics, the differential $dE$ cannot be integrated to an overall elastic energy. Instead, $dE = 0$ expresses the conditions for local mechanical balance.

The balance conditions are shared with fluid films in a foam~\cite{Weaire.etal2005}. Eq.~\eqref{eq:dE} thus describes a \emph{generalized foam}: in an ordinary fluid foam, all interfacial tensions are equal to twice the fluid's surface tension $\uptau_{ij} = 2\gamma$. In a generalized foam, the tensions are upgraded to independent, dynamical variables.
The idea of generalized foams has also been used to describe 3D cell clusters, like embryos during the cleavage stage~\cite{Ichbiah.etal2023}. These models posit a surface tension on the 2D interface between two cells $i, j$ and are the most natural 3D generalizations of ATNs. Many of the results presented here carry over to 3D (see Discussion).

\subsubsection{ATNs in the context of vertex models}

ATNs are a special type of ``vertex model'', in that the tissue is described as a polygonal tiling with vertices.
Vertex models form a broad spectrum, depending on the mechanical model used to determine the vertex positions.
While the ATN model shares the geometric degrees of freedom (vertex positions) of the much-studied area-perimeter vertex model~\cite{Farhadifar.etal2007,Bi.etal2015}, it fundamentally differs in the microscopic mechanics. The area-perimeter vertex model assumes a particular physical constitutive relation relative to a target cell perimeter and area.
These assumptions are encoded in the elastic energy $E_{\mathrm{AP}} = \tfrac12\sum_{i} \left[k_A (\upa_0-\upa_i)^2 + k_P(\Pi_0-\Pi_i)^2\right]$. Here, $\upa_0, \Pi_0$ are a cell's target area and perimeter, $k_A, k_P$ are generalized spring constants, and $\Pi_i = \sum_{j\sim i}\upell_{ij}$ is the cell perimeter ($i\sim j$ denotes all neighbors of cell $i$.)
The area-perimeter vertex model, therefore, describes a particular kind of spring network where a reference shape is defined through target lengths and areas~\footnote{For ``excess'' perimeter, there is a continuum of compatible reference shapes, so cells become floppy, which leads to fluid tissue behavior in the vertex model. However, in this regime interface tensions vanish, which is at odds with experimental observations, as noted in the introduction.}. This amounts to a constitutive relation $\uptau_{ij} = k_P (\Pi_i + \Pi_j -2\Pi_0)$ for edge tensions and $P(\upa_i) = k_A (\upa_0 - \upa_i)$ for pressure.
The effective stress tensor of the area-perimeter model has been analyzed in Refs.~\cite{Jensen.etal2020,Jensen.Revell2023}.
Several studies have added active dynamics, like self-propulsion or active tensions, on top of the passive area-perimeter energy~\cite{Bi.etal2016,Barton.etal2017,Sknepnek.etal2023,Rozman.etal2023,Yu.etal2026}.
In the self-propelled Voronoi~\cite{Bi.etal2016} and active vertex~\cite{Barton.etal2017} models, cell shapes are approximated by Voronoi diagrams, which, in particular, leads to efficient simulation of cell rearrangements. 

In the following, we instead focus on a setting where the $\uptau_{ij}$ are specified independently of the $\upell_{ij}$ (App.~\ref{app:perturbation} shows that adding a small perimeter-elasticity term does not change our results qualitatively). As we argue below, in such ATNs, an \emph{effective} reference configuration and a stress-strain relation (with non-zero shear modulus) emerge at large scales, even though microscopically, active tensions (stress) and edge lengths (strain) are independent.

\subsubsection{From tension to physical space}

\begin{figure}
    \centering
    \includegraphics[width=\linewidth]{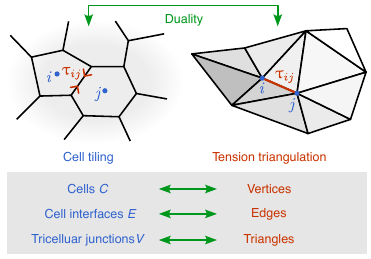}
    \caption{Duality between cell tilings and tension triangulations. $C$, $E$, and $V$ denote the number of cells, cell interfaces, and tricellular junctions, respectively.
    }
    \label{fig:duality}
\end{figure}

Our theory treats cell interfaces as force dipoles whose embedding in physical space -- the physical configuration of cells -- is constrained by force balance.
The key to our approach is to think of the tension configuration as a \emph{triangulation} with vertices $i,j,\dots$ for each cell, and edge lengths $\uptau_{ij}$ (Fig.~\ref{fig:duality}, top).
This construction is possible since cells generically meet in groups of three; therefore, their adjacency graph forms a triangulation \emph{dual} to the cell tiling (Fig.~\ref{fig:duality}, bottom).
The tension triangulation is a geometric representation of the network of the intrinsic active tensions in a tissue. 
By controlling actomyosin contractility, cells operate in ``tension space''.
Given the tension configuration, mechanical equilibration determines the resulting cell shapes and positions in ``physical space''.
This picture raises two questions. First, how does the tension configuration control the physical cell tiling? Second, what is the tissue's macroscopic mechanical state, and how does it respond to external forces?

To answer the first question, we construct the mechanically balanced cell tilings geometrically as an embedding of the tension triangulation into physical space.  Intuitively, every cell interface forms a contractile force dipole. The cell tiling defines an ``embedding'' that places these force dipoles in physical space. To answer the second question, we compute the macroscopic stress tensor $\boldsymbol{\sigma}$ by averaging over these force dipoles.
We identify the effective Airy stress function, the mathematical representation of all mechanically balanced stresses in 2D~\cite{Landau.Lifshitz1986}.

To carry out this program, we cast the constraints of force balance in geometric form, using the mathematical framework of discrete differential geometry. 
In the simplest case, the map from tension to physical space is realized by the well-known Voronoi/Delaunay duality, which constructs a cell tiling from a triangulation.
[As noted above, Refs.~\cite{Bi.etal2016,Barton.etal2017} use the Voronoi construction to approximate cell shapes in the area-perimeter model, adjusting the dual triangulation to minimize the elastic energy. Here, the role of the Voronoi construction is physically different -- it defines an exactly balanced state, and is constructed from the intrinsically prescribed tension triangulation.]

\subsubsection{Continuum mechanics of ATNs and conformal maps}
\label{sec:continuum_summary}

The present manuscript studies the cellular-scale ATN model, which is directly applicable to biological data~\cite{Brauns.etal2024}. The companion paper, Ref.~\cite{Claussen.etal2026}, presents a continuum theory for the behavior of the ATN model at large, supracellular length scales.
The geometric framework outlined above makes the transition to the continuum seamless, and we connect the ATN model to the effective continuum theory via a ``bottom-up'' coarse-graining analysis.

In the continuum limit, one approximates the tension triangulation as a smooth surface, a Riemannian manifold. A set of Lagrangian material coordinates $\xivecIt$ plays the role of the cell labels $i,j, \dots$. 
The distance $\uptau_{ij}$ between vertices of the tension triangulation becomes a Riemannian metric $\bm{g}(\xivecIt)$ on the tension manifold. (We use italic letters like $\bm{g}$ for continuous fields, and upright letters like $\uptau_{ij}$ for discrete objects).
In the continuum theory, the positions of the cells define an \emph{embedding} $\rvecIt(\xivecIt)$ of the tension manifold into physical space. 
The distortion of the tension metric by the embedding (analogous to the cartographic distortion in maps of the globe) then determines the physical stress.
To understand the resulting physics, the continuum theory draws on the power of complex analysis and conformal maps. A discrete version of conformal maps is also at the heart of the cell-level theory presented here.

\subsection{Outline}
\label{sec:coarse_graining_summary}

The present manuscript proceeds in three steps.
First, we parameterize the cell tilings compatible with microscopic force balance.
Second, we compute their macroscopic stress tensor.
Third, we identify how the physical control parameters, the tensions, determine the cell tiling, how the tissue responds to external forces,
and how tension dynamics lead to tissue flow and cell rearrangement. 
All steps rely on a geometric approach to mechanics, which represents the tension configuration as a triangulation.
This geometric formulation also allows connecting the discrete theory to its continuum limit.

Before developing the general theory, Sec.~\ref{sec:lattice} presents a self-contained, simplified setting where all three steps can be carried out with little mathematical complexity (we advise reading Sec.~\ref{sec:lattice} before any of Sec.~\ref{sec:Moebius_diagrams}--\ref{sec:continuum_limit}).
In brief, we show that via the Voronoi construction, the tension triangulation defines a macroscopically stress-free reference configuration of the tissue, deformations away from which generate (shear) stress. This constitutes an emergent form of elasticity. Slow, adiabatic tension dynamics deform the reference, allowing the tissue to morph while remaining solid against external forces.

We then generalize to arbitrary tension networks.
Sec.~\ref{sec:Moebius_diagrams} introduces \emph{M\"{o}bius diagrams} (MDs), which parametrize all possible force-balanced states of tension networks. Sec.~\ref{sec:stress} computes the stress tensor in an MD.
We then ask how the physical control parameters determine the tiling and stress. We begin in Sec.~\ref{sec:constant_pressure} with cell tilings where pressure is constant. We analyze the link between tiling deformations and stress and explain a duality between tension networks and granular materials.
Sec.~\ref{sec:pressure} incorporates pressure differentials, which we describe geometrically using (suitably discretized) conformal maps.
Sec.~\ref{sec:curvature} uses connections to conformal maps to complete the problem of computing the tissue configuration from the active tensions. This leads to a generalized von-Neumann law for the intracellular pressure.
Sec.~\ref{sec:governing} then derives governing equations for the macroscopic stress tensor and adiabatic tissue flow.
Sec.~\ref{sec:T1s} studies cell rearrangement (T1 processes), and introduces a geometric representation of the cell network topology, again using conformal maps. Sec.~\ref{sec:continuum_limit} determines the continuum limit of the cell-level theory, and summarizes the technical aspects of the work. 
The Discussion covers implications for morphogenesis and tissue mechanics.
Table~\ref{tab:notation} lists the symbols used in our notation.

\section*{Results}
\setcounter{subsection}{0}

\subsection*{Force-balance conditions for tension networks}\label{sec:force_balance_conditions}

\begin{figure}
    \centering
    \includegraphics[width=0.9\linewidth]{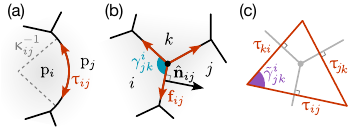}
    \caption{(a) Young--Laplace law relates inverse curvature radius $\upkappa_{ij}$ and pressure differences $\upp_i-\upp_j$. (b)
    A vertex formed by three cells $ijk$. Its geometry is described by the angles $\upgamma_{jk}^i$ and normals $\hat{\mathbf{n}}_{ij}$. The tensile forces $\mathbf{f}_{ij}$ must balance at the vertex.
    (c) The tensions of the interfaces $ij, jk, ki$ form a triangle. In mechanical balance, the corresponding triangle and tiling angles sum to $\pi$.}
    \label{fig:setup-force_balance}
\end{figure}

Given the active tensions $\uptau_{ij}$, what is the physical configuration of the cell tiling in force balance? This is the central question in the theory of active tension networks. 
We begin by casting the force balance condition Eq.~\eqref{eq:dE} in a geometric form. 
First, normal forces due to tension and pressure must balance along each interface $ij$. This leads to the Young--Laplace law:
\begin{align}
    \label{eq:Young--Laplace}
    \uptau_{ij} \upkappa_{ij} = \upp_i-\upp_j
\end{align}
where $\upkappa_{ij}$ is the interface curvature (signed inverse curvature radius; by convention,
$\upkappa_{ij}>0$ when the interface bulges into cell $j$).
Since intracellular pressures are uniform within each cell, interface curvature must be constant as well: interfaces are circular arcs. 

Second, the tensile forces at each tri-cellular vertex $\rvec_{ijk}$ must sum to zero. For reasons that will become clear in a moment, we write this in terms of the normal vectors $\hat{\nvec}_{ij}$ to the interfaces at vertex $ijk$
\begin{align}
    \uptau_{ij}{\hat{\nvec}}_{ij} + \uptau_{jk}{\hat{\nvec}}_{jk} + \uptau_{ki}{\hat{\nvec}}_{ki} = 0
    \label{eq:tension_balance}
\end{align}
(we use the convention that $\hat{\nvec}_{ij}$ points from cell $i$ to cell $j$; cells $ijk$ are ordered counter-clockwise; see Fig.~\ref{fig:setup-force_balance}).
Eq.~\eqref{eq:tension_balance} is the Young--Dupr\'{e} equation.
It also has a useful ``intrinsic'' version: Consider the ``tension triangle'' with side lengths $(\uptau_{ij}, \uptau_{jk}, \uptau_{ki})$ and angles $\tilde{\upgamma}_{ij}^k$ (the angles between tiling interfaces at vertex $ijk$ are denoted $\upgamma^k_{ij}$, without tilde).
Geometrically, Eq.~\eqref{eq:tension_balance} implies that
corresponding triangulation and tiling angles have to be complementary,
\begin{align}
    \label{eq:angle_complement}
    \upgamma_{jk}^i + \tilde{\upgamma}^i_{jk} = \pi
\end{align}
Thus, the three tensions $(\uptau_{ij}, \uptau_{jk}, \uptau_{ki})$ determine the relative orientations of cell-cell interfaces.

\section{The regular lattice}\label{sec:lattice}

We start with a simple setting that already captures much of the key physics of tension networks.
Consider a hexagonal lattice of identical cells with constant pressure. In such a lattice, all vertices are equivalent, so we pick a representative vertex and label the adjacent cells $i = 1,2,3$. The lattice geometry is defined by the three cell-edge vectors $\rvec_{ij}$, $ij=12,23,31$ (Fig.~\ref{fig:lattice-setup-main}).
In contrast to a passive foam, the tensions $\uptau_{ij}$ on these three interfaces can be distinct.
Geometrically, the tensions form a triangle, and the (identical) triangles at each vertex fit together to form a regular triangular lattice.
This lattice can be drawn in the plane using a set of 2D \emph{tension vertices} 
\begin{align}
    \label{eq:tension_embedding_lattice}
     \uptau_{ij}= p_0 |\tvec_{i}-\tvec_{j}|
\end{align}
We refer to this drawing as an \emph{embedding} of the tension triangulation. The \emph{reference pressure} $p_0$ converts units of line tension [N] to units of length [m], so that the $\tvec_i$ vertices have units of length. In the remainder of the paper, we non-dimensionalize so that $p_0 = 1$.

Eq.~\eqref{eq:tension_embedding_lattice} may seem somewhat abstract at first.
However, from any force-balanced cell tiling, one can reconstruct a drawing of the tension triangulation. The forces along each triangle, rotated
clockwise by $\tfrac{\pi}{2}$, read $\uptau_{ij} \hat{\mathbf{n}}_{ij}$. In equilibrium, forces sum to zero at every vertex and form a closed triangle. 
These triangles fit together to form the planar drawing of the tension triangulation, Eq.~\eqref{eq:tension_embedding_lattice}. This construction is
often referred to as the \emph{Maxwell--Cremona} force tiling~\cite{Maxwell1864}.
The Maxwell--Cremona construction is based solely on mechanical equilibrium and is also valid, for example, for a spring network in which tensions depend on the interface lengths. The key premise of the ATN model is that the tensions are instead determined by intrinsic motor molecule activity.

\subsection{From triangulation to tiling: Voronoi construction}\label{sec:Voronoi_lattice}

\begin{figure}
    \centering
    \includegraphics[width=0.6\linewidth]{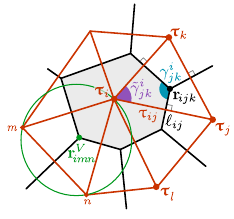}
    \caption{ATN and Voronoi geometry. Red: tension triangulation with vertices $\tvec_i$; Black: cell tiling with vertices $\rvec_{ijk}$. In mechanical balance, the angles $\tilde{\gamma}_{jk}^i+\gamma_{jk}^i=\pi$. Green: Voronoi vertices are triangle circumcircle centers.
    Cell interfaces and tension edges are orthogonal.  Interface length $\upell_{ij}$ is determined by a quadrilateral $ijkl$ of tension vertices.
    }
    \label{fig:setup-Voronoi}
\end{figure}

For this reason, we now proceed in the reverse direction. Starting from a drawing of the tension triangulation, we  construct a force-balanced cell tiling with constant pressure (the latter constraint will be relaxed below). 
To do so, we use the Voronoi dual of the tension triangulation. The Voronoi cell $C^\mathrm{V}_i$ is defined as the set of all points $\rvec$ whose closest triangulation vertex is $\tvec_i$:
\begin{align}
    \label{eq:Voronoi_definition}
    C^\mathrm{V}_i &= \{\rvec \, : |\rvec-\tvec_i|^2  \,  < |\rvec-\tvec_j|^2  \; \forall j\neq i \}
\end{align}
Voronoi cell vertices $\rvec_{ijk}^\mathrm{V}$ are thus equidistant from triangulation vertices $\tvec_i,\tvec_j,\tvec_k$ (they lie at the circumcenters of tension triangles, seeFig.~\ref{fig:setup-Voronoi}),
and the Voronoi cell edges are orthogonal to their duals in the triangulation. The normal to a cell edge carrying $\uptau_{ij}$ tension is $\hat{\nvec}_{ij} = (\tvec_j-\tvec_i) / |\tvec_i-\tvec_j|$. This orthogonality immediately implies that the Young--Dupr\'{e} law is fulfilled:
\begin{multline}
    \uptau_{ij}{\hat{\nvec}}_{ij} + \uptau_{jk}{\hat{\nvec}}_{jk} + \uptau_{ki}{\hat{\nvec}}_{ki} =  \\
     (\tvec_j-\tvec_i) + (\tvec_k-\tvec_j) + (\tvec_i-\tvec_k) = 0.
\end{multline}
The Voronoi tiling is thus mechanically balanced by construction.

We implicitly assumed that the Voronoi dual of the tension triangulation is a well-defined cell tiling with positive interface lengths $\upell_{ij}$; we will return to this issue in Sec.~\ref{sec:T1s_lattice}.

\subsection{Isogonal deformations}\label{sec:isogonal_lattice}

\begin{figure}
    \centering
    \includegraphics[width=\linewidth]{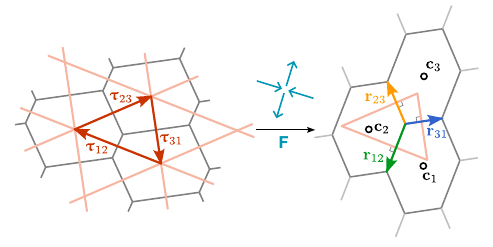}
    \caption{(Left)~Geometry of a periodic lattice formed from a single tension triangle and its dual Voronoi tiling.
    (Right)~Periodic cell tiling resulting from a uniform isogonal deformation $\Fiso$.
    Cell centroids are denoted $\mathbf{c}_i$ and cell edge vectors $\mathbf{r}_{ij}$.
    Both cell tilings have the same vertex angles and thus fulfill the Young--Dupr\'{e} law.
    }
    \label{fig:lattice-setup-main}
\end{figure}

The Young--Dupr\'{e} law fixes the orientation $\hat{\nvec}_{ij}$ of cell-cell interfaces, but that does not determine the cell tiling uniquely.
Displacing a cell interface along its normal leaves its orientation fixed.
Since these displacements preserve vertex angles, we refer to them as \emph{isogonal}~\cite{Noll.etal2017}.
We adopt the Voronoi configuration as a convenient reference with respect to which we measure isogonal deformations.
In the Voronoi configuration, the vertices $\tvec_i$ lie inside the corresponding cells $C_i$. In a regular lattice, the cell centroids are precisely the triangulation vertices $\tvec_i$.
We can then parametrize the isogonal deformations with a deformation tensor $\Fiso$ which maps the vectors $\tvec_{ij}:=\tvec_j-\tvec_i$ to the vectors connecting cell centroids $\mathbf{c}_{j} - \mathbf{i}_{j}$.
(A systematic parametrization for irregular cell tilings will be introduced below).
Let us denote the three edges in the lattice by $ij=12,23,31$.
Since tiling and triangulation vectors must be orthogonal, the cell edge vectors are 
$\rvec_{ij} = \upell_{ij} \tvec_{ij}^\perp / \uptau_{ij}$. 
A direct calculation (App.~\ref{app:batchelor_periodic}) shows that the isogonal deformation tensor of the periodic lattice reads
\begin{equation} \label{eq:isogonal-deformation-lattice}
    \begin{aligned}
    &\Fiso = \frac{1}{2\tilde \upa} \sum_{ij} \rvec_{ij} \otimes \tvec^\perp_{ij}
    = \frac{1}{2\tilde \upa} \sum_{ij} \upell_{ij} \uptau_{ij} \, \hat{\rvec}_{ij} \otimes \hat{\rvec}_{ij}, \\
    &\text{where } \left(\hat{\rvec}\otimes \hat{\rvec} \right)_{ab} = \hat{r}_{\!a}^{} \hat{r}_b^{}
    \end{aligned}
\end{equation}
where $\tilde a$ is the triangle area, $\otimes$ denotes the tensor product between two vectors, and we generally use boldface capital letters for matrices.
$\Fiso$ is symmetric because isogonal displacements induce no rotation.

Isogonal modes show that for the same tensions, multiple mechanically balanced states exist.
Isogonal modes also highlight the distinction between cell shape and tension configuration: 
Cells can be anisotropic even if tensions are isotropic (i.e., all equal), and vice versa.

\subsection{The macroscopic stress tensor and emergent elasticity}
\label{sec:stress_lattice}

The large-scale mechanical state of the tissue is defined by the stress tensor $\boldsymbol{\sigma}$, which we now relate to the microscopic configuration of tensions $\uptau_{ij}$ and isogonal deformation $\Fiso$. 
Within each cell, the local stress tensor is isotropic and equals $-p_0 \mathbb{I}$, while cell-cell interfaces contribute an anisotropic surface stress $\uptau_{ij} \hat{\rvec}_{ij} \otimes \hat{\rvec}_{ij}$.
The \emph{Batchelor formula}~\cite{Batchelor.Green1972} (also known as the virial stress formula~\cite{Irving.Kirkwood1950}), computes the total coarse-grained stress tensor, averaged over a given area $A$:
\begin{align}
\label{eq:batchelor_general}
\begin{aligned}
    \overline{\bm{\sigma}} &= \frac{1}{A}\int \bm{\sigma}(\rvec) \  da \\ &= 
    \frac{1}{A} \sum_{(ij)\in A} \upell_{ij} \uptau_{ij} \hat{\rvec}_{ij} \otimes \hat{\rvec}_{ij} - p_0 \mathbb{I}.
\end{aligned}
\end{align}
Each cell edge $(ij)$ contributes a force dipole of strength $\upell_{ij }\uptau_{ij}$. 
The physical stress depends both on the intrinsic tension $\uptau_{ij}$ and on how this force dipole is ``embedded'' in the plane (its length and orientation).
The total macroscopic stress is the sum of the tensile stress and the isotropic intracellular pressure (which we assume to be constant for the moment). 

The isogonal deformation tensor Eq.~\eqref{eq:isogonal-deformation-lattice} differs from the stress tensor Eq.~\eqref{eq:batchelor_general} only in the area factor $\tilde{\upa}$ vs $\upa$. Triangle and cell areas are related by $\upa = 2\tilde \upa \, \det \Fiso$, and therefore:
\begin{align}
\label{eq:isogonal_stress}
    \overline{\boldsymbol{\sigma}} &= p_0 \left( \frac{\Fiso}{\det \Fiso} - \mathbb{I} \right)
    \approx p_0\left( d\Fiso-\mathrm{Tr}[d\Fiso]\,\mathbb{I} \right)
\end{align}
For clarity, we restore here the dimensional factor $p_0$, which sets the overall stress scale, and expand for small deformation $\Fiso=\mathbb{I} +d \Fiso$ in the final step.

In the Voronoi configuration, $\Fiso=\mathbb{I}$ and Eq.~\eqref{eq:isogonal_stress} implies that $\overline{\bm\sigma}$ vanishes. 
The pressure $p_0$ cancels the local tensile stress, which is isotropic.
As a heuristic justification, consider the traction force $\mathbf{f}_{ij} = \uptau_{ij} \hat{\nvec}_{ij}$ across an interface between two cells with centroids $\mathbf{c}_i, \mathbf{c}_j$. In a regular lattice Voronoi configuration, $\mathbf{c}_i =\tvec_i$, so the momentum flux $|\mathbf{f}_{ij}|/|\mathbf{c}_i-\mathbf{c}_j| = \uptau_{ij}/|\tvec_i-\tvec_j|=1$ is isotropic and spatially constant.
This justifies the choice of $p_0$ as ``unit conversion factor'' from tension space to physical space. Indeed, by Eq.~\eqref{eq:dE} the pressure balances the contractile tensions, $p_0 \, \mathrm{d}  (\upell^2) = 2 p_0 \upell \, \mathrm{d}\upell  \sim  \uptau  \, \mathrm{d} \upell $. Hence, the pressure sets the overall scale between tension and physical space, $p_0 \sim  \uptau / \upell$.

We conclude that the Voronoi tiling, dual to the embedding $\tvec_i$ of the tension triangulation, is a macroscopically stress-free state that, given its uniqueness, may be adopted as a ``reference state''. A stress-free reference state is the first central element of conventional elasticity theory.
The second central element of conventional elasticity theory is a stress--strain relationship. 
Macroscopic stresses arise from isogonal deformations, which move and stretch active force dipoles (cell edges), even though microscopically, tensions are constant. 
These deformations can arise, for example, due to pulling forces applied at the edge of a tissue patch.
Equation~\eqref{eq:isogonal_stress} plays the role of an effective stress-isogonal strain relationship with shear modulus $p_0/2$, implying elastic macroscopic behavior. 
The stiffness is thus given by the mean edge tension divided by the cell size.
Notably, the macroscopic elastic response is \emph{isotropic} even if the microscopic tensions are anisotropic.
Thus, macroscopic (shear) deformation triggers a stress response, the hallmark of elastic behavior.
(The bulk modulus implied by Eq.~\eqref{eq:isogonal_stress} is negative, $-p_0/2$, because we assumed constant intracellular pressure. If cells resist compression, an additional contribution makes the total bulk modulus positive; see Sec.~\ref{sec:areas_isogonal}.)

For a hexagonal lattice of ordinary fluid foam cells, Eq.~\eqref{eq:isogonal_stress} reproduces known results: In a foam, all $\uptau_\mu$ are equal to the foam surface tension $\uptau$~\cite{Alexander1998,Weaire.etal2005,Davini2010}. The tension triangle area is $\tilde{\upa} = \sqrt{3} \, \uptau^2/4$ and the hexagonal cell area is $\upa = 3 \sqrt{3} \, \upell^2/2$.
This gives 
$p_0 = \uptau / (\sqrt{3} \, \upell) = \uptau \, \Pi /(4\mathrm{a})$
where $\Pi$ is the cell perimeter. For a foam bubble with radius $R$, this is the familiar Young--Laplace law for the pressure drop, $\Delta p = \uptau / 2R$.

A thought experiment on a conventional 2D fluid foam illustrates the importance of distinguishing macroscopic stress and the microscopic configuration of tensions. Each interface in the foam is under surface tension. Nonetheless, cutting a freely floating bubble raft is macroscopically stress-free: cutting it does not lead to macroscopic recoil. Hence, the macroscopic configuration can be stress-free while microscopic stresses are nonzero.
Importantly, the macroscopic stress in the Voronoi configuration vanishes, and the elastic response to deformation is isotropic independent of whether the tensions are equal or not, i.e., regardless of the microscopic tension anisotropy. 
In living tissues, stresses at different scales can be assessed by different types of laser ablation (or cutting) experiments (see Discussion).

\subsection{Adiabatic tissue flow}\label{sec:adiabatic_lattice}

During morphogenesis, cells adjust interfacial tensions to change tissue shape~\cite{Brauns.etal2024}.
How does a small change $d\uptau_{ij}$ in the tensions change the cell tiling? 
In the adiabatic limit, the cell tiling rapidly relaxes to force balance, compared to the timescale over which tensions change. Tissue flow is thus determined by how the tensions change the mechanical equilibrium state.

In a regular lattice, tension dynamics corresponds to the deformation of the (representative) tension triangle $(\uptau_{12},\uptau_{23}, \uptau_{31})$. The deformation rate of a triangle can be captured by a single matrix $\mathbf{M}$:
\begin{align}
    \frac{d\uptau_{ij}}{\uptau_{ij}} = \tfrac{1}{2} \hat{\tvec}_{ij}^T \cdot \mathbf{M} \cdot \hat{\tvec}_{ij},\quad   ij=12,23,31
\end{align}
To match the changed tensions, the tension vertices must move, $\tvec_i \mapsto \tvec_i +d\tvec_i$. We parametrize the motion via a continuous velocity field $\mathbf{v}(\tvecIt)$, with $\mathbf{v}(\tvecIt=\tvec_i) = d\tvec_i$.
The strain of the velocity field must match the tension triangle strain:
\begin{align}
    \label{eq:adiabatic_flow_lattice}
    \nabla\mathbf{v} + (\nabla\mathbf{v})^T - 2(\nabla\cdot\mathbf{v})\,\mathbb{I} = \mathbf{M} - \tfrac{1}{2}\mathrm{Tr}[\mathbf{M}]\,\mathbb{I}
\end{align}
(As we will explain in Sec.~\ref{sec:Beltrami}, it is only necessary to match the anisotropic tension strain -- isotropic strain corresponds to a rescaling, i.e., a change in pressure $p$).
Since the tension vertices are the centroids of the Voronoi cells, changes in the tensions deform the effective reference state of the tissue. This implies that the tissue can change shape (``flow'') due to tension dynamics, while resisting external forces like a solid.

\subsection{T1 transitions}\label{sec:T1s_lattice}

Large plastic deformations require cell rearrangements, known as T1-processes. A T1 process occurs whenever the length of a cell-cell interface vanishes, $\upell_{ij}=0$. 
In the Voronoi reference state, the length of interface $ij$ in a quadrilateral of cells $ikjl$ has a simple form (Fig.~\ref{fig:setup-Voronoi}, see also App.~\ref{app:Voronoi}):
\begin{align}
    \label{eq:cotan_length}
    \upell_{ij}^\mathrm{V} = |\rvec_{ijk}^\mathrm{V}-\rvec^\mathrm{V}_{ijl}| = \frac{\uptau_{ij}}{2} (\cot\tilde{\upgamma}_{ij}^k +\cot\tilde{\upgamma}_{ij}^l)
\end{align}
In the regular lattice, this simplifies to $\upell_{12} = \uptau_{12} \cot\tilde{\gamma}_{12}^3$.
By Eq.~\eqref{eq:cotan_length}, $\upell_{ij}^\mathrm{V} \geq 0$ entails 
\begin{align}
    \label{eq:Delaunay}
    \tilde{\upgamma}_{ij}^k + \tilde{\upgamma}_{ij}^l \leq \pi
\end{align}
This is called the \emph{Delaunay condition}~\cite{Delaunay_1934aa} (geometrically, at $\tilde{\upgamma}_{ij}^k + \tilde{\upgamma}_{ij}^l = \pi$, the circumcircles of triangles $ijk$ and $ijl$ coincide), and has previously been used in vertex-model simulations~\cite{Barton.etal2017}.
In the regular lattice, the Delaunay condition requires that $\tilde{\upgamma}_{12}^3\leq \pi/2$, i.e., non-obtuse triangles. In Sec.~\ref{sec:Voronoi_lattice}, we implicitly assumed the Delaunay condition is fulfilled. If the Delaunay condition is violated, a T1 transition is triggered that plastically modifies the connectivity of the tension triangulation.
Physically, Eq.~\eqref{eq:Delaunay} explains how tension dynamics control cell rearrangement, a question to which we will return in Sec.~\ref{sec:T1s}.

\section{Parameterizing force-balanced states of tension networks}\label{sec:Moebius_diagrams}

Let us now move from the regular lattice to arbitrary tension triangulations and cell tilings.
To this end, we use a generalization of Voronoi diagrams called \emph{M\"{o}bius diagrams} (MDs)~\cite{Boissonnat.etal2006}, introduced to tissue mechanics in Refs.~\cite{Moukarzel1997,Noll.etal2020} (the name is due to their close connection with M\"{o}bius transformations, which will be explained in Sec.~\ref{sec:pressure}).
The next two subsections focus on the method and formalism of MDs. Their physical meaning will be developed starting in Sec.~\ref{sec:constant_pressure}.

\subsection{M\"{o}bius diagrams}

We define a cell tiling from a set of \emph{vertex potentials} $\Phi_i(\rvec)\in\mathbb{R}$ as follows (see Fig.~\ref{fig:MD}):
\begin{align}
    C_i = \{\rvec \, : \Phi_i(\rvec) \,  \leq \Phi_j(\rvec) \; \forall j\neq i \}, \quad 
\end{align}
For the index of the cell containing point $\rvec$, we write:
\begin{align}
    \label{eq:i(r)_definition}
    i(\rvec) =\arg\min_i \Phi_i(\rvec) = i \text{ such that } \rvec\in C_i
\end{align}
Cell-cell interfaces $ij$ are the \emph{isosurfaces} $\Phi_i(\rvec) - \Phi_j(\rvec) = 0$. Thus, the interface normals are
\begin{align}
    \label{eq:interface_normals}
    \hat{\nvec}_{ij} = \frac{\nabla \Phi_i - \nabla\Phi_j}{|\nabla \Phi_i - \nabla\Phi_j|} 
\end{align}

In MDs, the vertex potentials are isotropic quadratic functions
\begin{align}
    \label{eq:iso_quadratic_potentials}
    \Phi_i(\rvec) = \upp_i\left(\tfrac{1}{2}|\rvec-\tvec_i|^2 - \uptheta_i \right)
\end{align}
We refer to $\upp_i>0$ as the scale factor, $\uptheta_i$ as the additive weight, and $\tvec_i$ as the seed point.
In the setting of Sec.~\ref{sec:lattice}, the seed points are the tension vertices. As we shall see below, the $\uptheta_i$ parametrize the isogonal deformations, and the $\upp_i$ correspond to the cellular pressures.
Since the potentials in Eq.~\eqref{eq:iso_quadratic_potentials} are isotropic, the interfaces are circular arcs, or, if $\upp_i = \upp_j$, straight lines.
MDs thus generate cellular tilings of circular-arc polygons \cite{Noll.etal2020}.

There are three special cases of MDs:
\begin{itemize}
    \item $\upp_i=1, \uptheta_i=0$: Voronoi diagrams. Each point $\rvec$ is assigned to the closest seed point $\tvec_i$.
    \item $\upp_i=1, \uptheta_i\neq 0$: Power diagrams. Like Voronoi diagrams, power cells have straight interfaces.
    \item $\upp_i\neq 1, \uptheta_i = 0$: Multiplicative Voronoi diagrams (MVDs).
\end{itemize}

\subsection{M\"{o}bius diagrams are force-balanced}

We now show that for any MD, there is a set of tensions $\uptau_{ij}$ and pressures $\upp_i$ so that Eq.~\eqref{eq:Young--Laplace} (Young--Laplace) and Eq.~\eqref{eq:tension_balance} (Young--Dupr\'{e}) are fulfilled. 
This is non-trivial: generic cell tilings are not force-balanced, no matter the values of tensions and pressures (App.~\ref{app:Noll_constraints}).

We set the interface tensions implied by the tiling to 
\begin{align}
    \label{eq:implied_tensions}
    \uptau_{ij} &= |\nabla \Phi_i - \nabla\Phi_j|
\end{align}
By Eq.~\eqref{eq:interface_normals}, Eq.~\eqref{eq:tension_balance} is then fulfilled \emph{automatically}:
\begin{multline}
    \left(\nabla \Phi_i - \nabla\Phi_j \right) + \left(\nabla \Phi_j - \nabla\Phi_k \right) + \left(\nabla \Phi_k - \nabla\Phi_i \right) \\
    =\uptau_{ij}{\hat{\nvec}}_{ij} + \uptau_{jk}{\hat{\nvec}}_{jk} + \uptau_{ki}{\hat{\nvec}}_{ki} = 0
\end{multline}
Explicitly, the implied tensions for Voronoi, power, and general M\"{o}bius diagrams read:
\begin{subequations}
    \label{eq:implied_tensions_explicit}
    \begin{align}
        \text{Voronoi \& Power }: \; \uptau_{ij}^2 &= |\tvec_i-\tvec_j|^2     \label{eq:implied_tensions_Voronoi} \\
        \text{Mult. Voronoi }: \; \uptau_{ij}^2 &= \upp_i\upp_j \, |\tvec_i-\tvec_j|^2 \label{eq:implied_tensions_mult_Voronoi} \\
        \text{General M\"{o}bius }: \; \uptau_{ij}^2 &= \upp_i\upp_j \, |\tvec_i-\tvec_j|^2 \label{eq:implied_tensions_MD} \\ 
        &+2\left(\upp_i - \upp_j\right) \left(\upp_i \uptheta_i - \upp_j\uptheta_j\right) \nonumber
    \end{align}
\end{subequations}
For Voronoi diagrams and power diagrams, one recovers Eq.~\eqref{eq:tension_embedding_lattice}: the seed points for a drawing of the tension triangulation. The meaning of the generalized Eqs.~\eqref{eq:implied_tensions_mult_Voronoi}-\eqref{eq:implied_tensions_MD} will be discussed in Sec.~\ref{sec:discrete_conformal}.

Next, the potentials in Eq.~\eqref{eq:iso_quadratic_potentials} ensure that tension and curvature are indeed constant along each interface.
If $\upp_i=\upp_j=1$, $\nabla \Phi_i - \nabla \Phi_j = \tvec_j-\tvec_i$: the interface is a straight line and has constant tension $\uptau_{ij} = |\tvec_i-\tvec_j|$. If $\upp_i\neq\upp_j$, the interface is a circular arc with center $\boldsymbol{\uprho}_{ij}$ and signed curvature radius $\mathrm{R}_{ij}$\footnote{
Their explicit form \cite{Noll.etal2020} is not needed here, but reads 
\begin{align}
\begin{aligned}
    \mathrm{R}_{ij}^{2} &= \frac{\upp_i\upp_j|\tvec_i-\tvec_j|^2 }{(\upp_i - \upp_j)^2} \!+2\frac{\upp_i\uptheta_i - \upp_j\uptheta_j}{\upp_i - \upp_j}, \;
    \boldsymbol{\uprho}_{ij} =  \frac{\upp_i\tvec_i - \upp_j \tvec_j }{\upp_i - \upp_j}
\end{aligned}
\end{align}}.
Direct computation shows
\begin{align}
    \label{eq:isosurface_gradients}
    \nabla \Phi_i - \nabla \Phi_j = (\upp_i - \upp_j) (\rvec - \boldsymbol{\uprho}_{ij})
\end{align}
In particular, the implied tension Eq.~\eqref{eq:implied_tensions} is well-defined since $|\rvec - \boldsymbol{\uprho}_{ij}|=\mathrm{R}_{ij}$, so $\uptau_{ij}$ is constant along the interface (for generic $\Phi_i$, the interfacial tension will vary).

Finally, we verify the Young--Laplace law. By Eq.~\eqref{eq:implied_tensions} and Eq.~\eqref{eq:isosurface_gradients}:
\begin{align}
    \upkappa_{ij} \uptau_{ij} = \frac{|\nabla \Phi_i - \nabla \Phi_j|}{\mathrm{R}_{ij}} = \upp_i - \upp_j
\end{align}
Thus, Eq.~\eqref{eq:Young--Laplace} is also fulfilled with pressures $\upp_i$, justifying our choice of notation.

\subsection{Forward and inverse problems for MDs}\label{sec:forward_inverse_MD}

All MDs are mechanically balanced. This raises the converse question: is every mechanically balanced cell tiling an MD?
First, note that the number of degrees of freedom of an MD precisely matches the number of mechanical variables.
We denote by $C, E$, and $V$ the numbers of cells, edges, and vertices of the cell tiling (Fig.~\ref{fig:duality}).
At each vertex, three edges meet, and each edge joins two vertices, so $ 3V = 2E$. Substituting into Euler's polyhedra formula $C-E+V=2$ yields $E\approx 3C, V\approx 2C$. MD have $4C$ parameters, $\tvec_i, \uptheta_i,\upp_i$, while pressures and tensions define $C+E = 4C$ variables.
This count suggests a one-to-one correspondence of MD and force-balanced cell tilings.
Indeed, App.~\ref{app:balanced_completeness} proves that, conversely, all tension-pressure force-balanced tilings admit an MD description. 
This proof also covers the area-perimeter model, whose energy minima are thus MDs\footnote{
In the area-perimeter literature, cells are often described by straight-edge polygons, neglecting the Young-Laplace law. The minima among straight-edge configurations are not MDs~\cite{Jensen.etal2020}.}.

The one-to-one correspondence between MDs and mechanical equilibria was originally pointed out in
Refs.~\cite{Moukarzel1997,Noll.etal2020} in the context of an \emph{inverse} problem: reconstructing pressures and tensions from given cell geometry.
This allows inferring pressures and tensions in living tissues, since cell shapes are readily observed under the microscope. In this work, we are interested in the corresponding \emph{forward} problem of determining the cell tiling given a set of (active) tensions.

The mechanical input data are the tensions $\uptau_{ij}$ and the equations of state $P(\upa_i)$.
The cell tiling is parametrized by MD parameters $(\tvec_i, \upp_i, \uptheta_i)$, from which the cell areas $\upa_i$ can be constructed.
The implied tensions and pressures of the MD must be consistent with the physical tension and equations of state, leading to the following nonlinear system of equations:
\begin{subequations}
\label{eq:MWPT-combined}
\begin{align}
    \uptau_{ij}^2 &= \upp_i \upp_j \big|\tvec_{i} - \tvec_{j}\big|^2  
     + 2(\upp_i-\upp_j) (\upp_i \uptheta_i- \upp_j\uptheta_j), \\
    \upp_i &= P\!\left(\upa_i\!\left[\{ \tvec_j, \upp_j, \uptheta_j\}_j\right]\right)
\end{align}
\end{subequations}
These are exactly $E+C=4C$ equations for $4C$ unknowns\footnote{
Along the boundary, there are two additional degrees of freedom per boundary cell. The boundary-facing interfaces depend on the weights $\upp_i, \uptheta_i$ of ``ghost cells'' along the boundary. These boundary degrees of freedom are needed to satisfy the mechanical boundary conditions -- either prescribed positions or traction forces.}. 
Throughout this manuscript, we will decode the physical and geometric meaning of Eq.~\eqref{eq:MWPT-combined}, which, at this point, is somewhat obscure. 
An efficient numerical implementation of Eq.~\eqref{eq:MWPT-combined} is an interesting and important challenge for future research.

\section{Stress tensor of balanced tilings}\label{sec:stress}

\begin{figure}
    \centering
    \includegraphics[width=0.7\linewidth]{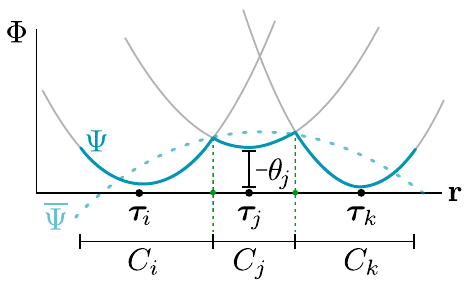}
    \caption{A M\"{o}bius diagram in one dimension with cells $C_i$, seed points $\tvec_i$, and additive weights $\uptheta_i$. The gray lines indicate the cell potentials $\Phi_i$, and the lower envelope $\Psi$ is shown in cyan. $\Psi$ has a kink across cell-cell interfaces and can be separated into cell-scale variations and a smooth part $\overline{\Psi}$.
    }
    \label{fig:MD}
\end{figure}

To characterize the mechanical state of an MD, we now compute its stress tensor.
The stress tensor of a cell tiling has two contributions. Within each cell $i$, pressure generates constant and isotropic stress $-\upp_i \mathbb{I}$. 
On cell-cell interfaces $ij$, there is tensile stress with magnitude $\uptau_{ij}$ and oriented along the cell interface. 
We write $\delta_{ij}(\rvec)$ for the $\delta$-function on cell-cell interface $ij$. 
The combined stress tensor density reads
\begin{align}
\begin{aligned}
    \boldsymbol{\sigma} &= -\upp_{i(\rvec)} \mathbb{I} + \sum_{ij}  \uptau_{ij}(\mathbb{I} - \hat{\nvec}_{ij} \otimes \hat{\nvec}_{ij}) \, \delta_{ij}(\rvec) \\
    &= -\upp_{i(\rvec)}\mathbb{I} - \mathbf{S} + \mathrm{Tr}[\mathbf{S}]\,\mathbb{I} \label{eq:stress_density}
\end{aligned}
\end{align}
where we define the tensor $\mathbf{S}$
\begin{align}
    \mathbf{S} &= \sum_{ij} \uptau_{ij}\hat{\nvec}_{ij} \otimes \hat{\nvec}_{ij} \, \delta_{ij}(\rvec) \label{eq:Sigma_definition}
\end{align}

\subsection{Airy stress function}\label{sec:envelope_airy}

We expect the stress tensor of a balanced tiling to be balanced as well, $\partial_a \sigma_{ab} = 0$. Any balanced stress can be represented by an Airy stress function
\begin{align}
    \sigma_{ab} = \epsilon_{ac}\epsilon_{bd} \partial_c \partial_d \psi = (\nabla^\perp\nabla^\perp \psi)_{ab}
\end{align}
where $\epsilon_{01}=-\epsilon_{10} =1$ is the antisymmetric symbol and $\nabla^\perp$ is the gradient, rotated by $90^\circ$.
Above, we showed that all balanced tilings are MDs. We now compute the Airy stress function from the MD data.

From the vertex potentials $\Phi_i$, one can define the \emph{lower envelope}  function (see Fig.~\ref{fig:MD}):
\begin{align}
    \label{eq:envelope}
    \begin{aligned}
    \Psi(\rvec) &= \Phi_{i(\rvec)}(\rvec) = \min_i \Phi_i(\rvec) 
\end{aligned}
\end{align}
The envelope function encodes much of the geometry of the tiling. In particular, the \emph{pressure-weighted displacement}
of cell points $\rvec$ from their seed points $\tvec_i$ reads:
\begin{align}
    \label{eq:seed_displacement}
    \mathbf{u}(\rvec) = \upp_{i(\rvec)}(\rvec-\tvec_{i(\rvec)}) = \nabla\Psi(\rvec)
\end{align}
Within each cell, the lower envelope is smooth, and its Hessian is
\begin{align}
    \nabla\nabla\Psi(\rvec) = \upp_i \mathbb{I}, \quad  \text{where} \quad \left(\nabla\nabla\Psi\right)_{ab} := \partial_a\partial_b \Psi
\end{align}
Across interfaces $ij$, $\Psi$ has a kink (see Fig.~\ref{fig:MD}). The discontinuity in the gradient is given by
\begin{align}
\begin{aligned}
    &\lim_{h\rightarrow 0}\nabla \Psi(\rvec - h \, \hat{\nvec}_{ij}) - \nabla \Psi(\rvec + h \, \hat{\nvec}_{ij})  \\
    &= \left(\nabla\Phi_i -\nabla\Phi_j \right)\rvert_{ij} = \uptau_{ij} \hat{\nvec}_{ij}
\end{aligned}
\end{align}
Therefore, on interfaces, the lower envelope's Hessian has a singular contribution:
\begin{align}
    \nabla\nabla\Psi(\rvec) = -\uptau_{ij} \hat{\nvec}_{ij} \otimes \hat{\nvec}_{ij} \,  \delta_{ij}(\rvec)
\end{align}
Comparing with Eq.~\eqref{eq:Sigma_definition}, and using a $2\times 2$ matrix identity, one concludes:
\begin{align}
    \label{eq:envelope_Airy}
    \boldsymbol{\sigma} = \nabla\nabla\Psi - \mathrm{Tr}[\nabla\nabla \Psi]\,\mathbb{I}
    = -\nabla^\perp \nabla^\perp \Psi
\end{align}
Thus, the envelope is (the negative of) the Airy function of the stress. In particular, the stress is automatically balanced. 
In terms of the seed-point displacement Eq.~\eqref{eq:seed_displacement}, one has
\begin{align}
    \label{eq:stress_seed_strain}
     \boldsymbol{\sigma} = \nabla\mathbf{u} - (\nabla\cdot \mathbf{u})\, \mathbb{I}
\end{align}

\subsection{Macroscopic stress}

The ``microscopic'' stress tensor Eq.~\eqref{eq:envelope_Airy} is defined in the sense of distributions, i.e., it must be integrated over a region $A$.  
We can split the microscopic envelope function into a smooth part (a moving average) and cell-scale, zero-average variations (Fig.~\ref{fig:MD}).
We can also directly compute the macroscopic stress by averaging:
\begin{align}
    \begin{aligned}
    \overline{\boldsymbol{\sigma}} &= \frac{1}{A} \int \boldsymbol{\sigma}(\rvec) d^2 r
     = \frac{1}{A} \int\left(\nabla\nabla\Psi - \Delta\Psi \,\mathbb{I} \right) d^2r \\
     &= \frac{1}{A} \oint_{\partial A} \left(\hat{\nvec}\otimes\nabla\Psi - (\hat{\nvec}\cdot\nabla \Psi) \,\mathbb{I} \right) d\ell
    \end{aligned}
    \label{eq:virial}
\end{align}

Thus, $\overline{\boldsymbol{\sigma}}$ is always a boundary integral -- only the cells that intersect $\partial A$ contribute. 
Non-zero macroscopic stress results from net displacement $\mathbf{u}$ between cells and their ``seed points.'' For example, $\overline{\bm{\sigma}}=0$ for any region $A$ over which $\mathbf{u}$ is periodic. More generally, Eq.~\eqref{eq:virial} leads to the bound:
\begin{align}
    \label{eq:stress_bound}
    |\overline{\boldsymbol{\sigma}}| \leq \sqrt{2} \, \frac{|\partial A|}{A} \, \max_{\rvec\in A} |\mathbf u (\rvec)|
\end{align}
where $|\partial A|$ is the perimeter of region $A$. If the seed-point displacement is bounded, the macroscopic stress thus vanishes as $A\rightarrow\infty$ since $|\partial A| /A \rightarrow 0$. 
In particular, Voronoi and multiplicative Voronoi diagrams are macroscopically stress-free: since $\tvec_i\in C_i$, the displacement is bounded by the cell size $\ell_C$, $|\mathbf{u}| \leq \upp_{\max} \ell_C$.
Below, we present a more detailed calculation that leads to a significantly tighter bound on the macroscopic stress.

\section{Tension networks with constant pressure}\label{sec:constant_pressure}

Sections~\ref{sec:Moebius_diagrams}--\ref{sec:stress} treated the tiling, defined by vertex potentials $\Phi_i$, as given, and the tensions $\uptau_{ij}$ as implicitly defined by Eq.~\eqref{eq:implied_tensions}.
Physically, however, the tensions are the control parameters, and the tiling results from mechanical equilibration. (As we will see, the pressures $\upp_i$ cannot be controlled independently. Instead, they are determined by passive (in)compressibility of the cells.) 
We now use the machinery of MDs to ask how a given set of tensions determines the cell tiling, generalizing the results obtained for the regular lattice in Sec.~\ref{sec:lattice}. We begin with the case of constant pressure.

As noted in the introduction, we consider the tension configuration as \emph{triangulation} dual to the cell tiling with vertices $i,j,\dots$ for each cell  (Fig.~\ref{fig:duality}). We will construct the cell tiling as an embedding of the tension triangulation, intrinsically defined by its edge lengths $\uptau_{ij}$, into ``physical'' $\rvec$-space.

\subsection{Flat triangulations and the Voronoi construction}
\label{sec:isogonal}

We begin with the simplest case: a flat triangulation that can be drawn in the 2D plane, like the regular lattice in Eq.~\eqref{eq:tension_embedding_lattice}:
\begin{align}
     \uptau_{ij}= |\tvec_i-\tvec_j|
\end{align}
We refer to the 2D vertices as an \emph{embedding} of the tension triangulation.
(Recall that by our choice of units, the $\tvec_i$ vertices have units of length and lie in the physical plane).
We will assume flatness throughout Sec.~\ref{sec:constant_pressure}-\ref{sec:pressure} and generalize in Sec.~\ref{sec:curvature}.

Interpolation on triangular meshes will be an important technical tool; for example, it defines discretizations of differential operators like the gradient (App.~\ref{app:interpolation}).
We use a basis of piecewise affine functions $\phi_i(\tvecIt)$, one per vertex, to interpolate a set of vertex values $\uph_i$:
\begin{align}
    h(\tvecIt) = \sum_i \uph_i \phi_i(\tvecIt),\quad (\nabla h) = \sum_i \uph_i\nabla\phi_i(\tvecIt)
    \label{eq:interpolation}
\end{align}

In Sec.~\ref{sec:Voronoi_lattice}, we saw that the Voronoi dual constructs a mechanically balanced cell tiling from the tension vertices $\tvec_i$.
We now recognize Eq.~\eqref{eq:Voronoi_definition} as a special case of MDs, Eq.~\eqref{eq:iso_quadratic_potentials}.
By Eq.~\eqref{eq:stress_bound}, the stress tensor in a Voronoi diagram is approximately zero; upon averaging over large areas $A$, tensile stress (localized to cell edges) and pressure (in cell interiors) cancel.
App.~\ref{app:stress} proves a stronger result: averaged over any tension triangle, the 2D Voronoi stress is exactly zero.
Thus, via the Voronoi construction, the tension triangulation defines a stress-free reference. As we show next, deformations away from the reference generate stress.

\subsection{Isogonal deformations and power diagrams}

\begin{figure}[b]
    \centering
    \includegraphics[scale=1.2]{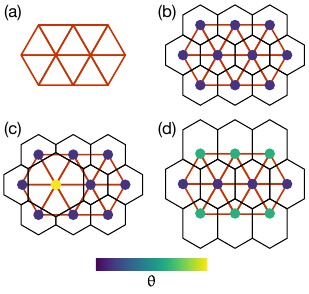}
    \caption{Internal and boundary isogonal modes. (a) tension triangulation. (b) reference cell tiling. (c) internal isogonal mode inflating a single cell without deforming the boundary of the cell patch. (d) global shear isogonal mode deforming the boundary.
    The color of the seed points corresponds to the magnitude of the isogonal mode.
    }
    \label{fig:iso-modes}
\end{figure}

The Young--Dupr\'{e} law Eq.~\eqref{eq:tension_balance} sets the orientations of cell interfaces. However, fixing the orientations of all cell interfaces does not fully determine cell tiling. 
Thus, for a single set of interfacial tensions $\uptau_{ij}$, there is a whole family of force-balanced cell tilings.

By Eq.~\eqref{eq:interface_normals}, the orientation of an interface normal is $\hat{\mathbf{n}}_{ij}\propto\nabla \Phi_i - \nabla\Phi_j$. Adding a constant $\uptheta_i$ to each vertex potential $\Phi_i$ therefore leaves interface orientations invariant, and generalizes Voronoi to \emph{power diagrams} (also known as Laguerre diagrams)~\cite{Aurenhammer1987}:
\begin{align}\label{eq:power_cell}
    \begin{aligned}
    C_i^\mathrm{P} &= \{\rvec \, : \, \Phi_i^\mathrm{P}(\rvec) < \Phi_j^\mathrm{P}(\rvec) \; \forall j\neq i \} \\
    \Phi_i^\mathrm{P} &=  \tfrac{1}{2} |\rvec-\tvec_i|^2 - \uptheta_i
\end{aligned}
\end{align}
Power diagrams are a special case of MDs in which $\upp_i=1$. Hence, the pressure is constant and
\begin{align}
    \uptau_{ij} = |\nabla\Phi_i^\mathrm{P}-\nabla\Phi_j^\mathrm{P}| =  |\tvec_i-\tvec_j|
\end{align}
App.~\ref{app:isogonal_exhaustive} shows that all constant-pressure equilibria are power diagrams.
Relative to the Voronoi reference $\uptheta_i=0$, these configurations are parametrized by the potential $\uptheta_i$.

How does $\uptheta_i$ deform the tiling? The deformation displaces, but does not rotate cell interfaces. This is precisely the definition of the isogonal deformations encountered in Sec.~\ref{sec:isogonal_lattice}. We thus refer to $\uptheta_i$ as the \emph{isogonal potential}. The simplest example is a linear potential, $\uptheta_i = \mathbf{b} \cdot \tvec_i$. It leads to a uniform displacement by $-\mathbf{b}$ of the power cells ($\mathbf{b} \cdot \tvec_i$ can be absorbed by $\rvec \mapsto\rvec +\mathbf{b}$ in $\Phi_i^\mathrm{P}(\rvec)$).
For a general isogonal mode, tricellular vertices are displaced relative to their Voronoi values by the gradient of $\uptheta_i$~\cite{DeGoes.etal2014}:
\begin{equation} \label{eq:isogonal_action}
    \rvec_{ijk}= \rvec_{ijk}^\mathrm{V} - (\nabla_{\!\tvec} \uptheta)_{ijk}.
\end{equation}
where $(\nabla_{\!\tvec} \uptheta)_{ijk}$ is the discrete gradient operator (see App.~\ref{app:interpolation}).
Intuitively, gradient displacements are curl-free, matching the fact that interfaces are not rotated.

Isogonal displacements are, however, \emph{not} conformal\footnote{In the literature, ``isogonal'' sometimes refers to maps that are either conformal or anti-conformal (angle-reverting). This is distinct from our usage of the term.}. In fact, isogonal deformations generally cause shear, as the examples in Fig.~\ref{fig:iso-modes} show.
Note that microscopically, i.e.\ on the level of the cell vertices, the isogonal displacements are highly non-affine.

\subsection{Cell areas and isogonal modes}\label{sec:areas_isogonal}

In the ATN model, the tensions $\uptau_{ij}$ are set intrinsically by motor molecule concentration. By contrast, the isogonal potential is determined by cell compressibility, encoded in the constitutive equation of state $\upp_i = P(\upa_i)$ that relates pressure $\upp_i$ and cell area $\upa_i$. For concreteness, we approximate the constitutive relation by 
\begin{align}
    \label{eq:constitutive_linear}
    P(\upa_i) = p_0\left( 1 - B(\upa_i-\upa_{0,i}) \right)
\end{align}
with non-dimensional bulk modulus $B$ and cell target areas $\upa_{0,i}$.
A limit case is cell incompressibility, $B\rightarrow\infty$, where the areas $\upa_i$ are fixed to $\upa_i=\upa_{0,0}$ for all cells $i$ and the pressures $\upp_i$ become Lagrange multipliers to enforce these constraints. These $C=\text{number of cells}$ equations determine the $C$ isogonal potentials $\uptheta_i$ (up to boundary conditions and the 3-dimensional subspace of isogonal modes that lead to rigid translation).

The isogonal mode can deflate and inflate cell areas under the constraint of vertex force balance (Fig.~\ref{fig:iso-modes}). We now compute how a change $d\uptheta_i$ affects the cell areas $\mathrm{a}_i$, following Ref.~\cite{DeGoes.etal2014} (see App.~\ref{app:Voronoi} for explicit formulas).
We break up the cell into pyramids with the seed point $\tvec_i$ as the tip and the interfaces $\upell_{ij}$ as the base, where $j$ denotes the neighboring cells. Since an isogonal deformation shifts the interfaces (Eq.~\eqref{eq:intersection_fraction}):
\begin{align}
    \label{eq:isogonal_area}
    d \upa_i = -\sum_{j \sim i} \upell_{ij} \frac{d \uptheta_j - d\uptheta_i}{\uptau_{ij}} := -(\Delta \uptheta)_i
\end{align}
Crucially, Eq.~\eqref{eq:isogonal_area} can be interpreted as a discrete Laplace operator $\Delta$. 
The finite-volume method defines the discrete Laplacian using the divergence formula, integrated over each cell $i$:
\begin{align}
 \label{eq:Laplace-from-div}
\begin{aligned}
    (\Delta \uptheta)_i :&=
    \int_{C_i} (\Delta \theta) d^2r = \int_{C_i} \div (\nabla \theta) d^2r \\
    &= \int_{\partial C_i} \nabla \theta \cdot \hat{\nvec} \, d\ell \approx \sum_{j \sim i}  \frac{\uptheta_j-\uptheta_i}{\uptau_{ij}} \, \upell_{ij}
\end{aligned}
\end{align}
since the cell normal $\hat{\nvec}$ is parallel to the triangulation edge vector $\tvec_{ij}$. 
Eq.~\eqref{eq:Laplace-from-div} plays an important role throughout this work. The continuous Laplace-Beltrami operator on a surface depends on its geometry; similarly, the discrete Laplacian depends on the tiling geometry through the interface lengths $\upell_{ij}$. In a Voronoi tiling, the finite-volume Laplacian is isotropic and coincides with a finite-element discretization (App.~\ref{app:discrete_Laplacian}); that is, the discrete Laplacian is the divergence of the discrete gradient Eq.~\eqref{eq:interpolation}.
Isogonal modes $\uptheta_i$ stretch interfaces and produce an anisotropic Laplacian (App.~\ref{app:isogonal_laplacian}; physically, stretching interfaces makes the cell-to-cell diffusivity anisotropic).

Overall, the vertex displacement equals the discrete gradient of $\uptheta_i$, Eq.~\eqref{eq:isogonal_action}, and the cell area change equals the discrete Laplacian of $\uptheta_i$, Eq.~\eqref{eq:isogonal_area}.
Fulfilling the constraint $\upa_i=\upa_{0,i}$ on cell areas becomes a Poisson problem\footnote{Since discrete Laplacian $\Delta$ itself depends on $\uptheta_i$, the Poisson problem is formulated differentially.}
\begin{align}
    \label{eq:Poisson_area}
    (\Delta d\uptheta)_i=\upa_i-\upa_{i,0}
\end{align}
Indeed, Ref.~\cite{Aurenhammer.etal1998} showed that for any set of target areas, a power diagram exists with corresponding cell areas.
If the area constraints are satisfied, $P(\upa_i) = p_0$ and the tiling still has constant pressure.

If, on the other hand, individual cells are fully compressible, $B=0$, the isogonal mode is an unconstrained soft mode. This case was studied in Ref.~\cite{Noll.etal2017} (a global, constant pressure $p_0$ is instead provided through a total area constraint).
However, while all isogonal deformations preserve the vertex angles $\upgamma_{ij}^k$, only deformations that are purely interior (i.e., leave the tissue boundary fixed) are true zero modes of the energy $dE = \sum_{ij} \uptau_{ij}d\upell_{ij} - p_0\sum_i d\upa_i$ (see Fig.~\ref{fig:iso-modes} for examples of purely interior and boundary isogonal modes). 
Isogonal deformations of the tissue boundary generically change the energy. As we will see next, the tissue acts like an elastic material with shear modulus $p_0/2$, and the boundary modes are determined by the stress boundary conditions, for example, stress-free boundary conditions or external traction forces.

\subsection{Stress and isogonal deformation}\label{sec:isogonal_stress_legendre}

Conceptually, the isogonal mode defines a \emph{map} from the triangulation (vertices $\tvec_i$) to the cell tilings (centroids $\rvec_i$), defined by Eq.~\eqref{eq:isogonal_action}.
This places the power diagrams in the context of our overall program of constructing balanced cell tilings from the tension triangulation.
For Voronoi diagrams, the macroscopic stress vanishes.
We now show that deformations relative to Voronoi generate stress, generalizing Eq.~\eqref{eq:isogonal_stress} (we assume that the pressure remains constant; the effect of cell (in)compressibility will be discussed below).

For the case of power diagrams, i.e.\ when $\upp_i=1$, the Airy stress function Eq.~\eqref{eq:envelope} reads:
\begin{align}
    \label{eq:envelope_legendre}
\begin{aligned}
    \Psi(\rvec) &= \tfrac{1}{2} |\rvec|^2 + \min_i \{ \underbrace{\tfrac{1}{2}|\tvec_i|^2 -\uptheta_i}_{\displaystyle =: \mathrm{g}(\tvec_i)} - \tvec_i\cdot\rvec \} \\
    &= \tfrac{1}{2}|\rvec|^2 - \mathrm{g}^*(\rvec)
\end{aligned}
\end{align}
where we introduce the function $\mathrm{g}(\tvec_i)$ and recognize its Legendre transform:
\begin{align}
    \label{eq:Legendre_definition}
    \mathrm{g}^*(\rvec) = \max_{\tvec_i} \{\rvec\cdot\tvec_i - \mathrm{g}(\tvec_i) \}
\end{align}
As we will show next, the functions $-|\rvec|^2/2$ and $\mathrm{g}^*(\rvec)$ are the Airy functions of pressure and the tensile stress, respectively.
The function $\mathrm{g}^*$ is piecewise linear. Since $\nabla\mathrm{g}^* = \tvec_{i(\rvec)}$, the seed-point displacement field reads
\begin{align}
    \label{eq:isogonal_deformation_general}
    \begin{aligned}
    \mathbf{u} &= \mathbf{r} - \nabla\mathrm{g}^* \approx -\nabla\uptheta,
    \end{aligned}
\end{align}
In the last line, we approximated for small isogonal displacement. The derivative of the discrete potential $\uptheta_i$ is defined by linear interpolation, Eq.~\eqref{eq:interpolation}.
By Eq.~\eqref{eq:stress_seed_strain}, the stress tensor reads:
\begin{align}
\begin{aligned}
    \label{eq:isogonal_strain_legendre}
    \boldsymbol{\sigma} &= \nabla\mathbf{u} - (\nabla\cdot\mathbf{u}) \mathbb{I}
    = \nabla^\perp\nabla^\perp \mathrm{g}^* - \mathbb{I} \approx \nabla^\perp\nabla^\perp \uptheta
\end{aligned}
\end{align}
We conclude that the isogonal potential defines the effective Airy function, linking the geometry of the tiling to its mechanics.
Eq.~\eqref{eq:isogonal_strain_legendre} is an effective stress-strain relationship, with shear modulus $p_0/2$.
It recovers the lattice result Eq.~\eqref{eq:isogonal_stress} and supports the picture of emergent elasticity for arbitrary tension triangulations. The isogonal deformation tensor due to the displacement Eq.~\eqref{eq:isogonal_deformation_general} is:
\begin{align}
    \Fiso^{-1} &= \nabla\nabla \mathrm{g}^* = (\nabla\nabla \mathrm{g})^{-1} \approx \mathbb{I} + \nabla\nabla\uptheta
\end{align}
since the Legendre transform inverts the Hessian of a function.
In terms of $\Fiso$, Eq.~\eqref{eq:isogonal_strain_legendre} becomes, using a 2D matrix identity: 
\begin{align}
    \label{eq:isogonal_strain_legendre_F}
    \boldsymbol{\sigma} &=-\Fiso^{-1} - (1 - \mathrm{Tr}[\Fiso^{-1}])\,\mathbb{I} = \frac{\Fiso}{\det \Fiso} - \mathbb{I}
\end{align}
Note that Eq.~\eqref{eq:isogonal_strain_legendre_F} can also be read as a ``stress transformation law'' that determines how the Batchelor stress Eq.~\eqref{eq:batchelor_general} changes as cell interfaces (force dipoles) are moved and stretched by the isogonal mode. 
Even though the isogonal deformation is highly non-affine, one finds a simple transformation law. It automatically results in a balanced stress, since $\nabla\cdot \left(\Fiso / \det \Fiso \right) =0$, a fact known as the Piola identity~\cite{Kupferman.Shachar2018} ($\Fiso / \det \Fiso$ is the so-called cofactor or adjugate matrix of the inverse map $\Fiso^{-1}$).

\subsubsection{Quadratic isogonal modes}

Let us consider a simple setting where $\mathrm{g}^*$ and the averaged stress $\overline{\boldsymbol{\sigma}}$ can be computed explicitly.
By Eq.~\eqref{eq:isogonal_action}, a constant $\uptheta_i$ does not alter the tiling, and a linear $\uptheta_i$ simply creates a rigid translation.
The simplest non-trivial isogonal potential is thus quadratic
\begin{align}
    \label{eq:quadratic_weights}
    \uptheta_i = \tfrac{1}{2} \tvec_i^T \cdot \bm{\Theta} \cdot \tvec_i, \quad \mathrm{g}(\tvec) = \tfrac{1}{2}\left(|\tvec|^2 - \tvec^T\cdot \bm{\Theta} \cdot \tvec \right)
\end{align}
with a constant matrix $\bm{\Theta}$. General isogonal potentials can be locally approximated as quadratic. 

For the macroscopic stress $\overline{\boldsymbol{\sigma}}$, we replace maximization over the seeds $\tvec_i$ in Eq.~\eqref{eq:Legendre_definition} with maximization over the continuous vector $\tvec$. This replaces $\mathrm{g}^*$ with its smooth part, discarding a cell-scale remainder term:
\begin{align}
\label{eq:isogonal_stress_quadratic}
\begin{aligned}
    \overline{\mathrm{g}}^*(\rvec) &\approx \tfrac{1}{2} \rvec^T\cdot(\mathbb{I}-\bm{\Theta})^{-1}\cdot \rvec, \quad  \bar{\Fiso} = \mathbb{I} - \bm{\Theta}, \\
    \overline{\bm{\sigma}} &\approx 
    -\bar{\Fiso}^{-1} - (1 - \mathrm{Tr}[\bar{\Fiso}^{-1}])\,\mathbb{I}
\end{aligned}
\end{align}

App.~\ref{app:isogonal_proof_new} shows that Eq.~\eqref{eq:isogonal_stress_quadratic} is, in fact, exact when the isogonal mode is quadratic.
{The reason for this is that the cells of a quadratic power diagram can be mapped to the cells of a (deformed) Voronoi diagram:
\begin{align}
    \label{eq:quadratic_power_to_Voronoi}
    \begin{aligned}
    \rvec' &= \Fiso^{-1/2}\cdot \rvec, \quad \tvec_i' = \Fiso^{1/2}\cdot \tvec_i \\
  \Rightarrow\;  C_i^\mathrm{P} &= \Fiso^{1/2} \cdot \left\{\rvec' : |\rvec'-\tvec_i'|^2 <   |\rvec'-\tvec_j'|^2
    \right\}  \\
    \end{aligned}
\end{align}
In the Voronoi tiling, the tensile stress is constant and isotropic. The ``stress-transformation law'' Eq.~\eqref{eq:isogonal_strain_legendre_F} arises from how individual force dipoles $\upell_{ij}'\uptau_{ij} \hat{\rvec}_{ij}'\otimes\hat{\rvec}_{ij}'$ (cell interfaces) are transformed by the map $\Fiso^{1/2}$.

\subsubsection{Convexity of the Airy function}

As it results from a Legendre transform, the Airy function $\mathrm{g}^*$ is automatically convex, so the tensile stress is positive definite.
Geometrically, convexity implies that cell interface lengths are positive, since $\nabla\nabla \overline{\mathrm{g}}^* \propto \upell_{ij}$.
Mechanically, $\upell_{ij}\geq0$ means that interfaces cannot exert a pushing force, which would lead to a buckling instability. T1 transitions, which are triggered when $\upell_{ij} = 0$, thus constitute a dynamical mechanism that ensures convexity.
As an aside, the convexity requirement leads to efficient algorithms for computing Voronoi and power diagrams from a set of seed points by means of convex hulls~\cite{Aurenhammer1987}.

\subsection{Duality between tension networks and granular materials}{\label{sec:legendre}}

The Legendre transform Eq.~\eqref{eq:envelope_legendre} is a manifestation of a deeper ``mechanical duality'' in which the roles of tension space and physical space are reversed.
One can consider the triangulation as a truss network or as the contact pattern of a granular assembly of frictionless disks, with a force $\upell_{ij}$ on each triangulation link. The force balance condition for the dual reads
\begin{align}
    \label{eq:mechanical_balance}
    \sum_{j\sim i} \upell_{ij} \hat{\tvec}_{ij} = 0
\end{align}
Because tiling polygons are closed, the dual is also in mechanical balance. Eq.~\eqref{eq:mechanical_balance} reflects the \emph{Maxwell--Cremona} correspondence~\cite{Maxwell1864} between the set of balanced link-forces on a graph and the graph's dual cell tiling (see Ref.~\cite{Erickson.Lin2020} for a modern account).
In the mechanical dual, the isogonal modes thus parameterize the set of local forces for a fixed material geometry (they are known as ``wheel moves'' in the granular materials literature~\cite{Bi.etal2015a}).
In a tension network, different isogonal modes instead correspond to different geometries (cell shapes and areas; Fig.~\ref{fig:iso-modes}) at fixed local forces.

The duality interchanges the roles of $\uptau_{ij}$ and $\upell_{ij}$ and can be thought of as a Legendre transform, since the elastic energy differential of the cell tiling $d{E} = \sum_{ij} \uptau_{ij} d\upell_{ij}$ is transformed to $dE^* = \sum_{ij} d(\uptau_{ij}\upell_{ij})-dE = \sum_{ij} \upell_{ij} d\uptau_{ij}$. 
The Legendre picture goes further: App.~\ref{app:duality} explains that the Airy stress function of the dual is $\mathrm{g}(\tvec)$~\cite{DeGoes.etal2013}, i.e., the Legendre transform of the tension network Airy function $\mathrm{g}^*(\rvec)$. Thus, the primal and dual stresses are inverses of one another.

Discrete stress functions and force tilings have an important tradition in granular materials~\cite{Bi.etal2015a}. In particular, Blumenfeld \emph{et al.} introduced a discrete stress potential for granular materials~\cite{Ball.Blumenfeld2002} and mapped it to solid foams~\cite{Blumenfeld2003}. DeGiuli \emph{et al.} developed a related scheme~\cite{DeGiuli.McElwaine2011,DeGiuli.Schoof2014}.
Ref.~\cite{Blumenfeld2003}'s calculus is closely related to ours\footnote{Up to a $\pi/2$ rotation, their ``loop forces'' correspond to the tension vertices $\tvec_i$, and their stress vector potential to the gradient of the envelope function $\nabla\Psi$.}.
However, physically, Ref.~\cite{Blumenfeld2003} allows the solid foam cell walls to support shear forces.
Similarly, in generic granular materials, grains can exert non-central forces onto each other. As a result, torque balance becomes a non-trivial constraint, and the stress tensor is not automatically symmetric. Torque balance has been treated in a mean-field approximation in Ref.~\cite{DeGiuli.Schoof2014}. 
In ATNs, cell interfaces only support tensile forces, but no bending. Thus, force balance automatically implies torque balance. The stress tensor is automatically symmetric and derives from an Airy function; no additional closure equation is needed.

On a more conceptual level, Refs.~\cite{Ball.Blumenfeld2002,DeGiuli.McElwaine2011} aim to reconstruct the local forces from a given granular geometry. Here, we aim to construct the cell geometry from given local tension forces. Moreover, the intracellular pressure $\upp_i$ has no counterpart in the mechanical dual. We next turn to the geometric and mechanical implications of pressure gradients, which are thus a feature unique to tension networks.

\section{Pressure differences and conformal maps}{\label{sec:pressure}}

So far, we constructed constant-pressure tilings. Recall that the overall length scale of the cell tiling is set by pressure--tension balance, $\upell \sim  \uptau /\upp$.
This suggests that a \emph{non-uniform} pressure field might act as a \emph{local} scale factor. Mathematically, local scale changes correspond to conformal deformations.
Indeed, force balance at the vertices of a tension network is a local angle constraint, Eq.~\eqref{eq:angle_complement}, invariant under scaling and rotation of the cell tiling, i.e., under conformal maps. 
However, a spatially varying scale factor bends cell interfaces. Normal force balance along the interface then requires a pressure difference across it (Young--Laplace law).

We now build on this intuition to show that a force-balanced cell tiling with pressure gradients can be transformed to a uniform-pressure cell tiling by a (suitably discretized) conformal map. 
We can therefore parametrize cell states with the same tensions, but different pressures, via a conformal map (just as curl-free isogonal maps parameterize states with the same tensions, but different macroscopic stress).
The Cauchy--Riemann equations for the conformal map turn out to be a continuum form of the Young--Laplace law~\cite{Claussen.etal2026}.
For fluid foams, the link between mechanics and conformal maps has been demonstrated in beautiful experiments~\cite{Drenckhan.etal2004}.

To make this idea rigorous, we deploy a new piece of mathematical machinery, \emph{discrete conformal maps}~\cite{Springborn.etal2008}. 
We will show that discrete conformal maps are closely related to M\"{o}bius diagrams, a new result that may be of independent mathematical interest.
Tilings with pressure differentials correspond to M\"{o}bius diagrams with non-trivial scale factors $\upp_i\neq 1$ which go beyond the Voronoi and power diagrams of Sec.~\ref{sec:constant_pressure}.

\subsection{Conformal invariance of force balance}

A conformal map $\rvec' = \boldsymbol{f}(\rvec)$ is locally a scaling plus a rotation:
its Jacobian is $\nabla \boldsymbol{f} = \lambda \mathbf{R}$, where $\lambda$ is the \emph{conformal factor} and $\mathbf{R}$ a rotation matrix.
To describe conformal maps, it can be useful to describe coordinates $(r_1, r_2)$ with complex numbers $z = r_1 + i r_2, \; \bar{z} = r_1 - i r_2$. Conformal maps then become analytic functions $f(z)$, and the conformal factor is $\lambda = |\partial f/\partial z|$.
We now show that applying a conformal map to a balanced cell tiling generates another balanced tiling with the same tensions $\uptau_{ij}$, but changed pressures~\cite{Weaire1999,Mancini2005}.
(Note that physically, conformal deformations are only possible if cell pressure depends on cell area ($B>0$); otherwise, the required pressure gradients cannot be generated.) 

\begin{figure}
    \centering
    \includegraphics[width=\linewidth]{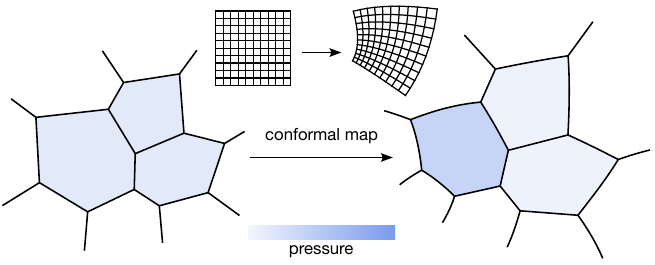}
    \caption{
    A discrete conformal map of a polygonal cell tiling produces a circular-arc polygonal tiling. The discrete conformal map is parametrized by a scale factor for each cell, which is related to intracellular pressure via the Young--Laplace law.
    }
    \label{fig:conformal-map}
\end{figure}

Tension force balance at vertices, Eq.~\eqref{eq:tension_balance},
is trivially invariant, since a conformal map only rotates the local interface normals $\hat{\nvec}_{ij}$. The Young--Laplace law is non-trivial, since cell-cell interfaces generally acquire line curvature $\upkappa_{ij}$. Indeed, consider a curve $\bm{\gamma}(s)$, parametrized by arclength $s$, with local normal $\hat{\bm{n}}(s)$ and curvature $\kappa(s)$. A conformal deformation changes the curvature by~\cite{wikipedia2026}:
\begin{align} \label{eq:curvature_transform}
    \kappa'(s) = \frac{\kappa(s)}{\lambda(\bm{\gamma}(s))} - \hat{\bm{n}}(s)\cdot \nabla\frac{1}{\lambda}\Big|_{\bm{\gamma}(s)}
\end{align}
where $\kappa(s) >0$ indicates bulging in the direction of $\hat{\bm{n}}(s)$. In fact, this result also holds for the mean curvature of a surface in 3D.

Normal force balance along the curved cell interfaces requires a pressure difference, determined by the Young--Laplace law
\begin{align}
    \upp_i(\bm{\gamma}(s))-\upp_j(\bm{\gamma}(s)) = \upkappa_{ij}(s) \uptau_{ij}.
    \label{eq:Young_Laplace}
\end{align}
We sum Eq.~\eqref{eq:Young_Laplace} around a vertex $ijk$. This computation can be thought of as a discrete ``curl''~\cite{Desbrun.etal2005} (see App.~\ref{app:balanced_completeness}). 
This results in a consistency equation for the curvatures evaluated \emph{at the vertex}:
\begin{align}
    \uptau_{ij}\upkappa_{ij} + \uptau_{jk}\upkappa_{jk} + \uptau_{ki}\upkappa_{ki} = 0
    \label{eq:Young_Laplace_vertex}
\end{align}
For example, curvatures that are all the same sign (a ``spiral'' shape) are not possible. Eq.~\eqref{eq:Young_Laplace_vertex} implies that the curvatures are curl-free and must derive from a potential, namely, pressure. 
Eq.~\eqref{eq:curvature_transform} is the key to the conformal invariance of force balance, since it transforms as:
\begin{align}
    \label{eq:Young_Laplace_vertex_transformed}
    \sum_\mathrm{cyc.} \uptau_{ij} \upkappa_{ij}' = \frac{1}{\lambda_{ijk}}\sum_\mathrm{cyc.} \uptau_{ij} \upkappa_{ij} - \left[ \sum_\mathrm{cyc.} \uptau_{ij} \hat{\bm{n}}_{ij} \right] \cdot \nabla \frac{1}{\lambda} \Big|_{ijk} = 0
\end{align}
where the subscript ${ijk}$ indicates that $\lambda$ and its gradient are evaluated at the vertex.
The second term vanishes due to tension force balance Eq.~\eqref{eq:tension_balance}. We conclude that Eqs.~\eqref{eq:Young_Laplace_vertex} and \eqref{eq:tension_balance} are invariant under conformal maps, which thus preserve force balance at vertices.

We now check how conformal maps change cell tiling and mechanics. We begin with an approximate continuum analysis, supposing also that the undeformed tiling has constant pressure $p_0=1$.
First, the local scaling changes cell areas. By expanding $\lambda(\rvec)$ around the cell seed point $\tvec_i$, one finds that the conformally deformed cell area is:
\begin{align}
    \label{eq:cell_areas_conformal}
    \begin{aligned}
    \mathrm{a}_i' &= \int_{C_i}\lambda^2(\rvec)d^2r \\ & \approx 
    \lambda(\tvec_i)^2 \mathrm{a}_i \,\left(1 + 2(\nabla\log\lambda)_i \cdot \mathbf{u}_i + \dots \right)
    \end{aligned}
\end{align}
where $\mathbf{u}_i$ is the cell centroid-seed point displacement.

Second, the cellular pressures must change, since the deformed interfaces are curved. We combine Eq.~\eqref{eq:curvature_transform} with the Young--Laplace law Eq.~\eqref{eq:Young_Laplace} for an interface $ij$:
\begin{align}
    \begin{aligned}
    \hat{\bm{n}}(s) \cdot  \nabla_{\!\rvecIt} \lambda^{-1}(\rvecIt)
    &= -\kappa'(s) \\
    &= \frac{\upp_j' - \upp_i'}{\uptau_{ij}} 
    \approx \hat{\bm{n}}(s) \cdot \nabla_{\!\tvecIt} p(\tvecIt)
    \end{aligned}
\end{align}
In the last line, we approximated the pressure differences by the gradient of a continuous pressure field $p(\tvecIt)$ on the triangulation. 
In the Voronoi state, cell centroids and triangulation vertices coincide, so $\rvecIt\approx\tvecIt$. We conclude:
\begin{align}
    \label{eq:pressure_transform_Voronoi}
    p(\rvecIt)=1/\lambda(\rvecIt)
\end{align}
The pressure is the inverse of the conformal factor (isogonal deformations lead to an additional correction term\footnote{In general, tension space $\tvecIt$ and cell space $\rvecIt$ differ due to isogonal deformation, Eq.~\eqref{eq:isogonal_deformation_general}. Since $\nabla_{\!\tvecIt} = \Fiso \,\nabla_{\!\rvecIt}$, one has 
\begin{align}
    \label{eq:pressure_transform}
    \Fiso \cdot \nabla_{\rvecIt} p =  \nabla_{\rvecIt} \lambda^{-1}
\end{align}
The general solution to Eq.~\eqref{eq:pressure_transform} is $p(\rvec) = \lambda^{-1}(\rvec+\nabla\theta)$. For small isogonal deformations:
\begin{align}
    p(\rvec) \approx \frac{1}{\lambda} \left( 1 - \nabla \log\lambda \cdot \nabla \theta \right).
\end{align}
The isogonal displacement ``displaces'' the pressure away from its Voronoi value $p=1/\lambda$.}).

Finally, let us compute how the conformal deformation changes the macroscopic stress.
Let $\nabla \bm{f} = \lambda \cdot \mathbf{R}$ be the Jacobian of the map. This scale-rotation transforms the stress in Eq.~\eqref{eq:batchelor_general}.
Area and edge lengths are rescaled by $\lambda$, $\mathbf{R}$ rotates the interfaces $\hat{\rvec}_{ij}$, and pressure changes to $p_0/\lambda$. The tensions remain the same.
Hence, the stress in the conformally deformed tiling is
\begin{align}
    \label{eq:stress_conformal_qualitative}
   \overline{\boldsymbol{\sigma}}' &= \frac{1}{\lambda^2 A} \,\mathbf{R}\cdot\left(\sum_{(ij)\in A} (\lambda \upell_{ij}) \uptau_{ij} \hat{\rvec}_{ij} \otimes \hat{\rvec}_{ij} \right) \cdot \mathbf{R}^T - \frac{p_0}{\lambda} \mathbb{I} \nonumber \\
   &=\frac{1}{\lambda} \mathbf{R}\cdot \overline{\boldsymbol{\sigma}} \cdot \mathbf{R}^T
\end{align}
Conformal deformations rotate and scale the stress -- they do not create any stress on their own.

\subsection{Möbius transformations}

Overall, these results suggest that force-balanced cell tilings with a fixed set of tensions, but different cell pressures, can be mapped into one another via conformal transformations.
However, for a general conformal map, the curvature along a mapped interface is not constant (Eq.~\eqref{eq:curvature_transform}). The Young--Laplace law Eq.~\eqref{eq:Young_Laplace} would therefore require intracellular pressure gradients. 

To resolve this issue, we consider a special type of conformal maps: Möbius transformations (MTs) $M(\rvec)$. These are compositions of scale-rotations, translations, and inversions $\rvec \mapsto \rvec / |\rvec|^2$.
A generic conformal map can be locally approximated by an MT to second order.
MTs are the only conformal maps which map circles to circles, as one can easily verify for these ``elementary'' transformations (straight lines are considered as generalized circles through $\rvec=\infty$, the ``point at infinity'').

Algebraically, this is a consequence of the following important identity. Consider the inversion $M(\rvec) = \rvec/|\rvec|^2$, whose conformal factor is $\lambda(\rvec) = 1/ |\rvec|^2$. An inversion satisfies 
\begin{align}
\label{eq:inversion_identity}
\begin{aligned}
    |M(\mathbf{x}) - M(\mathbf{y})|^2 &= \left\lvert \frac{\mathbf{x}}{|\mathbf{x}|^2} -\frac{\mathbf{y}}{|\mathbf{y}|^2} \right\rvert^2 = \frac{|\mathbf{x}-\mathbf{y}|^2}{|\mathbf{x}|^2 |\mathbf{y}|^2}  \\
    &= \lambda(\mathbf{x})\lambda(\mathbf{y}) \, |\mathbf{x}-\mathbf{y}|^2
\end{aligned}
\end{align}
In fact, Eq.~\eqref{eq:inversion_identity} holds for any MT.

\begin{figure}
    \centering
    \hspace{-1.1cm}
    \includegraphics[width=\linewidth]{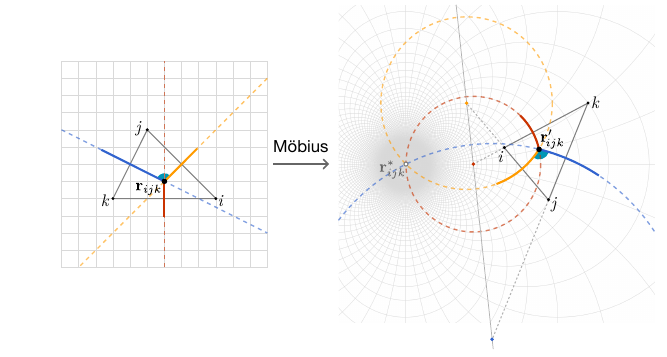}
    \includegraphics[width=\linewidth]{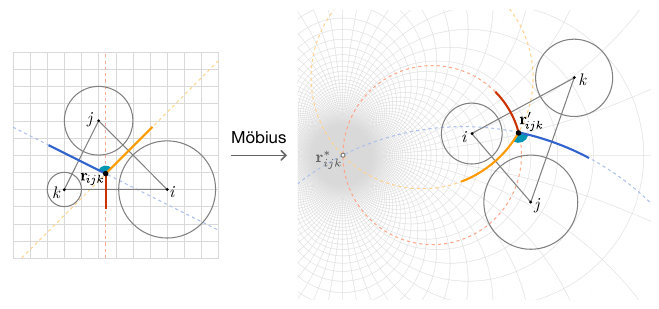}
    \caption{
    (Top)~Möbius transformation of a triangle and the dual Voronoi edges (colored) and vertex $\rvec_{ijk}$.
    The straight Voronoi edges become circular arcs. When completed, these circles meet in $\rvec_{ijk}^*$, the image of the point at infinity. Since the three circles intersect in two points, their centers lie on a straight line. Vice versa, any circular-arc diagram can be (locally) mapped to a straight-edge diagram by the inversion about $\rvec_{ijk}^*$.
    (Bottom)~Möbius transformation of a single triangle decorated by power circles of radius $\sqrt{2\uptheta_i}$ with the corresponding vertex and edges of the power diagram. The Möbius transformation generates a new set of power circles, and the power diagram is mapped to a general M\"{o}bius diagram.}
    \label{fig:Moebius-triangle}
\end{figure}

Due to circle preservation, applying an MT to a Voronoi tiling yields a tiling of circular arc polygons with the same vertex angles as the original tiling (Fig.~\ref{fig:Moebius-triangle}). 
By the argument of the preceding section, this tiling is in mechanical balance. Because the interfaces are circular arcs, the implied pressure inside each cell is constant.
Therefore, MTs are \emph{bona fide} force-balance modes at the discrete level.

\subsubsection{M\"{o}bius covariance of M\"{o}bius diagrams}

Since the conformally deformed tiling is in force-balance, it should admit a description as an MD (Sec.~\ref{sec:forward_inverse_MD}).  
Crucially, MDs are ``covariant'' under MTs: applying an MT to an MD results in another MD with transformed parameters $\tvec_i', \upp_i', \uptheta_i'$.

The reason for this lies in the geometry of power cells $C_i^\mathrm{P}$, defined by potentials $\Phi_i^\mathrm{P} = \tfrac{1}{2}|\rvec-\tvec_i|^2 - \uptheta_i $, Eq.~\eqref{eq:power_cell}.
Geometrically, $\Phi_i^\mathrm{P}$ measures half the squared tangential distance from $\rvec$ to a \emph{power circle} circle $S_i$ of radius $\sqrt{2\uptheta_i}$ around $\tvec_i$~\cite{Aurenhammer1987,DeGoes.etal2014} (Fig.~\ref{fig:Moebius-triangle}, top left).
We write $\Phi^\mathrm{P}(\rvec, S_i) = \Phi^\mathrm{P}_i(\rvec)$ for this \emph{power distance}.

Applying an MT to the set of power circles $S_i$ yields a new set of power circles $S_i' = M(S_i)$ (Fig.~\ref{fig:Moebius-triangle}, top right). These new power circles define a deformed MD:
\begin{subequations}
\label{eq:power_transform}
\begin{align}
    \Phi_i'(\rvec') &= \upp_i \,\Phi^\mathrm{P}\big(\rvec', M(S_i)\big) = \upp_i\left(\tfrac{1}{2}|\rvec'-\tvec_i'|^2-\uptheta_i'\right) \\
    \tvec_i' &= \text{center}\big(M(S_i)\big) \\
    \upp_i^{-1} &= \text{radius}\big(M(S_i)\big) / \text{radius}\big(S_i\big) \\
    \uptheta_i' &= \tfrac{1}{2}\,\text{radius}^2\big(M(S_i)\big) =  \upp_i^{-2} \uptheta_i 
\end{align}
\end{subequations}
A generalization of Eq.~\eqref{eq:inversion_identity} entails that the deformed MD precisely matches the image of the original tiling under the MT.
See App.~\ref{app:discrete_conformal} for details and App.~\ref{app:minkowski} for a more abstract proof of M\"{o}bius covariance. 

\subsection{Discrete conformal maps}\label{sec:discrete_conformal}

\subsubsection{Multiplicative Voronoi diagrams}

Let us consider the transformation law Eq.~\eqref{eq:power_transform} for the special case $\uptheta_i=0$: the undeformed tiling is a Voronoi diagram, and, as we will show in a moment, the deformed tiling is a multiplicative Voronoi diagram (MVD). An MVD cell $i$ is the set of points $\rvec$ in the plane closest to vertex $\tvec_i'$, using a ``conformally'' scaled distance
\begin{align}
\label{eq:MVT_definition}
\begin{aligned}
    C_i^\mathrm{C} &= \{\rvec \, : \, \Phi_i^\mathrm{C}(\rvec) < \Phi_j^\mathrm{C}(\rvec) \; \forall j\neq i \}, \\
    \Phi_i^\mathrm{C} &= \frac{\upp_i}{2} |\rvec-\tvec_i'|^2
\end{aligned}
\end{align}
The notation $\tvec_i'$ anticipates that the seed points will be the image of the original tension vertices $\tvec_i$ under a conformal map. The tensions implied by Eq.~\eqref{eq:MVT_definition} can be read off from Eq.~\eqref{eq:implied_tensions}:
\begin{align}
    \label{eq:tension_multiplicative_Voronoi}
    \uptau_{ij}  = |\nabla\Phi_i-\nabla\Phi_j| = \sqrt{\upp_i \upp_j}\,|\tvec_i'-\tvec_j'| = \sqrt{\upp_i \upp_j}\, \uptau_{ij}'
\end{align}
The cell pressures are $\upp_i$, so the cell interfaces are circular arcs\footnote{
More precisely, interfaces are defined by the locus $\upp_i|\rvec-\tvec_i'|^2=\upp_j|\rvec-\tvec_j'|^2$. Thus, the interfaces are so-called Apollonian circles that intersect $\tvec_{ij}' = \tvec_i'-\tvec_j'$ orthogonally.} with curvature:
\begin{align} \label{eq:curvature-discrete-conformal}
    \upkappa_{ij} = \frac{\upp_i - \upp_j}{\uptau_{ij}} = \frac{1}{\sqrt{\upp_i \upp_j}} \,\frac{\upp_i - \upp_j}{|\tvec_i'-\tvec_j'|}
\end{align}

Applying an MT $M$ to an ordinary Voronoi diagram results in a MVD, and the transformation law Eq.~\eqref{eq:power_transform} specializes to:
\begin{align}
    \label{eq:Moebius_multiplicative_Voronoi}
    \tvec_i' = M(\tvec_i), \quad  \upp_i = 1/\lambda_M(\tvec_i), \quad \rvec_{ijk}'  = M(\rvec_{ijk})
\end{align}
Eq.~\eqref{eq:curvature-discrete-conformal} for the interface curvature in an MVD now becomes a discrete version of Eq.~\eqref{eq:curvature_transform}, the change of line curvature under a conformal map.
The pressure is the inverse of the conformal factor, confirming the approximate analysis above
\begin{align}
    \label{eq:pressure_inverse_conformal}
    \upp_i = 1/\lambda_M(\tvec_i)
\end{align}
For this reason, we equivalently refer to the $\upp_i$ as either pressures or (inverse) ``conformal factors''.
We emphasize that Eq.~\eqref{eq:pressure_inverse_conformal} is not a constitutive equation but a geometric consequence of the Young--Laplace law.

Let us verify Eq.~\eqref{eq:Moebius_multiplicative_Voronoi}. The undeformed cells are $C_i = \{\rvec :\,\, |\rvec-\tvec_i|^2 < |\rvec-\tvec_j|^2\}$. We use Eq.~\eqref{eq:inversion_identity} to apply the MT $M$ to both sides of this inequality:
\begin{align}
    \label{eq:inversion_identity_cell}
    \begin{aligned}
     \frac{\left\lvert M(\rvec) - M(\tvec_i) \right\rvert^2}{\lambda_M(\tvec_i)} 
    < \frac{\left\lvert M(\rvec) - M(\tvec_j) \right\rvert^2}{\lambda_M(\tvec_j)}
        \end{aligned}
\end{align}
which matches the MVD cells Eq.~\eqref{eq:MVT_definition} with parameters given by Eq.~\eqref{eq:Moebius_multiplicative_Voronoi}.
Under the MT, both the seed points and the scale factors change, 
\begin{align}
    \uptau_{ij}' = |M(\tvec_i)-M(\tvec_j)| = \frac{\uptau_{ij}}{\sqrt{\upp_i \upp_j}} 
\end{align}
But by Eq.~\eqref{eq:tension_multiplicative_Voronoi}, the physical tensions are invariant.

\subsubsection{Discrete conformal maps}

Thus, MTs give rise to MVDs.
However, Eq.~\eqref{eq:MVT_definition} defines a force-balanced tiling for \emph{any} set of pressures $\upp_i$. A global MT, defined by 6 parameters, is clearly insufficient to parametrize all multiplicative diagrams. 
Above, we noted that MTs are a special case of conformal maps.
We now argue that a generic set of pressures can be thought of as a generic conformal map with local conformal factor $1/\upp_i$. To do so, we draw on the notion of \emph{discrete conformal maps} from the mathematics and computer graphics literature~\cite{Springborn.etal2008,Crane.etal2013}.

A continuous conformal map locally rescales lengths by an isotropic conformal factor $\lambda$. Discrete conformal maps define an analogous notion for maps between triangulations. Such a map is a set of per-vertex scale factors $\uplambda_i$ (often parameterized as $\uplambda_i =e^{\mathrm{u}_i}$) which rescale the length $\uptau_{ij}$ of each triangulation edge~\cite{Springborn.etal2008}:
\begin{equation} \label{eq:discrete_conformal_definition}
    \uptau_{ij} \mapsto \uptau_{ij}'= \sqrt{\uplambda_i\uplambda_j} \, \uptau_{ij}, \quad \uplambda_i = 1/ \upp_i
\end{equation}
We will identify the scale factors with the inverse pressure.
Note that an arbitrary set of scale factors $\upp_i^{-1}$ results in a non-planar triangulation (one that cannot be drawn in the plane without distortion) -- we will return to this in Sec.~\ref{sec:curvature}. Discrete conformal maps reproduce many features of their smooth counterparts. For example, the composition of two discrete conformal maps is again discrete conformal. However, a discrete conformal map does \emph{not} preserve angles in a triangulation since the internal angles uniquely specify a triangulation up to a global scale factor.

A key example of a discrete conformal map is an MT $M$ of the plane, $\tvec_i' = M(\tvec_i), \; \upp_i^{-1} = \lambda_M (\tvec_i)$, which immediately follows from Eq.~\eqref{eq:discrete_conformal_definition}.
A general conformal map defines a set of triangle-wise MTs $\{M_{ijk}\}$ that fit together across edges (see Fig.~\ref{fig:discrete_conformal} and App.~\ref{app:discrete_conformal}).
Overall, discrete conformal maps furnish a discretization of smooth conformal maps, analogous to the approximation of a continuous function by linear interpolation (Eq.~\eqref{eq:interpolation}).

\subsubsection{Multiplicative diagrams as discrete conformal maps}

We now recognize that Eq.~\eqref{eq:tension_multiplicative_Voronoi} for the tensions in a MVD precisely matches the definition of a conformal map, Eq.~\eqref{eq:discrete_conformal_definition}.
Thus, the triangulation of the seed points $\tvec_i'$ is the image of the tension triangulation $\uptau_{ij}$ under a conformal map with scale factors $\upp_i^{-1}$. On the triangulation, the discrete conformal map acts as a rescaling $\uptau_{ij}^2 \mapsto\upp_i^{-1}\upp_j^{-1}\uptau_{ij}^2$, while on the dual tiling, it acts as a local M\"{o}bius transformation (to our knowledge, the notion of discrete conformal maps of dual cell tilings has not been studied in the mathematics literature).
Hence, a multiplicative Voronoi diagram is the image of the ordinary Voronoi diagram dual to the tension triangulation under a discrete conformal map.
While triangulation angles change under discrete conformal maps, the tiling's vertex angles $\upgamma_{jk}^i$ are preserved \emph{exactly}, as required by the Young--Dupr\'{e} law, Eq.~\eqref{eq:angle_complement}. To accommodate this constraint, the cell interfaces become curved, Eq.~\eqref{eq:curvature-discrete-conformal}. 

Taken together, we generalized the global Möbius invariance of force balance to a local invariance under arbitrary discrete conformal maps defined by a set of scale factors. This makes the intuition of pressure as a local scale factor precise, and constitutes another step towards our goal of constructing cell tilings from maps of the tension triangulation to physical space.

\subsubsection{Discrete conformal maps of power diagrams}{\label{sec:MWPT}}

We now generalize discrete conformal maps to the case $\uptheta_i\neq 0$, known in the computer graphics literature as \emph{decorated} discrete conformal map~\cite{Bobenko.Lutz2024} (the power circles are said to ``decorate'' the triangulation).
The simplest example of a decorated map is an MT $M$ applied to a power diagram with vertices $\tvec_i$ and isogonal mode $\uptheta_i$.
Explicitly, the power circle transformation law Eq.~\eqref{eq:power_transform} corresponds to a deformation of the seed points $\tvec_i'$: 
\begin{subequations}
\label{eq:decorated_conformal}
\begin{align}
    |\tvec_i'-\tvec_j'|^2 &= \frac{\uptau_{ij}^2}{\upp_i\upp_j} + 2\left( \frac{1}{\upp_i}- \frac{1}{\upp_j}\right) \left( \frac{\uptheta_i}{\upp_i}- \frac{\uptheta_j}{\upp_j}\right), \\
    \uptheta_i'  &= \upp_i^{-2} \,\uptheta_i 
\end{align}
\end{subequations}
Eq.~\eqref{eq:decorated_conformal} is the definition of a decorated discrete conformal map, defined by a set of scale factors $\upp_i^{-1}$.
Compared to undecorated maps, one now transforms power circles instead of vertex points, leading to a modified transformation law.
Compared to Eq.~\eqref{eq:discrete_conformal_definition}, it is not quite as obvious why Eq.~\eqref{eq:decorated_conformal} deserves to be called a discrete conformal map.
However, it matches the behavior of decorated triangulations under MTs, the ``archetypal'' conformal map, and, in the dual cell tiling, exactly preserves the vertex angles $\upgamma_{jk}^i$.
Comparing Eq.~\eqref{eq:decorated_conformal} and the implied tensions in an MD, Eq.~\eqref{eq:implied_tensions_MD}, confirms that the physical tensions $\uptau_{ij}$ are invariant (the scaled isogonal potentials $\upp_i^{2}\, \uptheta_i'$ are also invariant).

In summary, general MDs can be thought of as the image of a power diagram under a discrete conformal map\footnote{Any MD can be locally mapped back to a constant-pressure diagram using an MT. For a given tricellular vertex $\rvec_{ijk}$, the inversion that maps the ``ghost vertex'' $\rvec_{ijk}^*$ to infinity ``straightens'' out the cell interfaces, so that $\upp_i=\upp_j=\upp_k$ (see Fig.~\ref{fig:Moebius-triangle}).} with scale factors $1/\upp_i$. This completes the argument that cellular pressure differences are geometrically parametrized by conformal deformations.

\begin{figure*}[t]
    \centering
    \includegraphics{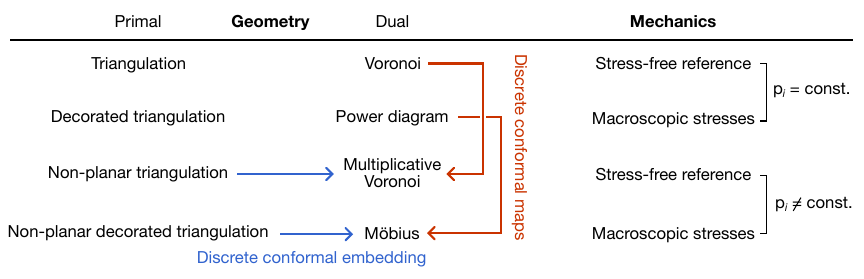}
    \caption{Connections between discrete geometry (M\"{o}bius diagrams, generalized Voronoi tilings) and the mechanical states they represent. Discrete conformal maps and embeddings are piecewise Möbius, parametrized by scale factors at triangle vertices.
    }
    \label{fig:overview}
\end{figure*}

\subsection{Macroscopic stress with pressure gradients}

To compute the stress in a general MD, we check how an MT $M$ with conformal factor $\lambda$ changes the stress of a power diagram (any MD can be mapped back to a power diagram by an MT, at least locally). 
Eq.~\eqref{eq:stress_conformal_qualitative} implicitly assumed $\lambda$ to be constant over the area $A$ over which the stress is averaged. We now use the machinery of Sec.~\ref{sec:stress} to make it rigorous. 
First, recall Eq.~\eqref{eq:stress_bound}: the stress is set by the seed-point displacement field $\mathbf{u} = \upp_i(\rvec - \tvec_{i(\rvec)})$. In MVDs, $|\mathbf{u}|$ is bounded by the cell diameter $\ell_C$ since $\tvec_{i(\rvec)}\in C_i$. Like in ordinary Voronoi diagrams, the macroscopic stress vanishes. More generally, the lower envelope $\Psi$ transforms as
\begin{align}
    \Psi'(\rvec'=M(\rvec)) = \lambda(\rvec) \cdot \Psi(\rvec)
\end{align}
This follows directly from Eq.~\eqref{eq:power_transform}.
The envelope $\Psi'$ is the Airy stress function of the deformed tiling:
\begin{align}
    \bm{\sigma}' = \nabla'\nabla' \left((\lambda\Psi)\circ M^{-1} \right) - \Delta' \! \left((\lambda\Psi)\circ M^{-1} \right) \mathbb{I} 
\end{align}
As a brief sanity check, for a uniform scaling $\lambda(\rvec) =\lambda_0$, $\nabla' =\lambda_0^{-1}\nabla$, so $\bm{\sigma}' = \lambda_0^{-1}\bm{\sigma}$ as in the scaling analysis of Eq.~\eqref{eq:stress_conformal_qualitative}.
Since the isogonal mode is the non-trivial part of $\Psi$, it determines the stress not just in power diagrams, but in general MDs.
Compared to the scaling analysis in Eq.~\eqref{eq:stress_conformal_qualitative}, one finds additional pressure gradient terms in the stress.

Taking a step back, we have constructed a series of successively more general force-balanced tilings: Voronoi-, power-, multiplicative Voronoi and, finally, M\"{o}bius diagrams.
By connecting MDs to (discrete) conformal maps and computing their macroscopic stress tensors, we elucidated the connections between cell geometry and mechanics, summarized in Fig.~\ref{fig:overview}.
The cell tilings are constructed from an embedding of the tension triangulation in the physical plane.
We emphasize, however, that the $\tvec_i'$ are an intermediate step without direct physical significance~\footnote{
Note that Eq.~\eqref{eq:implied_tensions_MD} for the tension in an MD depends on $\uptheta_i$. However, it is \emph{incorrect} to conclude that the isogonal mode \emph{changes} the edge tensions when pressure differences are nonzero. Instead, the isogonal mode changes $\tvec_i'$ and $\upp_i$ (which are the dependent variables) so that the tensions $\uptau_{ij}$ remain preserved.
Keeping $\tvec_i'$ fixed is not physically meaningful.}.
Next, we show how to compute the embedding from the physical control parameters, the tensions $\uptau_{ij}$ and the isogonal potential $\uptheta_i$.

\section{Non-planar tension triangulations}
\label{sec:curvature}

\begin{figure*}
    \centering
    \includegraphics[scale=1]{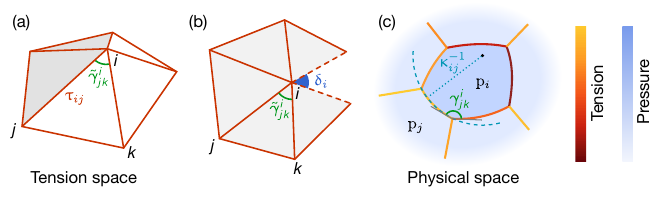}
    \caption{Curvature of the tension triangulation (a) is revealed through the angle deficit that appears when ``flattening'' a plaquette (b). In physical space, this implies pressure gradients because the angle deficit must be accommodated by line curvatures $\upkappa_{ij}$ in the plane (c).
    (The curved triangulation (a) is drawn in 3d space for illustration purposes; the theory does not use an embedding in 3D).}
    \label{fig:angle-deficit}
\end{figure*}

\subsection{Discrete Gaussian curvature}

\subsubsection{The angle deficit}

So far, we have considered flat tension triangulations, which can be drawn in the plane without distortion. However, in general, a triangulation specified through its adjacency and the edge lengths cannot be isometrically embedded in the plane. The elementary obstruction to flatness is the \emph{angle deficit} of the plaquette around a vertex (Fig.~\ref{fig:angle-deficit}b)
\begin{align} \label{eq:tri-curvature}
    \updelta_i = 2\pi-\!\!\sum_{(jk)\sim i} \tilde{\upgamma}^i_{jk}
\end{align}
This angle deficit measures a concentration of Gaussian curvature at the vertex. A triangulation drawn in the plane automatically has zero angle deficit.

Mechanically, a non-planar tension triangulation means that tensions alone cannot be in global force balance in the plane -- pressure differences are required. Force balance at the vertices implies that the angles at which cell interfaces meet are complementary to the triangle angles, $\tilde\upgamma^i_{jk}=\pi-\upgamma^i_{jk}$ (Eq.~\eqref{eq:angle_complement}). Hence,
\begin{align}
    \updelta_i = 2\pi- \sum_{jk\sim i} \tilde{\upgamma}^i_{jk} = 2\pi - n\pi + \sum_{jk\sim i} \upgamma^i_{jk}
\end{align}
where $n$ is the number of neighbors of $i$. In a conventional 2D fluid foam, interfaces meet at $2\pi/3$ angles, so that $\updelta_i = \tfrac{6-n}{3}\pi$ -- the angle deficit vanishes for 6-sided cells.
For a straight-edge polygon, $\sum_{jk} \upgamma^i_{jk}=(n-2)\pi$, so the only way to compensate for the angle deficit in the plane is to curve the cell interfaces: 
\begin{align}
    \updelta_i = \oint_{\partial C_i} \kappa \mathrm{d}\ell= \sum_{j\sim i} \upell_{ij} \upkappa_{ij}
    \label{eq:angle_deficit_kappa}
\end{align}
since $\upell_{ij} \upkappa_{ij}$ is the rotation angle of the tangent vector along the (circular arc) interface $ij$.

In this section, we systematically consider the implications of the angle deficit for tension networks, using the machinery of (discrete) conformal maps developed in the preceding section. 

\subsubsection{Angle deficit sources pressure differentials}

By Eq.~\eqref{eq:angle_deficit_kappa}, Gaussian curvature of the tension triangulation $\updelta_i \neq 0$ implies line curvature of the cell interfaces and thus pressure differentials~\cite{Noll.etal2017}. In a conventional 2D fluid foam, a 5-sided cell is under higher pressure while a 7-sided cell is under lower pressure (the angle deficit being $\pm \pi/3$). Diffusion of gas driven by these pressure differences gives rise to the famous von Neumann law governing the coarsening of a 2D foam. 

Active tension networks are generalized foams, and we now generalize the von Neumann law to a (discrete) Poisson equation, with angle deficit as a source term. We start from Eq.~\eqref{eq:angle_deficit_kappa}. The curvatures $\upkappa_{ij}$ are determined by the pressures via the Young--Laplace law:
\begin{align} \label{eq:Poisson_pressure_discrete}
    \updelta_i &= \sum_{j \sim i} \upell_{ij}  \upkappa_{ij} =  \sum_{j \sim i} \upell_{ij} \frac{\upp_i-\upp_j}{\uptau_{ij}} = - (\Delta \mathrm{p})_i 
\end{align}
where we recognize the discrete Laplace operator from Eq.~\eqref{eq:Laplace-from-div}.
Recall that the Laplacian depends on the tiling geometry, and hence on the $\upp_i$ through the lengths $\upell_{ij}$,  making Eq.~\eqref{eq:Poisson_pressure_discrete} implicitly non-linear.

By Eq.~\eqref{eq:Poisson_pressure_discrete}, the pressure field has no bulk degrees of freedom -- it is determined by its boundary values.
Put another way, a cell cannot change its pressures and tensions independently in mechanical equilibrium.
Geometrically, pressure corresponds to discrete conformal maps of the cell tiling. Eq.~\eqref{eq:Poisson_pressure_discrete} means that for non-planar tension triangulations, some degree of conformal deformation is unavoidable.
While we have assumed that the physical cell tiling lies in a flat plane, Eq.~\eqref{eq:Poisson_pressure_discrete} immediately generalizes to tissues on curved surfaces, e.g., the ellipsoidal shape of the early \emph{Drosophila} embryo. In this case, the angle deficit must be computed with respect to the curvature of the physical target surface.

\subsection{Discrete conformal embeddings of non-planar triangulations}

We now fill in this geometric picture. For simplicity, we first consider the case $\uptheta_i=0$.
In Sec.~\ref{sec:constant_pressure}, we used Voronoi and power diagrams to construct balanced cell tilings from an isometric embedding $\tvec_i, \; \uptau_{ij} = |\tvec_i-\tvec_j|$ of the tension triangulation.
If the angle deficit (discrete curvature) is non-zero, an isometric embedding does not exist: drawing the tension triangulation in a flat plane necessarily induces some distortion. 
This fact is familiar from cartographic projections of curved surfaces, like the globe.
However, not all distortion is created equal. In particular, so-called conformal or isothermal embeddings only distort area, but preserve angles (an example is the Mercator projection of the globe).

Sec.~\ref{sec:pressure} showed that force balance only constrains the angles in the cell tiling, and is thus invariant under (discrete) conformal maps. This suggests conformally deforming the non-flat tension triangulation to embed it into the flat 2D plane. We proceed in three steps:
\begin{subequations}
 \label{eq:flattening_problem_overall}
\begin{enumerate}
    \item Given the tensions $\uptau_{ij}$, find a set of scale factors $\upp_i$ so that the rescaled triangulation is flat:
    \begin{align}
        \label{eq:flattening_problem}
        \uptau_{ij}' = \uptau_{ij} /  \sqrt{\upp_i\upp_j} \;\; \text{s.th.} \;\;\updelta_i' = 0
    \end{align}
    The condition $\updelta_i'=0$ defines $C$ equations to be solved using the $C$ variables $\upp_i$.
    ($\updelta_i'=0$ must hold in the tissue interior; the boundary values $\upp_i$ are set by the boundary conditions for the cell pressures).
    \item Find a set of 2D embedding vertices $\tvec_i'$ so that 
    \begin{align}
        \label{eq:flattening_problem_embedding}
        |\tvec_i'-\tvec_j'| = \uptau_{ij}' = \uptau_{ij} /  \sqrt{\upp_i\upp_j}
    \end{align}
    This is possible since $\updelta_i'=0$.
    \item Given the embedding, construct a dual cell tiling using MVDs:
    \begin{align}
        \Phi_i^\mathrm{C}(\rvec) = \frac{\upp_i}{2} |\rvec-\tvec_i'|^2
    \end{align}
    Per Eq.~\eqref{eq:implied_tensions_mult_Voronoi}, the corresponding tensions are $\uptau_{ij} = |\tvec_i'-\tvec_j'|\,\sqrt{\upp_i\upp_j}$.
\end{enumerate}
\end{subequations}
We have thus arrived at our goal: a mechanically balanced tiling with the given tensions, now constructed from a discrete conformal embedding of the tension triangulation.

The next question is how to solve Eq.~\eqref{eq:flattening_problem}. The problem of finding a discrete conformal map of a triangulation to the flat 2D plane has been studied in the computer graphics literature, motivated by the need to parametrize triangular meshes for texture mapping.
By calculating how the angles of a triangle change as its side lengths are rescaled,
Ref.~\cite{Springborn.etal2008} showed that a small discrete conformal deformation of a triangulation changes its angle deficit as:
\begin{align}
    d\updelta_i = (\Delta^\mathrm{\!V} d\log \upp)_i = \sum_{j\sim i} (\upell_{ij}^\mathrm{V}) \frac{d\log \upp_j-d\log \upp_i}{\uptau_{ij}}
    \label{eq:yamabe_discrete}
\end{align}
We write $\Delta^\mathrm{\!V}$ to emphasize that it is the Laplacian of the triangulation and its Voronoi dual.
By contrast, Eq.~\eqref{eq:Poisson_pressure_discrete} used the physical tensions and interface lengths
(both equations are correct, but correspond to two slightly different statements at nonlinear order in $(\upp_i-1)$).

By integrating Eq.~\eqref{eq:yamabe_discrete}, one can prove a discrete \emph{uniformization theorem}. For any triangulation\footnote{We implicitly assume that the triangulation is topologically equivalent to a disk and that the Delaunay condition Eq.~\eqref{eq:Delaunay} is fulfilled in both the $\uptau_{ij}$ and the $\uptau_{ij}'$ triangulations. In general, finding a conformal embedding can require T1 transitions.}, one can find a set of vertex scaling factors/pressures $\upp_i$ so that the rescaled triangulation has zero angle defect and can be drawn in the plane with 2D vertices $\tvec_i'$.
Physically, Eq.~\eqref{eq:yamabe_discrete} is the geometric equivalent of the Poisson Eq.~\eqref{eq:Poisson_pressure_discrete} for the pressure field.
This means that for any set of active tensions, a set of intracellular pressures exists that brings the system into mechanical balance.

\subsubsection{Embedding decorated triangulations}\label{sec:embedding_decorated}

\begin{figure}
    \centering
    \includegraphics[width=\linewidth]{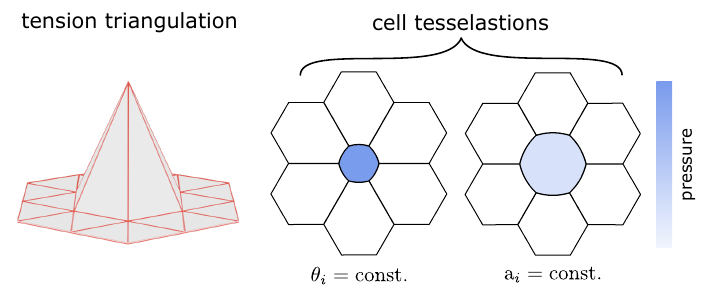}
    \caption{Interplay of pressure and isogonal mode. A non-planar tension triangulation (left) is discrete-conformally mapped to a planar cell tiling (center, right). The resulting pressure differential depends on the isogonal mode: compared to the Voronoi case (center), isogonally inflating the central cell so all cell areas are equal (right) lowers the pressure gradient. With longer cell-cell interfaces, a smaller interface curvature suffices to compensate for the angle deficit. 
    }
    \label{fig:weighted-Laplacian}
\end{figure}

We now generalize to $\uptheta_i\neq 0$, where discrete conformal maps become the decorated conformal maps of Sec.~\ref{sec:MWPT}. 
Physically, computing a discrete conformal embedding requires finding pressures that compensate for the angle deficit of the prescribed tensions, Eq.~\eqref{eq:Poisson_pressure_discrete}. These pressures depend on the isogonal mode $\uptheta_i$. For instance, if a cell is isogonally inflated, its interfaces elongate. Smaller interface curvatures, and thus smaller pressure differences, suffice to compensate for the same angle deficit (see Fig.~\ref{fig:weighted-Laplacian}).
Mathematically, the discrete Laplace operator Eq.~\eqref{eq:Poisson_pressure_discrete} depends on interface lengths $\upell_{ij}$, and thus on $\uptheta_i$. 

Indeed, Ref.~\cite{Bobenko.Lutz2024} generalized the discrete uniformization theorem to decorated triangulations\footnote{
Ref.~\cite{Bobenko.Lutz2024} assumes that $\uptheta_i\geq 0$ and that the power circles of neighboring cells do not intersect.
Physically, this means that the isogonal deformation remains bounded.}
. For a given decorated triangulation $(\uptau_{ij}, \uptheta_i)$, one can find a set of scale factors so that the transformed triangulation $\uptau_{ij}'$, Eq.~\eqref{eq:decorated_conformal} can be drawn in the plane with vertices $\tvec_i'$. The conformal factors/pressures obeys 
\begin{align}
d\updelta_i = (\Delta^\mathrm{\!P} d\log \upp)_i
    \label{eq:yamabe_discrete_decorated}
\end{align}
where the Laplacian $\Delta^\mathrm{P}$ uses the rescaled triangulation and its power dual, not the physical tensions and interface lengths. 
Given the embedding $\tvec_i'$ and the pressures $\upp_i$, one can then construct the cell tiling as an MD. This tiling is mechanically balanced and, per Sec.~\ref{sec:MWPT}, has the required interfacial tensions.

\begin{figure*}
    \centering
    \includegraphics{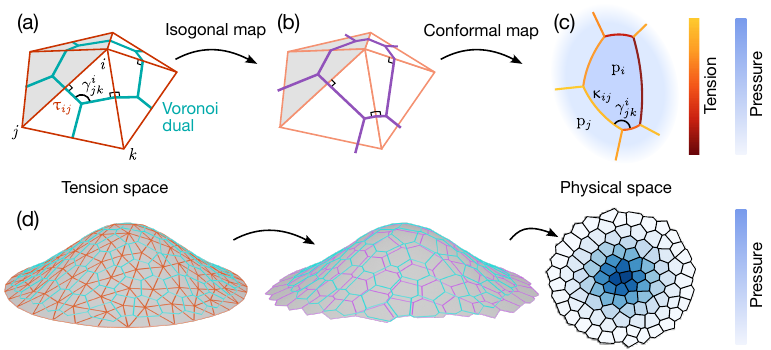}
    \caption{(a)~Non-Euclidean tension triangulation (red) with its Voronoi dual (cyan) constructed from the triangle circumcenters in the plane of each triangle. 
    (The drawing as an embedding in 3D is only for illustration purposes. Only the intrinsic geometry, encoded in the edge lengths $\uptau_{ij}$, matters.)
    (b)~Isogonal modes displace the cell vertices in the plane of the respective tension triangle.
    (c)~The conformal map trades Gaussian curvature of the tension triangulation for line curvatures $\kappa_{ij}$ of cell interfaces. Physically, this implies pressure differences between cells (Young--Laplace law).
    (d)~Example for a larger tissue patch (tension triangulation in red, Voronoi in cyan, and isogonal tiling in purple). The flat tiling is an MD obtained by numerical optimization of 2D seeds and conformal factors $(\tvec_i',\upp_i)$ to fulfill Eq.~\eqref{eq:MWPT-combined} for the given tension triangulation and isogonal mode.
    }
    \label{fig:curved-tri-iso-conf}
\end{figure*}

\subsection{Intrinsic Voronoi and power diagrams}\label{sec:intrinsic}

While the decorated conformal map systematically constructs the force-balanced cell tiling, it requires the auxiliary seed points $\tvec_i'$, which do not have a direct physical significance. 
We now present an alternative picture that is physically more transparent.
In short, we will generalize the logic of planar tension triangulations (Secs.~\ref{sec:constant_pressure} and~\ref{sec:pressure}) by constructing non-planar Voronoi and power diagrams \emph{in the curved space defined by a non-planar triangulation} first, and only then map into the plane.

Our starting point is the interpretation of the triangulation, specified via the edge lengths $\uptau_{ij}$, as a curved surface. 
For a planar triangulation, this surface is trivially (a subset of) the Euclidean plane.
For a non-planar triangulation, each quadrilateral $ikjl$ (i.e., a pair of adjacent triangles $ijk$ and $ijl$) can be isometrically mapped to the plane by flattening the ``hinge'' $ij$~\cite{Crane2025}. 
Mathematically, the hinge map makes the triangulation into a Riemannian manifold. Such a manifold is defined by a set of local coordinate charts and transition functions between them~\cite{Lee2012}. Here, the hinge map defines the charts. Within each chart, the metric is Euclidean. The transition maps defined on pairs of overlapping quadrilaterals are affine.
Even though the triangulation appears ``kinked'' when drawn in 3D, intrinsically, it forms a well-defined Riemannian manifold\footnote{Except at the vertices, which can form cone singularities. These singularities pose no problem since they are isolated and can be removed by local smoothing.}.

Via the hinge map, the construction of the Voronoi dual generalizes easily to non-flat triangulations: 
one can perform the Voronoi construction in the plane of each triangle separately, as illustrated in Fig.~\ref{fig:curved-tri-iso-conf}. Each Voronoi vertex is the circumcenter in the plane of the corresponding triangle. Thanks to the hinge map, the resulting tiling of ``kinked polygons'' [cyan lines in Fig.~\ref{fig:curved-tri-iso-conf}(a)] is consistent across neighboring triangles.
The Voronoi tiling can be thus defined purely \emph{intrinsically}. Indeed, the Voronoi interface lengths and cell areas $\upa_i^\mathrm{V}$ can be expressed in terms of the edge lengths $\uptau_{ij}$ and angles $\tilde{\upgamma}_{ij}^k$ of the tension triangulation alone~\cite{Meyer.etal2003} (see App.~\ref{app:Voronoi}).

This non-planar Voronoi tiling serves as a stress-free reference configuration in the (discretely) curved space of the tension triangulation. The power diagram can also be constructed intrinsically~\cite{DeGoes.etal2014}. The isogonal displacement acts locally in the tangent plane\footnote{Strictly speaking, this is well-defined only when the isogonal displacement is sufficiently small compared to the scale of the tension triangulation curvature. However, the areas and edge lengths of the intrinsic power diagram are always well-defined.}, yielding a non-planar power diagram [purple lines in Fig.~\ref{fig:curved-tri-iso-conf}(b)]. The isogonal potential is the Airy stress function for the tensile stress tensor in the local tangent space.

Finally, to obtain the physical cell tiling, one conformally maps the non-planar power diagram into the plane [Fig.~\ref{fig:curved-tri-iso-conf}(c); Fig.~\ref{fig:curved-tri-iso-conf}, bottom shows a bigger tissue patch]. Since conformal maps preserve force balance, the resulting planar tiling is balanced. The conformal map eliminates discrete Gaussian curvature at the vertices and ``converts'' it into line curvature of the cell edges. 
It is worth noting that the conformal map rescales the intrinsic to the physical cell areas by a factor $\uplambda_i^2 = \upp_i^{-2}$ (approximately, to leading order in $\nabla\log\uplambda$).
The above construction is not just more intuitive, but also useful for passing to the continuum limit.

In summary, incorporating pressure gradients generalizes the results for a flat tension triangulation to the generic case in which the tensions $\uptau_{ij}$ form a non-planar triangulation. The tension triangulation acts as a mechanical reference state, now with non-zero intrinsic curvature. 

\section{Governing equations for stress and adiabatic tissue flow}\label{sec:governing}

\subsection{Discrete biharmonic equation for Airy function}

Above, we computed a discrete conformal embedding, and hence cell tiling, given tensions and isogonal potentials.
However, physically, the isogonal potential is determined by cell compressibility.
The implied pressures of the MD must be compatible with the constitutive equation between pressure and cell area, $\upp_i = P(\upa_i)$.
For a planar tension triangulation, we can first find the embedding $\tvec_i$, and then adjust the isogonal mode to obtain the required cell areas, Eq.~\eqref{eq:Poisson_area}. Unfortunately, for non-planar tension triangulations, the embedding $\tvec_i'$ itself depends on the isogonal mode. The non-linear consistency Eq.~\eqref{eq:MWPT-combined} generally requires an iterative approach. 

Fortunately, for small triangulation curvature, the situation simplifies considerably.
We consider small deviations from an ordinary Voronoi tiling and work to linear order around $\upp_i=1, \, \uptheta_i=0$. We furthermore assume that $\upp_i$ varies slowly compared to the cell scale $\ell_C$.
First, we expand Eq.~\eqref{eq:implied_tensions_MD} to linear order, $\uptau_{ij}^2 \approx (\upp_i\upp_j)\,|\tvec_i'-\tvec_j'|^2$. Thus, the conformal embedding, scale factors $\upp_i$ and vertices $\tvec_i'$, are approximately $\uptheta_i$-independent.
Next, we expand Eqs.~\eqref{eq:isogonal_area},\eqref{eq:cell_areas_conformal} for the cell area, the linearized constitutive Eq.~\eqref{eq:constitutive_linear}, and the Poisson Eq.~\eqref{eq:Poisson_pressure_discrete} for pressure. 
We obtain the following system of equations\footnote{Here, $\Delta$ is the discrete Laplacian Eq.~\eqref{eq:Laplace-from-div} of the intrinsic Voronoi tiling defined by the $\uptau_{ij}$. Recall that $\Delta$, in general, depends on $\uptheta_i$ through the cell edge lengths, but this can be neglected at linear order.}:
\begin{subequations}
\label{eq:linearized_self_consistency}
\begin{align}
    (\Delta d\upp_i) &= -\updelta_i \\ 
    \mathrm{a}_i &= \mathrm{a}_{0,i} - B^{-1} d\upp_i \\
    (\Delta d\uptheta)_i &=(1-2d\upp_i)\mathrm{a}_i^\mathrm{V} - \mathrm{a}_i
     \label{eq:linearized_self_consistency_area}
\end{align}
\end{subequations}
Eq.~\eqref{eq:linearized_self_consistency} can be solved in three decoupled steps, top to bottom. It can also be combined into a single \emph{biharmonic equation}:
\begin{align}
    \label{eq:biharmonic}
    (\Delta^2 d\uptheta)_i = -B^{-1} \updelta_i + \left(\Delta \left((1-2d\upp_i) \mathrm{a}^\mathrm{V} - \mathrm{a}_0 \right) \right)_i
\end{align}
In the limit $B\rightarrow\infty$ of incompressible cells, Eq.~\eqref{eq:biharmonic} reduces to Eq.~\eqref{eq:linearized_self_consistency_area}; the $\uptheta_i$ are determined by $\upa_i=\upa_{0,i}$.

Recall that $\uptheta_i$ is the discrete Airy function. 
In conventional 2D linear elasticity, a biharmonic equation for the Airy function arises from the requirement that stress be \emph{compatible} with a planar strain field~\cite{Landau.Lifshitz1986}.
Thus, Eq.~\eqref{eq:biharmonic} is in line with the physical picture of effective elasticity.
However, the stress tensor in a tension network is, generically, not compatible: Eq.~\eqref{eq:biharmonic} has non-zero source terms, leading to non-zero stress even in the absence of external forces. 
Incompatibility can arise both from tension curvature $\updelta_i$ and differences between the target cell areas and the local ``tension density'' (the $\upa_i^\mathrm{V}$, which are functions of the tensions $\uptau_{ij}$).
Eq.~\eqref{eq:biharmonic} requires two boundary conditions, for $\upp_i$ and for $\uptheta_i$. These two boundary values suffice to accommodate arbitrary boundary conditions, either given cell positions or traction forces.

\subsection{Adiabatic tension dynamics move cells}\label{sec:Beltrami}

Up to this point, we considered the problem of finding the tissue's rest state and elastic response for a given, fixed set of tensions.
However, the tissue can also remodel \emph{plastically} by changing internal tensions $\uptau_{ij} \mapsto \uptau_{ij} + d\uptau_{ij}$, i.e.\ deforming the tension triangulation (as we already discussed in Sec.~\ref{sec:adiabatic_lattice}).
Physically, we now compute the velocity $\dot{\rvec}_i$
due to time-dependent tensions, i.e., $d\uptau_{ij} = \dot{\uptau}_{ij} dt$ for a time increment $dt$. In the adiabatic limit, the velocity is determined by the change of the mechanical equilibrium state.

Geometrically, we need to compute the dynamics of the embedding $\tvec_i$, the cell seed points. The tension update defines a strain of the tension triangulation. Let us define the embedding \emph{strain operator} $\mathcal{D}$ 
by 
\begin{align}
    (\mathcal{D}\, d\tvec')_{ij} = \hat{\tvec}_{ij}' \cdot \frac{d\tvec_i'-d\tvec_j'}{\uptau_{ij}'}
\end{align}
$\mathcal{D}$ maps seed point displacement $d\tvec_i'$ to edge length changes.
For flat tension triangulations,  $|\tvec_i-\tvec_j|=\uptau_{ij}$. A first idea is thus to match the displacement strain and the strain of the tensions:
\begin{align}
    \label{eq:strain_operator_naive}
    (\mathcal{D}\, d\tvec')_{ij} \stackrel{?}{=} \frac{d \uptau_{ij}}{\uptau_{ij}} = d \log \uptau_{ij}
\end{align}
However, Eq.~\eqref{eq:strain_operator_naive} is generally overconstrained: the $d\tvec_i'$ are $2C$ variables, but Eq.~\eqref{eq:strain_operator_naive} are $E=3C$ equations.
Geometrically, the tension changes can create discrete curvature that cannot be matched by seed point displacement. 

Note that the flattening problem Eq.~\eqref{eq:flattening_problem_overall} only requires that the $\tvec_i$ define a conformal embedding, i.e., an embedding faithful up to local scaling.
Thus, locally rescaling tensions does not require updating the embedding, and the displacement strain $(\mathcal{D}\, d\tvec)_{ij}'$ need only match the anisotropic part of the tension strain.
Physically, locally rescaling tensions changes the pressure $p$ and does not require cell movement.
To project out the isotropic part, we define the \emph{cross-ratio operator}~\cite{Springborn.etal2008}.
For an edge $ik$, sitting as diagonal within a quadrilateral $ijkl$, it reads:
\begin{align}
    (\mathcal{P} \, \upv)_{ijkl} = \upv_{ij} - \upv_{jk} + \upv_{kl} - \upv_{li}
\end{align}
$\mathcal{P}$ derives from the logarithm of the \emph{cross-ratio} $(\upell_{ij}\upell_{kl})/(\upell_{jk}\upell_{li})$, which is conformally invariant~\cite{Springborn.etal2008}.
$\mathcal{P}$ annihilates any isotropic strain, i.e., strain resulting from a discrete conformal map of the tensions: 
\begin{align}
\begin{aligned}
    d\log\uptau_{ij} &= -\tfrac{1}{2}\,d(\log \upp_i + \log\upp_j)\\ 
    \Rightarrow \;\; (\mathcal{P} d\log \uptau)_{ijjl} &= 0
\end{aligned}
\end{align}
Composing $\mathcal{P}\circ \mathcal{D}$ maps seed point displacements to anisotropic strain. In particular, $\mathcal{P}\circ \mathcal{D}$ annihilates displacements from conformal maps (MTs), since the latter induce isotropic strain~\cite{Springborn.etal2008}.
Using $\mathcal{P}$, we can fix the naive Eq.~\eqref{eq:strain_operator_naive}:
\begin{align}
    \label{eq:discrete_Beltrami}
    ((\mathcal{P}\circ \mathcal{D}) d\tvec')_{ijkl} = (\mathcal{P}\, d\log\uptau)_{ijkl}
\end{align}
which can be confirmed by expanding Eq.~\eqref{eq:flattening_problem_embedding}, $|\tvec_i'-\tvec_j'|^2 =  \uptau_{ij}^2 / (\upp_i\upp_j)$, in $d\tvec_i$ and $d\uptau_{ij}$.
Since $\mathcal{P}$ annihilates the $C$-dimensional space of isotropic strains, Eq.~\eqref{eq:discrete_Beltrami} only defines $E-C=2C$ equations for the $2C$ variables $d\tvec_i'$ (to completely fix $d\tvec_i'$, boundary conditions are required).
For reasons that will be explained in Sec.~\ref{sec:continuum_limit}, we refer to Eq.~\eqref{eq:discrete_Beltrami} as a \emph{discrete Beltrami} equation.
Physically, it describes \emph{adiabatic flow}: after a small change in the energy parameters $\uptau_{ij}$, the system configuration converges to a new mechanical equilibrium in which cell centers have moved.

The discrete Beltrami equation for the adiabatic tissue flow concludes our discussion of cell mechanics. We next turn to the topology of the cell network and cell rearrangements.

\section{Topology, tension anisotropy, and T1 transitions}
\label{sec:T1s}

Large plastic deformations require topological modifications of the cell network by neighbor exchange (T1), cell extrusion (T2), and cell division events.
In the previous section, we explained that the tension triangulation defines a (piecewise-linear) manifold.
Different triangulations can realize the same tension manifold (e.g., different triangulations of the same planar domain, Fig.~\ref{fig:circle_pack}, top).
Therefore, the triangulation contains additional information/degrees of freedom, namely its adjacency graph (two cells $i,j$ are adjacent if they share an interface in real space). It defines \emph{how} the tension manifold is ``tiled'' by individual tension triangles, and encodes the local tension configuration. As we will see below, this information is essential for determining when T1s occur.

\subsection{Geometric representation of triangulation topology by circle packs}\label{sec:circle_packs}

\begin{figure}
    \centering
    \includegraphics[width=\linewidth]{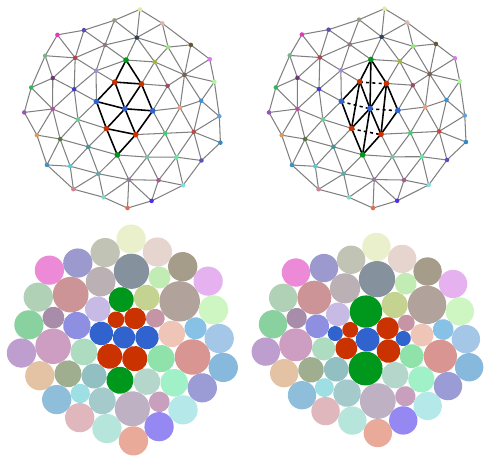}
    \caption{
    Two distinct tension triangulations (top) and the circle packs corresponding to their adjacency graph.
    Both triangulations triangulate the same region of the plane, and thus encode the same tension metric $\bm{g}$ and domain $\Omega$. However, their adjacency graphs differ due to T1 transitions in the highlighted region. This difference is reflected in the corresponding circle packings, computed using the uniform neighbor algorithm~\cite{Stephenson2005}.}
    \label{fig:circle_pack}
\end{figure}

The (graph) topology of a triangulation is encoded by its adjacency matrix $A_{ij}$, with $A_{ij}=1$ if two cells share an edge, and $A_{ij}=0$ otherwise. 
The Koebe–Andreev–Thurston circle-packing theorem~\cite{Stephenson2005} shows that every adjacency graph (a maximal planar graph) can be represented by a circle packing: a planar drawing with one circle per cell $i$ so that two circles are tangent if and only if the corresponding cells are adjacent (Fig.~\ref{fig:circle_pack}, bottom). 
With the radii of the boundary circles fixed to unity as in Fig.~\ref{fig:circle_pack}, the circle packing is rigid. The packing thus yields a unique geometric representation of the adjacency relations. (Alternatively, one can fix the boundary circles to be tangent to the unit disk, which yields a packing unique up to Möbius transformations of the disk~\cite{Stephenson2005}; see Fig.~\ref{fig:Thurston-in-disk} in App.~\ref{app:LTC}).

Thurston introduced circle packings as discrete approximations of conformal maps~\cite{Stephenson2005}. Indeed, one can reformulate circle packings in terms of the decorated discrete conformal maps from Sec.~\ref{sec:MWPT}~\cite{Bobenko.Lutz2024}.
Consider the adjacency graph as a triangulation where the geometric information is discarded and all edges have unit length.
Next, decorate each vertex with a circle of radius $\sqrt{2\uptheta_i} = 1/2$ so neighboring vertex circles are tangent. By Eq.~\eqref{eq:decorated_conformal}, a discrete conformal map to the plane, with scale factors $\Lambda_i$, hence results in a circle packing: a pattern of circles in the plane with radii $\Lambda_i/2$ so that two circles $i,j$ are tangent if cells $i,j$ are adjacent.
We denote the circle centers by $\boldsymbol{\upzeta}_i$.
The boundary circle radii define the boundary condition for the discrete conformal map.

Overall, circle packings convert the ``combinatorial'' information in the adjacency matrix $A_{ij}$ into geometric fields: the embedding $\boldsymbol{\upzeta}_i$ and the conformal factor $\Lambda_i$. 
This representation may also be useful for analyzing cell tilings extracted from experimental data.
Numerically, circle packs can be efficiently computed using the ``uniform neighbor algorithm''~\cite{Stephenson2005}.

\subsection{Local tension configuration parameter and Beltrami coefficients}

\begin{figure}
    \centering
    \includegraphics[scale=1.2]{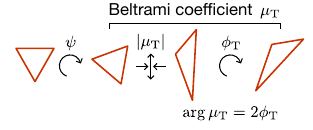}
    \caption{
     The map from a reference equilateral triangle to a tension triangle defines the local tension configuration (LTC). The singular-value decomposition decomposes the map into a rotation, a shear, and a second rotation. The last two operations determine the anisotropy magnitude and orientation, and are combined into the complex-valued Beltrami coefficient $\muT = |\muT|e^{2i\phiT}$.
    }
    \label{fig:LTC}
\end{figure}

We can also use the adjacency triangulation and the associated circle packing to describe the local tension configuration (LTC) that parametrizes the tension triangle shapes in a fashion suitable for coarse-graining \cite{Brauns.etal2024}.
One component of the LTC parameter is the \emph{tension anisotropy} quantifying the elongation of tension triangles.
In the present context, we first introduce the affine map $\mathrm{T}_{ijk}$ that transforms a reference equilateral triangle -- the corresponding element of the adjacency triangulation -- into the given tension triangle $ijk$.
To simplify notation, we suppress the index $ijk$ in the following. 

Since T is affine, its Jacobian can be represented by an (asymmetric) matrix $\mathbf{T}$ acting on triangle edge vectors in the plane. $\mathbf{T}$ relates the adjacency metric to the tension metric and is conveniently parametrized via a singular value decomposition $\mathbf{T} = \mathbf{R}(\phiT) \cdot \bm{\Sigma}_\mathrm{T} \cdot \mathbf{R}(-\psi)$ (Fig.~\ref{fig:LTC}). 
We previously used this construction to define a set of LTC order parameters~\cite{Brauns.etal2024} (App.~\ref{app:LTC}). 
The first SVD angle $\psi$, the ``LTC phase'', determines if a triangle is acute or obtuse. 
Acute triangles correspond to ``tension cables'', chains of high-tension interfaces. The opposite case (obtuse triangle) has been termed a ``tension bridge'', corresponding to an alternating pattern of high- and low tensions~\cite{Brauns.etal2024}.
The diagonal matrix $\bm{\Sigma}_\mathrm{T} = \mathrm{diag} \, (s_1, s_2)$ of singular values and the second angle $\phiT$ determine the direction and magnitude of the tension anisotropy. Both are conveniently combined into the complex-valued Beltrami coefficient, normalized to be independent of the triangle area:
\begin{align}
    \label{eq:beltrami_LTC}
    \muT = \frac{s_1^2-s_2^2}{s_1^2+s_2^2} \, e^{2i\phiT}
\end{align}
(In Refs.~\cite{Claussen.Brauns2025,Claussen.etal2024} we used the notation $q = |\muT|$.) The Beltrami coefficient $\muT$ is independent of the scale and orientation of the reference triangle.

\subsubsection{Tension configuration relative to circle pack reference}

So far, we used an equilateral triangle with a fixed ``Eulerian'' orientation as reference. 
To define the LTC parameters in a purely intrinsic form and connect them to circle packs, we now use the (equilateral) adjacency triangulation as reference.
Since the embedding $\boldsymbol{\upzeta}_i$ of the adjacency triangulation is discrete conformal, it (approximately) preserves triangle shapes: for each vertex $ijk$, the triangle $(\bm{\upzeta}_i, \bm{\upzeta}_j, \bm{\upzeta}_k)$ is (approximately) equilateral and can serve as reference.
Similarly, the tension embedding $\boldsymbol{\uptau}_i$ is discrete conformal: the embedded triangle $(\bm{\uptau}_i, \bm{\uptau}_j, \bm{\uptau}_k)$ is a faithful representation of the tension triangle up to scale.
The map $\mathrm{T}_{ijk}$ now maps circle pack centers $\bm{\upzeta}_i$ to tension vertices $\bm{\uptau}_i$.
Its Jacobian $\mathbf{T}_{ijk}$ transforms reference edge vectors to target edge vectors: $\mathbf{T}_{ijk} \cdot(\bm{\upzeta}_i - \bm{\upzeta}_j) =\tvec_{i} - \tvec_{j}$.
In complexified notation,  $\zeta=\zeta_1 +i\zeta_2, \bar{\zeta} = \zeta_1-i\zeta_2$, the affine map $\mathrm{T}_{ijk}$ becomes:
\begin{align}
\begin{aligned}
    \mathrm{T}(\zeta,\bar\zeta) &= \mathrm{T}_0 + \mathbf{T}_{ijk}\cdot\boldsymbol{\upzeta} \\
    &= \mathrm{T}_0 + \frac{(\mathrm{T}_{11}+\mathrm{T}_{22}) + i (\mathrm{T}_{21} -\mathrm{T}_{12})}{2} \zeta  \\
    &\qquad + \frac{(\mathrm{T}_{11}-\mathrm{T}_{22}) + i (\mathrm{T}_{21}+\mathrm{T}_{12}) }{2} \bar\zeta
\end{aligned}
\end{align}
where we have suppressed the index $ijk$ to simplify notation. 
Eq.~\eqref{eq:beltrami_LTC} is closely related to the standard Beltrami coefficient in complex analysis~\cite{Ahlfors1966}:
\begin{align}
    \label{eq:beltrami_zBar}
   \frac{\partial \mathrm{T} /\partial \bar \zeta}{\overline{\partial \mathrm{T} /\partial \zeta}} = \frac{s_1-s_2}{s_1+s_2} \, e^{2i\phiT} \approx \frac{1}{2} \muT
\end{align}
We now upgrade the collection of maps $\mathrm{T}_{ijk}$, defined separately for each  triangle, to a globally defined map $T(\zeta,\bar\zeta)$
from the circle-packing of the adjacency graph to the 2D tension triangulation.
At vertices, $T(\boldsymbol{\upzeta}_i) = \boldsymbol{\uptau}_i$. Elsewhere, $T(\boldsymbol{\zeta})$ is defined by linear interpolation (App.~\ref{app:interpolation}), so that on each triangle, the Jacobian $(\nabla T)_{ijk}$ is constant.
Locally, $T(\zeta,\bar\zeta)$ maps (approximately) equilateral adjacency triangles to tension triangles.
The Jacobian $(\nabla T)_{ijk}$ is equivalent to the single-triangle LTC matrix $\mathbf{T}_{ijk}$. 

The adjacency triangulation provides an intrinsically defined gauge field (or connection) that relates the orientations of adjacent reference triangles.
Locally, the LTC phase $\psi$ measures the triangle's hexatic phase \emph{relative to the nematic phase} $\phiT$. Spatial correlations in $\psi$ can, therefore, be used to define a ``hexanematic'' order parameter \cite{Claussen.etal2024}.
For a network with coordination number $6$ for every cell, the adjacency triangulation is a periodic lattice of equilateral triangles. Thus, the reference triangles are globally aligned, and the hexatic phase in real space is given by $\phiT + \psi$.
Topological defects (cells with coordination number $\neq \, 6$) will distort the adjacency triangulation and therefore act as an obstruction to hexatic order in physical space.

The tension anisotropy encoded in $\muT$ describes how tension $\uptau_{ij}$ depends on interface orientation. Importantly, it is \emph{distinct} from the macroscopic stress anisotropy, which, by  Eq.~\eqref{eq:batchelor_general}, additionally depends on the interface lengths $\upell_{ij}$ (a short interface contributes less to the stress). Indeed, we showed that the macroscopic stress is independent of the local tension configuration.
Instead, the tension triangulation acts much like a finite-element mesh onto which the Airy function $\Psi$ is discretized.

\begin{figure*}
    \centering
    \includegraphics{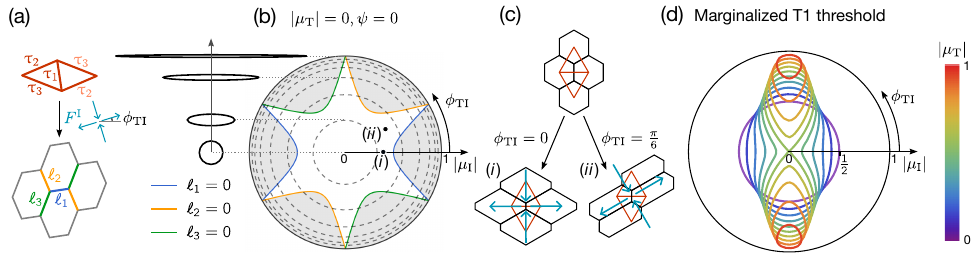}
    \caption{
    T1 threshold for a periodic lattice.
    (a)~Geometry setup and labels. The tension triangle is oriented such that the anisotropy axis points along the $x$-axis (i.e.\ $\phiT = 0$). The angle $\phiTI = \phiI - \phiT$ measures the isogonal deformation axis relative to this tension anisotropy axis.
    (b)~Locus where one of the edge lengths $\upell_\mu$ vanishes, as a function of the isogonal deformation anisotropy $|\muI|$ and orientation $\phiTI$ for an equilateral tension triangle $|\muT| = 0$ with (relative) lattice orientation $\psi = 0$. Areas shaded in gray are unphysical configurations in which one of the edge lengths is negative.
    (c)~Illustration of isogonal compression (\textit{i}) and stretching (\textit{ii}) along an edge. In the latter case, the T1 threshold approaches $|\muI| = 1$ at which the deformation becomes degenerate.
    (d)~T1 threshold ``marginalized'' over the shape phase $\psi$ for a range of magnitudes of tension anisotropy $|\muT|$. 
    }
    \label{fig:T1-thresh}
\end{figure*}

\subsection{T1 threshold}

Physically, a cell interface must have non-negative length, $\upell_{ij} \geq 0$.
When the length vanishes, $\upell_{ij}=0$, a T1 cell rearrangement event takes place, corresponding to an edge flip in the tension triangulation. The $\upell_{ij}$ are determined by the local geometry of the tension triangulation and the isogonal mode (Eq.~\eqref{eq:isogonal_length} in App.~\ref{app:Voronoi}). For Voronoi diagrams, $\upell_{ij}^\mathrm{V} \geq 0$ leads to the Delaunay condition Eq.~\eqref{eq:Delaunay}. For a non-zero isogonal mode, we refer to the condition $\upell_{ij}\geq 0$ as the \emph{generalized Delaunay} condition.
Since $\upell_{ij} = 0$ is invariant under local scaling, conformal deformations do not affect the Delaunay condition. 
Mechanically, T1s define the ``yield strain'' from the tension network microstructure.

We now obtain a tractable version of the generalized Delaunay condition, a ``T1-threshold'', in terms of the LTC parameters and the isogonal strain tensor $\Fiso$. 
Recall that Eq.~\eqref{eq:quadratic_power_to_Voronoi} shows that for constant $\Fiso$, power diagrams can be mapped to Voronoi diagrams. This allows mapping the generalized Delaunay condition to the ordinary one.

The length $\upell_{ij}$ depends on the two tension triangles $ijk$, $ijl$ meeting at the edge $ij$~\footnote{The T1 threshold is thus fully \emph{local}: it depends on the configuration at interface $ij$ only, in contrast to the energy barriers for T1 transition in ``elastic'' vertex models, like the area-perimeter model.}. Due to their shared edge, adjacent triangles are strongly correlated~\cite{Claussen.Brauns2025}. We therefore consider symmetric kites in which the $ijk$ and $ijl$ are identical, effectively, by working in the simplified setting of the regular lattice from Sec.~\ref{sec:lattice} (Fig.~\ref{fig:lattice-setup-main}).
Appendix~\ref{app:T1_threshold} shows that
\begin{align}
    \upell_{ij}\propto   -\tvec_{jk}\cdot\Fiso\cdot \tvec_{ki}
\end{align}
where $ij,jk,ki$ are the edges in clockwise order.
This allows calculating the threshold $\upell_{ij} = 0$ from the isogonal deformation $\Fiso$ and the shape of a tension triangle $ijk$, described by the map $\mathbf{T}$.

Up to a scale factor, the isogonal deformation $\Fiso$ is characterized by its Beltrami coefficient $\muI$, encoding anisotropy magnitude $|\muI|$ and orientation $\arg \muI = 2\phiI$ (see App.~\ref{app:T1_threshold}).
Using Eq.~\eqref{eq:isogonal-deformation-lattice}, we compute the three lengths $\upell_\mu(\muT, \muI, \psi)$ corresponding to the triangle (Fig.~\ref{fig:T1-thresh}a).
By rotation invariance, $\ell_\mu$ only depend on the relative angle $\phi_\mathrm{TI} = \phiI-\phiT$.
A T1 event happens whenever one of the lengths hits zero, defining a critical manifold
\begin{equation}
    \mathcal{C}_\mathrm{T1} : \min_{ij} \upell_{ij}(|\muI|, |\muT|, \phi_\mathrm{TI}, \psi) = 0
\end{equation}
Fig.~\ref{fig:T1-thresh}b shows $\mathcal{C}_\mathrm{T1}$ in the polar $(|\muI|,\phi_\mathrm{TI})$ plane for the case of an equilateral tension triangle ($|\muT| = 0$) with phase $\psi = 0$.
Additional plots for different triangle shapes $(|\mu_\mathrm{T}|,\psi)$ are shown in Fig.~\ref{fig:T1-threshold_q-sweep}.

The T1 threshold manifold $\mathcal{C}_\mathrm{T1}$ defines a yield strain magnitude $|\muI^\mathrm{T1}|(|\muT|, \phi_\mathrm{TI}, \psi)$ -- the maximum amount of isogonal strain that can be sustained.
Notably, T1 transitions can be triggered by changes in the tensions $\uptau_{ij}$ even in the absence of isogonal deformation.
Tension dynamics can drive one of the cell interfaces to length zero in the stress-free Voronoi reference configuration; see Fig.~\ref{fig:T1-threshold_q-sweep}(b).
There is, hence, a ``spectrum'' of T1 transitions with the limiting cases of \emph{passive} T1s, driven via the isogonal mode by external forces, and \emph{active} T1s driven by local tension dynamics~\cite{Brauns.etal2024}.

\subsubsection{Effect of the LTC phase}

In the periodic lattice, this T1 threshold for both active and passive T1s  sensitively depends on the LTC phase $\psi$.
The yield strain is minimal for shear perpendicular to a cell interface (Fig.~\ref{fig:T1-thresh}c(i)).
This effect is captured by the (hexatic) lattice orientation $\phiT + \psi$ relative to the principal shear orientation $\phiI$, as illustrated in Fig.~\ref{fig:T1-thresh}c \footnote{We define the phase $\arg \muT$ as $2\phiT$, not vice versa. This is important in the case $|\muT| = 0$ where $\arg \muT$ is a priori indeterminate but $\phiT + \psi$ determines the hexatic phase, so $\phiT$ is defined even if tension is isotropic.}. 
For active T1s,  the Voronoi edge length sensitively depends on whether the triangle is obtuse ($\psi \approx 0$) or acute ($\psi \approx \pi/6$): for obtuse triangles, a smaller tension anisotropy suffices to drive T1s.

This effect is not limited to a periodic lattice. Significant bias (or ``polarization'') in the distribution of $\psi$ is also possible in an amorphous tissue without hexatic order. This is because the hexatic phase in physical space is given by $\psi + \phiT$, so sufficient disorder in the nematic orientation $\phiT$ (with a standard deviation around $\pi/6$) is sufficient to destroy hexatic order even if the $\psi$-distribution is sharply peaked.
Capturing this subtle interplay between nematic and hexatic order and their roles for T1 transitions (both active and passive) poses an interesting challenge for building continuum theories. It has important applications in understanding the emergence of hexatic order in tissues such as the \textit{Drosophila} pupal wing~\cite{Classen.etal2005} and eye epithelia~\cite{Couturier.etal2025}.

\subsubsection{The marginal T1 threshold}

A significant simplification of the T1 threshold can be obtained when $\psi$ is uniformly distributed. One can then define a ``marginal'' T1 manifold 
\begin{equation}
    \label{eq:C_tilde_T1}
    \tilde{\mathcal{C}}_\mathrm{T1} : \min_{\mu=12,23,31} \min_\psi \upell_\mu(|\muI|, |\muT|, \phi_\mathrm{TI}, \psi) = 0
\end{equation}
which tracks when a T1 will first happen for any value of $\psi$.
App.~\ref{app:marginal_T1} shows that Eq.~\eqref{eq:C_tilde_T1} leads to
\begin{equation}
    \label{eq:marginal_T1_exact}
    \left(1-|\muI|^2\right)\left(1-|\muT|^2\right)
    =\frac{3}{4}\left(1+|\muI|\,|\muT|\cos2\phiTI\right)^2
\end{equation}
as defining equation for $\tilde{\mathcal{C}}_\mathrm{T1}$.
A family of T1 thresholds in the polar $(|\muI|, \phi_\mathrm{TI})$ plane is shown in Fig.~\ref{fig:T1-thresh}(d).
For $|\muT| = 0$, this marginalized T1 threshold is simply given by $|\muI| = 1/2$. The same is true when exchanging $\mathrm{I}$ and $\mathrm{T}$. In fact, the entire marginal T1 manifold is symmetric under this exchange (Fig.~\ref{fig:T1-thresh_psi-marginalized}a) as is evident from Eq.~\eqref{eq:marginal_T1_exact}.
In other words, macroscopic stress anisotropy and microscopic tension anisotropy play identical roles in the $\psi$-marginalized setting.

One can define a ``composite'' Beltrami coefficient that combines tension and isogonal-strain anisotropy:
\begin{equation} \label{eq:Beltrami-composition}
    \mu_\mathrm{TI} := \frac{\muT + \muI }{1 + \bar\mu_\mathrm{T} \muI},
\end{equation}
The exact, geometric T1 threshold Eq.~\eqref{eq:marginal_T1_exact} is well approximated by the combined Beltrami coefficient of tension and isogonal anisotropy: The locus $|\mu_\mathrm{TI}| = 1/2$ matches $\tilde{\mathcal{C}}_\mathrm{T1}$; see Fig.~\ref{fig:T1-thresh_psi-marginalized}(c).

In summary, using the link between geometry (isogonal mode $\Fiso$) and mechanics (stress tensor $\overline{\boldsymbol{\sigma}}$), one can calculate the yield stress of a ``generalized foam'' with arbitrary tension configurations from the microscopic geometry. Yielding occurs via T1 transitions, which reduce the isogonal strain. Related questions have been previously studied in the literature on 2D foams~\cite{Marmottant.etal2008,Raufaste.etal2010} via phenomenological approaches.

\section{Continuum limit of tension networks}
\label{sec:continuum_limit}

\begin{figure}
    \centering
    \includegraphics{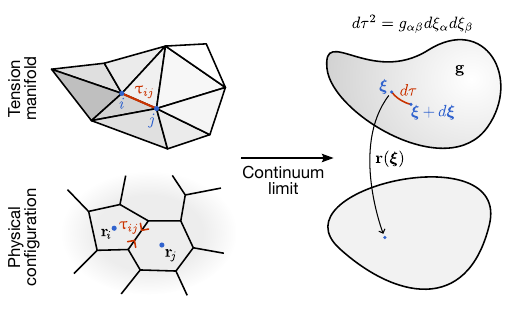}
    \caption{A triangulation (left) is a discrete Riemannian surface.
    In the continuum limit, the tension triangulation defines a Riemannian metric. An embedding of the metric into physical space is the continuum equivalent of the centroids in the cell tiling. Adapted from Ref.~\cite{Claussen.etal2026}.
    }
    \label{fig:tri-to-mfd}
\end{figure}

The discussion of T1 transitions concludes our inquiry into the mechanics, topology, and dynamics of active tension networks on the cell scale. We now explain how an effective continuum description emerges when the cell-scale theory is coarse-grained. This ``bottom-up'' analysis complements the continuum theory, which Ref. \cite{Claussen.etal2026} derives ``top-down'' without reference to the particular microstructure assumed here. Table~\ref{tab:placeholder_discrete_continuum} summarizes the discrete-continuum dictionary.

\subsection{Coarse-graining using discrete conformal geometry}

The key to linking microscopic and macroscopic scales lies in formulating mechanics in geometric terms. 
One can then think of the cell tiling's discrete geometry as an approximation of a smooth, differential geometry, an idea that goes back to Regge's work on general relativity~\cite{Regge1961}.
In particular, we draw on the powerful mathematics of discrete conformal maps, which goes back to the work of Thurston and Koebe on circle packings.
We showed that M\"{o}bius diagrams, previously introduced to parametrize mechanically balanced cell tilings~\cite{Moukarzel1997,Noll.etal2020}, can be understood as the image of the tension triangulation under discrete conformal maps. This ``dual'' formulation of discrete conformal maps is a new result that may be of independent mathematical interest. Discrete conformal maps also underlie powerful numerical algorithms for surface processing, developed by the computer graphics community \cite{Crane.etal2013}, which may now be harnessed for numerical simulations of tissue mechanics.

The literature on foam mechanics~\cite{Weaire1999,Drenckhan.etal2004,Mancini2005} previously noted the invariance of force balance under Möbius transformations, and demonstrated it in beautiful experiments~\cite{Elias.etal1999}. We generalized this result: first, from an ordinary foam with constant surface tensions to an arbitrary tension triangulation, second, from a global to a local Möbius symmetry, which approximates arbitrary conformal maps in the continuum limit, and lastly, by computing the resulting pressure change.

\subsection{From tension triangulation to tension manifold}

One can treat the tension triangulation as a continuous surface, the tension manifold (see Sec.~\ref{sec:intrinsic}). In the continuum limit, a set of Lagrangian ``material coordinates'' $\xivecIt$ plays the role of the cell labels $i,j, \dots$. The triangulation edge lengths define a Riemannian \emph{tension metric} $\bm{g}$. The metric measures the tension between nearby cells at coordinates $\xivecIt_i, \, \xivecIt_j = \xivecIt_i + d\xivecIt_{ij}$:
\begin{align}
    \uptau_{ij}^2 \approx g_{ab}(\xivecIt)d\xi_{ij,a} d\xi_{ij, b}
\end{align}
The physical cell tiling defines an embedding $\rvecIt(\xivecIt)$ of the tension manifold into physical space, with $\rvecIt(\xivecIt_i)$ corresponding to the centroid of cell $\rvec_i$ (Fig.~\ref{fig:tri-to-mfd}).

We showed that the (intrinsic) Voronoi dual of the tension triangulation is stress-free, and that stress arises from deformation away from this reference. Accordingly, in the continuum theory of Ref.~\cite{Claussen.etal2026}, the tensile stress is determined by the deformation of the metric $\bm{g}$ by the embedding $\rvecIt(\xivecIt)$.

\subsection{From discrete to continuous conformal maps}

In the continuum limit, a discrete conformal map with scale factors $\uplambda_i  =1/\upp_i$ converges towards a conformal map with factor $\lambda(\xivecIt)$~\cite{Springborn.etal2008}. The 2D drawing $\tvec_i$ of the tension triangulation becomes a conformal embedding of the tension manifold, denoted $\bm{z}(\xivecIt)$ in Ref.~\cite{Claussen.etal2026}. 
Like in discrete Voronoi diagrams, in the $\bm{z}(\xivecIt)$-configuration, the continuum tensile tensor is isotropic, $\sigma_{ab} = \lambda(\bm{z})^{-1} \delta_{ab}$, so the effective pressure equals the conformal factor\footnote{Ref.~\cite{Claussen.etal2026} uses the \emph{inverse} convention for the conformal factor of the metric, $\lambda \leftrightarrow\lambda^{-1}$.}.
The latter obeys the Liouville equation:
\begin{align}
    \label{eq:liouville}
    K = \lambda^{2} \Delta \log \lambda
\end{align}
where $K$ is the Gaussian curvature of the tension metric. 
Eq.~\eqref{eq:liouville} is the equivalent of the discrete Eq.~\eqref{eq:Poisson_pressure_discrete}, a generalization of the von Neumann law for the pressure in a foam bubble.
In both discrete and continuous settings, a conformal embedding of the tension manifold defines a stress-free reference state for the tissue. Gaussian ``tension curvature'' acts as a source of conformal deformation, which physically implies pressure gradients.

The link between conformal factor (geometry) and pressure (mechanics) ultimately arises from the Cauchy-Riemann (CR) equations for conformal maps. In particular, the CR equations link the conformal factor $\lambda$ and the vorticity $\omega$ of the map:
\begin{align} \label{eq:Cauchy-Riemann_pressure}
    \partial_a \log \lambda = \epsilon_{ab} \partial_b \omega.
\end{align}
Eq.~\eqref{eq:Cauchy-Riemann_pressure} plays the role of a continuum form of the Young--Laplace law: curvature (a gradient in vorticity) is balanced by a gradient in pressure (conformal factor) along the orthogonal direction.

\subsubsection{Adiabatic Beltrami flow}

We showed that adiabatic tension dynamics deforms the stress-free reference state of the cell tiling.
In the continuum theory, tension dynamics correspond to changes in the tension metric, and the equivalent of the reference state is the conformal embedding $\bm{z}(\xivecIt)$. The change of a conformal embedding is governed by a so-called Beltrami equation~\cite{Ahlfors1966}, which precisely reproduces Eq.~\eqref{eq:adiabatic_flow_lattice} for the adiabatic tissue flow.
(The cross-ratio operator $\mathcal{P}\circ \mathcal{D}$ of Sec.~\ref{sec:Beltrami} is the discrete equivalent of the continuum $\partial / \partial\bar{z} = \tfrac{1}{2}\left( \partial/\partial r_1 +i\partial/\partial r_2\right)$; both annihilate conformal maps.)

\subsection{Isogonal mode and effective stress-strain relationship}

Via the Voronoi construction, the tension triangulation defines an effective reference state for the cell tiling. Isogonal modes displace cell vertices by a discrete gradient, Eq.~\eqref{eq:isogonal_action}. By interpolation, Eq.~\eqref{eq:interpolation}, the discrete isogonal potentials $\uptheta_i$ define a continuous function $\theta(\tvecIt)$. In the continuum, the isogonal mode corresponds to curl-free deformations of the embedding from the reference state $\bm{z}(\xivecIt)$, $\bm{z} \mapsto \rvecIt = \bm{z} - \nabla\theta$, with strain tensor $F_{ab} = \partial r_b / \partial z_a$.
Like in the discrete Eq.~\eqref{eq:isogonal_strain_legendre},
Ref.~\cite{Claussen.etal2026}  finds an effective stress- (curl-free) strain relation: the potential $\theta$ is the effective Airy function, $\sigma_{ab} = \epsilon_{ac}\epsilon_{bd}\partial_c\partial_d \theta$.
(App.~\ref{sec:macroscopci_stress_SVD} shows that the continuum and discrete stress-strain relations also match at the non-linear level).
Due to the constitutive equation for pressure, the continuum Airy function obeys a biharmonic equation analogous to Eq.~\eqref{eq:biharmonic}.

\subsubsection{Deformation modes of tension networks}

In general, compatibility with continuum force balance requires that any deformation be composed of a curl-free part and a conformal part. Thus, continuum force balance reproduces the two deformation modes we find at the cell-level.
These modes characterizing the response of tissue to external forces determine how the active force dipoles (cell edges) are embedded into physical space. They thus ``mediate'' between the ``microscopic'' and macroscopic mechanical state of the tissue. As a result, we find an emergent stress-strain relationship, even though individual cell edges are far from being passive elastic objects.
The isogonal mode sets the cell interface length and the macroscopic stress tensor.
By contrast, the conformal mode sets cell edge orientation, so that force balance in the conformal ``sector'' can be thought of as torque balance. 
In our framework, torque balance is automatic and does not impose additional conditions~\cite{Jensen.etal2020}.
One may also consider the conformal mode as a ``Goldstone mode'' corresponding to the scale-invariance of the force-balance constraint.

\subsubsection{Link to the finite-element method}

The interpolation method used to pass between discrete and continuous functions is known as the linear Lagrange element in the finite-element literature~\cite{Brenner.Scott2008}. The discrete Laplace operator Eq.~\eqref{eq:Laplace-from-div} also arises in finite-element discretizations
(the power diagram Laplace operator $\Delta^\mathrm{P}$ corresponds to an anisotropic Laplacian in the continuum, App.~\ref{app:nonflat_isogonal_continuum}).
Strikingly, in this interpolation scheme, the discretization of a continuous stress tensor reproduces the microscopic force-balance equations (App.~\ref{app:stress_discretization})~\cite{Desbrun.etal2013}. 
The cell-level triangulation appears as a finite-element-like discretization of the continuum theory.
The tension triangulation only determines the ``interpolation nodes'' $\tvec_i$ at which the effective Airy function $\uptheta_i=\theta(\tvec_i)$ is evaluated. The continuum field $\theta$, and thus the macroscopic stress tensor, are independent of the triangulation structure. This result reveals emergent simplicity: the macroscopic stress is independent of details of the local tension configuration.

\subsection{Cell topology in the continuum limit}

Overall, the tension triangulation encodes two distinct pieces of information: first, the surface it defines, and second, the cell network's adjacency graph (i.e., which pairs of cells are adjacent). Different triangulations can realize the same surface (e.g., different triangulations of the plane), ultimately leading to the distinction between macroscopic stress and microscopic tension configuration. In the continuum, the cell network topology is represented by a second Riemannian metric $a_{ab}(\xivecIt)$, defined so that the distance between adjacent cells is $1$. The adjacency metric $\bm{a}$ corresponds to a triangulation where every edge has length 1.

To parametrize the discrete adjacency information, we introduced the Thurston circle packing $\bm{\upzeta}_i$, again using discrete conformal maps. The ``Thurston coordinates'' translate topology into geometry and provide a privileged parametrization that reveals the local tension configuration (tension triangle shape) central to the description of active T1 transitions.
They may also be a useful tool to analyze the adjacency patterns and dynamics of tissues in experimental data~\cite{Merkel.etal2017}.
For such data analysis, circle packs of the unit disk may be convenient, since they automatically register different datasets into a common reference frame. 

In the continuum limit, the circle pack $\bm{\upzeta}_i$ of the adjacency graph becomes a conformal embedding $\bm{\zeta}(\xivecIt)$ of the adjacency metric $\bm{a}$. The map between the two embeddings $\bm{\zeta}(\xivecIt)$ and $\bm{z}(\xivecIt)$ is the equivalent of the map $\mathbf{T}$ from the reference, equilateral, to the tension triangle. This map thus encodes the tension anisotropy via its Beltrami coefficient, denoted $\mu_{\zeta|z}$.
The adjacency metric and the ``Thurston embedding'' $\bm{\zeta}(\xivecIt)$ thus define a systematic way of accounting for cell topology in the continuum limit, with a direct connection to the underlying discrete theory.

Cell rearrangement (T1 transitions) dynamically changes the adjacency graph, and thereby the adjacency metric $\bm{a}$ (by increasing the ``adjacency distance'' between cells that lose contact).
Here, in stark contrast to macroscopic stress, the triangulation ``microstructure''~\cite{Claussen.Brauns2025} becomes relevant. We showed that the threshold for T1s can be approximated by the combined tension- and isogonal strain anisotropy. This approximation underlies the phenomenological continuum description of the T1 rate in Ref. ~\cite{Claussen.etal2026}.

\begin{table*}[]
    \renewcommand{\arraystretch}{1.4}
    \centering
    \begin{tabular*}{\textwidth}{l@{\extracolsep{\fill}}l}
    \toprule
    \multicolumn{2}{@{}c@{}}{%
      \begin{tikzpicture}[>={Stealth[length=5pt]}]
        \def\xL{0.}   
        \def\xR{0.443}   
        \def\aL{0.09}   
        \def\aR{0.34}   
        \path (0,0) (\textwidth,0);
        \node at (\xL\textwidth,0) {Discrete};
        \node at (\xR\textwidth,0) {Continuum};
        \draw[->] (\aL\textwidth, 0.12) -- node[above=-1pt]{\small continuum limit}
                  (\aR\textwidth, 0.12);
        \draw[->] (\aR\textwidth,-0.12) -- node[below=-1pt]{\small piecewise-linear discretization}
                  (\aL\textwidth,-0.12);
      \end{tikzpicture}} \\
    \midrule
    Cell labels $i,j,k, \cdots$ & Material coordinates $\xivecIt$ \\
    Tension triangulation $\uptau_{ij}$ & Riemann surface with tension metric $g_{ab}$ \\
    Tension vertices $\tvec_i$ and their Voronoi dual & Conformal embedding $\bm{z}(\xivecIt)$ of tension metric \\
    Discrete conformal factor $\uplambda_i=1/\upp_i$ & Conformal factor $\lambda(\xivecIt)$\\
    Power displacements $(\nabla \uptheta)_{ijk}$ & Isogonal/curl-free mode $\nabla \theta$ \\
    Discrete conformal (piecewise Möbius) map $M_{ijk}$ $\quad \quad$ & Conformal mode  \\
    Adjacency graph $A_{ij}$ and circle packing $\bm{\upzeta}_i$ & Adjacency metric $a_{ab}$ and its isothermal embedding $\zeta(\xi)$\\
    Local tension configuration $\mathbf{T}$ & Quasi-conformal map from adjacency- to tension-embedding $\mu_{\zeta|z}$ \\
    T1 transitions (remodeling of adjacency graph $A_{ij}$) & Dynamics of the adjacency metric \\
    \bottomrule
    \end{tabular*}
    \caption{Dictionary for the transition between the discrete and continuum theories.
    }
    \label{tab:placeholder_discrete_continuum}
\end{table*}

\section*{Discussion}
\setcounter{subsection}{0}

Unlike conventional elastic materials, many living tissues have no fixed reference state and control active stresses via motor molecule activity, rather than a constitutive relationship. 
By studying the geometry of force balance, we showed that the configuration of active tensions in 2D epithelia -- interpreted as a triangulation dual to the cell tiling -- gives rise to an emergent, reference state where tensions balance against a uniform pressure.
This explains how cells can define a tissue's macroscopic shape by controlling the microscopic tensions.

\subsection{Force balance and morphing flow}

Overall, we find that the mechanics of active tension networks can be described as ``emergent elasticity'': the combination of active tensions and pressure allows tissues to resist external (shear) forces, in contrast to models where the tissue shear modulus can be tuned to vanish~\cite{Farhadifar.etal2007,Bi.etal2015}. Tissue flow emerges through the adiabatic dynamics of active tensions, ``locked in'' by topological cell rearrangement.
We refer to this phenomenology -- plastic deformation through internal tension dynamics, while remaining rigid to external forces -- as ``morphing flow''.
Plastic flow can be driven both internally, by tension dynamics, and by external forces. A rigorous link between the tissue's mechanical state and local plasticity is a key advance of our geometric framework. It underlies an elegant formulation of plastic flow in the continuum~\cite{Claussen.etal2026}, using quasiconformal maps to represent tension anisotropy and isogonal strain.

\subsection{Macroscopic stress versus local tension anisotropy}

We showed that in tissues where cell-scale mechanics is dominated by active interfacial tensions, the macroscopic stress is \emph{independent} of the local tension configuration. In geometric terms, the macroscopic mechanics is determined by the ``coarse shape'' of the tension surface alone, independently of how it is triangulated. Specifically, in the Voronoi reference configuration, anisotropic local tensions lead to anisotropic cell shapes, but macroscopic stress remains isotropic.
Macroscopic stress results from forces applied at the boundary, which deform the tissue away from the Voronoi reference state.

In many hydrodynamic models for tissue flow, tension anisotropy is identified with an anisotropic active stress~\cite{Streichan.etal2018,Ibrahimi.Merkel2025,Ioratim-Uba.etal2023,Serra.etal2023}.
Our results argue that the relationship between local tensions and macroscopic stress can be more subtle, since the latter depends on how tension-bearing cell interfaces are ``embedded'' in real space. The relation between stress and embedding underlies the continuum theory presented in the companion manuscript~\cite{Claussen.etal2026}.

Tension and stress anisotropy are thus two distinct observables. Experimentally, they can be assessed by laser ablation of individual interfaces and of macroscopic tissue regions, respectively (Fig.~\ref{fig:laser-ablation}).
Macroscopic stress can be isotropic, even when tensions are anisotropic.
For instance, during convergent extension in gastrulating \emph{Drosophila} embryos, interfacial tension is highly anisotropic~\cite{Bertet.etal2004,Brauns.etal2024}. We predict that, if external pulling on the tissue is abolished (as in a \emph{twi} mutant~\cite{Gustafson.etal2022}), macroscopic stress will be isotropic despite anisotropic interfacial tensions. Overall, our results have direct implications for interpreting experimental measurements of tissue mechanics.
The macroscopic stress tensor is an important additional diagnostic, in addition to the cell-level tensions and their anisotropy.
Both local tensions and macroscopic stress can be inferred directly from microscopy data~\cite{Noll.etal2017}.

\begin{figure}
    \centering
    \includegraphics[width=0.5\textwidth]{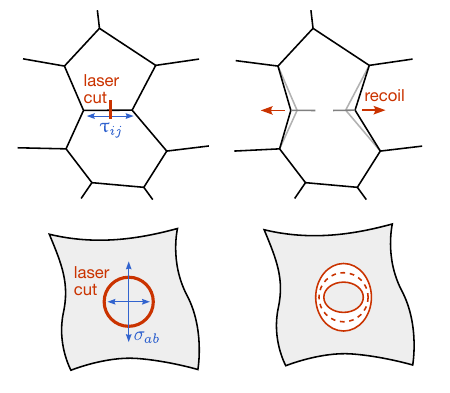}
    \caption{Top: Laser ablations of individual interfaces measure the tension $\uptau_{ij}$ on an individual cell-cell interface ($i,j$ label adjacent cells).
    Bottom: Laser ablations on the tissue scale measure the macroscopic stress tensor $\sigma_{ab}$ (where $a,b$ are coordinate indices).
    }
    \label{fig:laser-ablation}
\end{figure}

\subsection{Modeling morphogenesis with tension networks}

As outlined in the Introduction, we believe that actively generated forces dominate in many morphogenetic systems. This makes models based on an assumed reference shape (i.e., generalized spring networks) inapplicable. 
We believe that active tension networks represent a step towards a theory of morphogenesis that puts active forces first, and reflects the fundamental differences between inert and biological matter.

ATNs are highly idealized descriptions of epithelia, neglecting the effects of traction forces, apico-medial myosin, intermediate filaments, or multilayered tissues.
Nevertheless, they can be applied to concrete experimental systems.
In past work~\cite{Noll.etal2017,Brauns.etal2024,Claussen.etal2024}, we have used the ATN framework to understand a specific episode of \emph{Drosophila} development, namely germ-band extension, a paradigmatic example of tissue elongation through convergent extension.
We hope that the improved understanding of active mechanics presented here will inspire the ``physics of morphogenesis'' community to adapt ATNs to new systems.
For example, the gastrulating avian embryo represents an interesting opportunity. Like in \emph{Drosophila}, the embryonic epithelium is suspended on a fluid yolk (so no traction forces), and tissue flow is driven by junctional actomyosin activity~\cite{Saadaoui.etal2020,Serra.etal2023,Caldarelli.etal2024}.
Settings that feature more localized cell rearrangement could also be of interest, for example, the generation of hexagonal order in the \emph{Drosophila} wing~\cite{Classen.etal2005}.

The present framework starts from prescribed tensions and derives the resulting cell tiling.
To model biophysical dynamics, this mechanical formalism must be combined with a concrete model for tension dynamics.
For instance, in Ref.~\cite{Claussen.etal2024}, we proposed a model for convergent extension tissue flow based on a positive mechanical feedback loop, where tension mediates recruitment of motor molecules. 
This leads to non-linear dynamics for the tensions:
\begin{align}
    \label{eq:tension_dynamics_positive_feedback}
    \frac{1}{\uptau_{ij}}\frac{d \uptau_{ij}}{dt} = k_\mathrm{on} \uptau_{ij} - k_\mathrm{off}\!\left(\upell_{ij}^{-1}{\dot \upell_{ij}},\, \{\upp_i, \upp_j\} \right)
\end{align}
In Ref.~\cite{Claussen.etal2024}, the off-rate $k_\mathrm{off}$ served as a Lagrange multiplier to conserve the total motor-molecule mass (i.e., the average tension).
More generally, the off-rate can implement stabilizing negative feedback loops via mechanosensitive dependence on the interface strain rate $\upell_{ij}^{-1}{\dot \upell_{ij}}$~\cite{Noll.etal2017,Gustafson.etal2022}, or pressures $\upp_i, \upp_j$ (see below).
Future ATN models could include stochastic fluctuations~\cite{Kim.etal2021} or control by morphogens~\cite{Ibrahimi.Merkel2025}, as well as additional mechanical effects like traction on a substrate.

Tension dynamics deform the tension triangulation, and hence the tissue's rest shape.
In Ref.~\cite{Claussen.etal2024}, we used this geometric picture to investigate what mechanical feedback loop is required to drive tissue shear flow and cell rearrangement.
We introduced a classification of tension dynamics by their ``geometric phase portrait'' which represents how they deform individual tension triangles.
The idealized nature of ATNs thus allows a very precise understanding of how local active force dynamics translate into morphogenesis, which has been a challenge in more complex models combining active and passive components.

Within the ATN framework, mechanical information can be inferred from microscopy data~\cite{Noll.etal2020, Brauns.etal2024}. The inferred tensions allow direct comparison between model and experiment, and can be incorporated into models, for example as initial conditions. One can also imagine fitting dynamical models directly to the inferred experimental tensions using statistical methods~\cite{Brunton.Kutz2019}.

Finally, microscopic tension dynamics can be coarse-grained into effective continuum dynamics. We took a first step towards this goal in Ref.~\cite{Claussen.Brauns2025}, where we characterized the mechanical state of a mesoscopic tissue patch by a probabilistic ensemble of tension triangle shapes in a mean-field setting. 
We hope that the continuum theory for ATNs, briefly outlined in Sec.~\ref{sec:continuum_limit} and developed in detail in Ref.~\cite{Claussen.etal2026}, will lead to deeper theoretical insight into emergent elasticity and active plastic flow on the tissue scale.

\subsubsection{Mechanosensation and feedback loops}

Recent work has highlighted the importance of mechanical feedback loops in morphogenesis~\cite{Heisenberg.Bellaiche2013,Lenne.etal2021}. 
Just like experimentalists, cells need distinct tools to sense and respond to local tensions and macroscopic stress.
Interfacial tension can be measured locally by force-bearing molecules embedded in the actomyosin cortex, like vinculin, $\alpha$-catenin, or myosin itself, whose binding rates depend on tension~\cite{Petridou.etal2017}.
By contrast, interfacial tension alone is not informative of macroscopic stress, which instead must be sensed from deformations of the cell ``bulk''. For example, the nuclear envelope is tethered to the cytoskeleton and is capable of mechanosensation~\cite{Mejat2010}.

This distinction between tension- and bulk stress feedback loops should also be reflected in cell-scale and continuum models.
Properly accounting for mechanical feedback loops is particularly important because they control the active stress, which is dominant in living systems. These feedback loops replace the constitutive relation on the cell scale, determining, for instance, the response to external force~\cite{Noll.etal2017}.

The structure of the ATN model provides a clean decomposition between tensions and (isogonal) bulk stress. This provides a systematic way to incorporate both cell-interior stresses (e.g.\ due to stress fibers or intermediate filaments) and junctional forces into cell- or continuum-level models.

\subsubsection{Control of intracellular pressure}

In our model, the key variable characterizing the mechanics of the cell bulk is the intracellular pressure.
In mechanical balance, cells cannot independently change pressure and tensions. A generic change in the tensions $\uptau_{ij}$ leads to a non-zero angle deficit $\updelta_i$ (discrete Gaussian curvature).
By Eq.~\eqref{eq:Poisson_pressure_discrete}, tension curvature leads to pressure differentials.
In a conventional foam, pressure differentials slowly equilibrate via diffusion, leading to foam coarsening. In cells, coarsening is prevented by osmotic pressure, since ions and proteins cannot diffuse through the membrane.

Over longer timescales, cells actively regulate their volume and osmotic pressure via directed transport of ions and other small osmolytes. This means that the equation of state for pressure is under biological control.
Since the pressures in force balance are determined by their values at the tissue boundary, pressure feedback may enable local sensing of large-scale deformations and overall tissue size \cite{Irvine.Shraiman2017}.

Pressure regulation must also be coupled to the tension in the actomyosin cytoskeleton.
For example, cells could reduce pressure gradients by modulating their overall tension level
\begin{align}
    \frac{1}{\uptau_{ij}}\frac{d\uptau_{ij}}{dt}=  k_{\mathrm{off}, \Delta p} \cdot \big[ (\Delta \upp)_i + (\Delta \upp)_j \big]
    \label{eq:Ricci_flow_discrete}
\end{align}
Eq.~\eqref{eq:Ricci_flow_discrete} is a discrete Ricci flow~\cite{Tao2008} that smooths out curvature in the tension manifold.
In Eq.~\eqref{eq:tension_dynamics_positive_feedback}, it represents an additional feedback term in the off-rate that prevents the buildup of large pressure gradients, e.g., cells collapsing due to excess active tension.
Eq.~\eqref{eq:Ricci_flow_discrete} operates quasi-statically and requires a non-zero cellular bulk modulus to build up pressure gradients. In a compressible tissue, tension curvature causes secular elongation or contraction of cell-cell interfaces. 
Interface strain rate feedback in Eq.~\eqref{eq:tension_dynamics_positive_feedback} restores mechanical stability, by virtue of having a Lyapunov function~\cite{Noll.etal2017}.
Such ``dynamic recruitment'' of myosin was detected experimentally using optogenetic perturbations~\cite{Gustafson.etal2022}. 
It will be interesting to design analogous tests of the hypothesized tension-pressure feedback Eq.~\eqref{eq:Ricci_flow_discrete}.

\subsection{Generalizations and extensions}

\subsubsection{Curved and folding tissue sheets}

We have assumed that the tissue is constrained to lie in a fixed 2D surface, for example due to a rigid constraint like an eggshell.
If the target surface is not fixed, out-of-plane force balance must be accounted for in addition to the in-plane force balance.
It would be interesting to understand how the intrinsic tension geometry and the ``extrinsic'' geometry of a dynamically folding tissue sheet interact. For example, can in-plane tensions be used to control out-of-plane curvature? We leave it as a question for future research.
The connections to computer graphics and discrete differential geometry~\cite{Springborn.etal2008,Crane.etal2013} we described could also prove useful for simulations of curved tissue sheets~\cite{Silliman.etal2026}, which are more numerically challenging than flat 2D systems.

\subsubsection{3D tension networks}\label{sec:3D}

The most natural 3D generalization of a 2D tension network is a 3D foam with arbitrary active surface tensions $\uptau_{ij}$ on the 2D interface between two cells $i, j$. Such ``generalized foam'' models have recently been used to describe 3D cell clusters, like embryos during the cleavage stage~\cite{Ichbiah.etal2023}. Generalizing foam physics like the von Neumann law to 3D has also received attention in the soft matter literature~\cite{Hilgenfeldt.etal2001,MacPherson.Srolovitz2007}.

Many of our results carry over to 3D (or indeed, $N$D) tension networks.
Like in 2D, force balance is governed by the Young--Dupr\'{e} law Eq.~\eqref{eq:tension_balance} (along junction lines where 3 cells meet) and the Young--Laplace law Eq.~\eqref{eq:Young--Laplace} (across interfaces where 2 cells meet; the line curvature $\upkappa_{ij}$ must be replaced by twice the mean curvature of a 2D interface)~\cite{Weaire.etal2005}.
M\"{o}bius diagrams remain well-defined and mechanically balanced in 3D, and the formula for their (balanced) macroscopic stress tensor Eq.~\eqref{eq:envelope_Airy} also generalizes. Indeed, the arguments of Secs.~\ref{sec:Moebius_diagrams}--\ref{sec:stress} do not make use of two-dimensionality.

However, in higher dimensions, MDs no longer parametrize all possible shapes of generalized foams. While the interfaces of 2D foams are always circular arcs, 3D foam interfaces can be any surface of constant mean curvature.
The complexity of such surfaces (for example, nodoids, unduloids, or the famous Kelvin foam~\cite{Weaire.etal2005}) is not captured by MDs, whose interfaces are always spherical. Nevertheless, general 3D foams can often be approximated by the more restricted class of MDs \cite{Xu.etal2018}.
Moreover, MDs no longer span all possible macroscopic stress configurations.
The macroscopic stress of an MD is determined by a single, scalar Airy function. A general 3D stress tensor, by contrast, requires three distinct Maxwell stress functions~\cite{DeGiuli2018}.
The Maxwell--Cremona duality between tension networks and granular materials also appears to be specific to 2D.

The special properties of our theory in 2D are closely related to the power of conformal maps and complex analysis in 2D. 
Any 2D surface can be conformally mapped to the plane, which we exploit to ``flatten'' the tension triangulation and construct balanced cell tilings. 
In higher dimensions, generic manifolds can not be conformally ``flattened''~\cite{Lee2018} and conformal maps are significantly more rigid, a fact that carries over to discrete conformal maps~\cite{Pinkall.Springborn2021}.
The 3D counterpart of the tension triangulation is a tetrahedral tension network. Discrete conformal maps are, generally, insufficient to map this tetrahedral network into Euclidean 3D space to construct a dual cell tiling\footnote{
The 3D equivalent of the angle deficit $\updelta_i$ is the dihedral angle deficit on edges~\cite{Regge1961}. In a 3D foam, there are $E\approx 7C$ edges~\cite{Hilgenfeldt.etal2001}, so the $C$ conformal factors are insufficient to remove the curvature.}. 
A more careful investigation of 3D tension geometry, notably of the connection between tension network curvature and physical shape, is an interesting avenue for future research.

\subsubsection{Other tissue types}

Taking a step back, the overarching question we have addressed is how intrinsically controlled stresses define cell and tissue geometry. On a technical level, we asked how intrinsically prescribed force dipoles must be embedded in physical space for force balance to hold. We addressed this question in the specific context of epithelia dominated by interfacial tensions and isotropic intracellular pressure. This setting allowed the use of powerful geometric relationships. 

We expect that our framework can be generalized to account for traction forces with a substrate (e.g.\ extracellular matrix)~\cite{Harris.etal1981}, and for structures which generate anisotropic stresses in the cell interior (e.g., the nucleus~\cite{Kim.etal2024}, microtubules~\cite{Singh.etal2024}, or stress fibers~\cite{Lopez-Gay.etal2020}).
A perturbative treatment in the limit of small traction force and bulk anisotropy is a promising approach, potentially converting these additional forces into effective interfacial tensions or pressures~\cite{Lin.etal2023}.

Like many others~\cite{Odell.etal1981,Farhadifar.etal2007,Bi.etal2016}, we have focused on confluent epithelia, whose tightly constrained geometry makes for an appealing modeling target.
However, epithelia are not the only tissue type of importance.
In mesenchyme, internal forces are generated along protrusions through which cells (transiently) pull on one another~\cite{Oster.etal1983,Wallingford.Harland2002}.
Cells control the formation and retraction of protrusions based on local geometric and mechanical cues (e.g., through ``contact-inhibition of locomotion''~\cite{Stramer.Mayor2017}). 
The architecture of the force network therefore changes more dynamically and is less geometrically constrained than in an epithelium.
Developing a systematic framework that links these fluctuating, cell-scale configurations to stresses and coarse-grained mechanical behavior remains an important outstanding challenge~\cite{Grigas.etal2025}.

\subsubsection{Granular materials}

The geometric formalism we developed may also prove applicable to other amorphous materials. In particular, we described a mechanical Legendre duality that maps tension networks to (frictionless) granular materials. The duality interchanges the roles of edge tensions and lengths, and we showed that the Airy functions of the tension network and the granular material are Legendre duals.
Thus, certain of our results also apply to granular materials. For example, the geometric criterion for T1 transitions bounds the admissible stress states compatible with a given grain contact network; $\upell_{ij}=0$ corresponds to the loss of contact between two grains as the pushing force they exert on one another vanishes.

In granular materials, force tilings and discrete Airy functions have a long tradition~\cite{Ball.Blumenfeld2002,Blumenfeld2003,DeGiuli.McElwaine2011,Bi.etal2015a,Behringer.Chakraborty2019}, going back to the work of Maxwell and Cremona~\cite{Maxwell1864,Cremona1879}.
The duality connects our results to this existing literature. However, there are also two important distinctions. 

First, in tension networks, the Airy function is connected to physical cell shape changes, and thus determined by mechanics, namely the target cell areas and external forces. In granular matter, the Airy function leaves the system geometry invariant and is typically indeterminate. One can thus adopt a statistical-physics approach where one constructs ensembles over different stress configurations and computes correlation functions~\cite{Henkes.Chakraborty2009,DeGiuli2018}. 
Such a statistical approach also underlies a notion of  ``emergent elasticity'' introduced for granular materials in Ref.~\cite{Nampoothiri.etal2022}, which is therefore distinct from the elasticity exhibited by ATNs (see App.~\ref{app:elasticity_granular}).
Applying the statistical ideas from the granular materials literature to tension networks is an interesting avenue for future research.

Second, intracellular pressures and the resulting conformal mode have no counterpart in granular matter and are unique to tension networks. Vice versa, (frictional/non-spherical) grains can exert non-central forces on each other, leading to torques. This is not possible in a tension network where forces are always central and act along cell-cell interfaces, since the model's cell interfaces are tensile ``strings'' that do not resist bending (real cell interfaces may have a small bending resistance). Thus, torque balance is automatic in tension networks, but a key ``sticking point'' in granular materials~\cite{Ball.Blumenfeld2002,DeGiuli.Schoof2014}.

Our formalism can be generalized to describe torques. In brief, the key object of our theory is the displacement field $\mathbf{u}$ between the physical configuration and the Maxwell--Cremona force tiling. Non-central forces imply a relative rotation between the two, i.e., a solenoidal displacement field $\mathbf{u}_s =\nabla^\perp \chi$. The field $\chi$ is closely connected to the so-called Mindlin couple stress potential, a continuum description of microscopic torques~\cite{Mindlin1965}.
A generic displacement field can be decomposed into a potential gradient, a solenoidal term, and a conformal term, $\mathbf{u} = \nabla\theta + \nabla^\perp \chi + \bm{f}$.
In tension networks, mechanical balance requires $\chi = \mathrm{const.}$, which leaves the isogonal and conformal modes discussed in this paper. 
The generalization to non-trivial solenoidal modes, and the application to granular materials, is a potential direction for future research.

\subsection{Conclusion}

Together, this work and the companion manuscript~\cite{Claussen.etal2026}, lay out a framework for ``tension-first mechanics'' of epithelia, seamlessly connecting the cell scale with the tissue-scale continuum.
An immediate next step is to develop a cross-scale description of active morphing flows in experimental model systems such as fly, chick, and zebrafish embryos.
Beyond epithelia, it will be interesting to explore active solids whose shape is determined by the requirement to balance an intrinsic, active stress.
For example, plant growth is controlled by active pressure balanced by (passive) tension in the cell walls.
Our work shows that it is key to understand how the microscopic structure determines the transformation of active stress under deformations. Systematic study of such ``stress transformation laws'' may uncover rich geometry and exotic mechanical phases.

\vspace{1cm}

\begin{acknowledgments}
N.H.C.\ is supported by a fellowship from Princeton University through the Princeton Center for Theoretical Science, and by the National Institute of General Medical Sciences (NIGMS) of the National Institutes of Health (NIH) under award \#R01GM082938. 
F.B.\ acknowledges support from the Max Planck Society and the Gordon and Betty Moore Foundation postdoctoral fellowship (grant \#2919).
B.I.S.\ acknowledges support of the NSF Physics (PoLS) grant \#2210612. 
The content is solely the responsibility of the authors and does not necessarily represent the official views of the National Institutes of Health or National Science Foundation.

\paragraph*{AI usage acknowledgment.}
We acknowledge use of Claude Code by Anthropic (Model Opus 5.5) for language editing
and as a proof assistant (in particular, in Appendices~\ref{app:minkowski},~\ref{app:balanced_completeness}, and~\ref{app:stress}).

\end{acknowledgments}

\bibliography{GBE.bib}

\appendix

\setcounter{figure}{0}
\renewcommand{\thefigure}{S\arabic{figure}}

\clearpage

\section{Perturbation by a weak target-perimeter term}
\label{app:perturbation}

Consider a work differential combining dominant junctional tensions with a weak perturbation by a perimeter elasticity term
\begin{equation}
    dE = \sum_{i\sim j} \Big[\uptau_{ij} + \frac{k_P}{2} (\Pi_i+\Pi_j - 2\Pi_0)\Big] d\upell_{ij} - \sum_i\upp_i \, d\upa_i.
\end{equation}
In this appendix, we perform a perturbation expansion in the limit of small perimeter elasticity and show that the behavior in the limit $k_P \to 0$ is regular. 
In other words, weak perimeter elasticity only weakly changes the mechanical ground state and effective shear modulus.

To simplify the calculations, we specialize to a periodic lattice of rectangular cells (Fig.~\ref{fig:rect_foam}). The work differential then reads 
\begin{equation}
    dE_\mathrm{rect} = \sum_{\mu = 1,2} \big[\tau_\mu + k_P (\Pi - \Pi_0)\big] d\upell_\mu - p \, da,
\end{equation}
where the index $\mu = 1,2$ labels the two sides of the rectangle and $\Pi = 2\upell_1 + 2\upell_2$ is its perimeter.
Moreover, we specialize to the incompressible setting where $p$ is a Lagrange multiplier for the area constraint $a = \upell_1 \upell_2 = \upa_0$. 

In the absence of traction forces at the boundary, the force balance equations implied by $dE_\mathrm{rect} = 0$ read
\begin{align}
\label{eq:rect-fb}
\begin{aligned}
    \tau_{1} + k_P (2\tilde{\upell}_1 + 2\tilde{\upell}_2 - \Pi_0) &= \tilde{p} \tilde{\upell}_{2} \\
    \tau_{2} + k_P (2\tilde{\upell}_1 + 2\tilde{\upell}_2 - \Pi_0) &= \tilde{p} \tilde{\upell}_{1} 
\end{aligned}
\end{align}
In this appendix, $\tilde{\cdot}$ indicates the traction-free steady state, and the ``reference pressure'' $\tilde{p}$ needs to be solved for using the constraint $\tilde{\upell}_1 \tilde{\upell}_2 = \upa_0$.
The macroscopic stress tensor reads
\begin{align}
\begin{aligned}
    \overline{\bm{\sigma}} &= \frac{1}{\upell_1 \upell_2} \begin{pmatrix}
        T_1 \upell_1 & 0 \\ 0 & T_2 \upell_2 
    \end{pmatrix} - p \mathbb{I} \\
    &= \begin{pmatrix}
        T_1/\upell_2 - p & 0 \\ 0 & T_2 / \upell_1 - p
    \end{pmatrix},
\end{aligned}
\end{align}
where we introduced the shorthand $T_\mu = \tau_\mu + k_P (\Pi - \Pi_0)$. 
In the absence of traction forces, the macroscopic stress tensor must vanish since we consider a uniform, periodic array of cells. Indeed, this follows directly from the force balance equations \eqref{eq:rect-fb}. The traction-free steady state therefore plays the role of a macroscopically stress-free reference configuration, and we call $\tilde{p}$ the ``reference pressure''.

Consider now a deformation parametrized by strains $\varepsilon_\mu$ relative to the stress-free reference: $\upell_\mu = \tilde{\upell}_\mu (1 + \varepsilon_\mu)$.
We only consider configurations that maintain the rectangular cell shapes, so traction forces at the boundaries must be oriented perpendicular to the boundary so that the stress tensor is diagonal $\overline{\bm{\sigma}} = \mathrm{diag}(\overline{\sigma}_{11}, \overline{\sigma}_{22})$.
For a deformed configuration, the stress' (11)-component reads
\begin{align}
    \overline{\sigma}_{11} &= \frac{T_1}{\tilde{\upell}_2 (1 + \varepsilon_2)} - p \\
    &= \tilde{p} (1 + \varepsilon_1) - p + 2k_P \frac{\tilde{\upell}_1 \varepsilon_1 + \tilde{\upell}_2 \varepsilon_2}{\tilde{\upell}_2 (1 + \varepsilon_2)} \\
    &\approx \tilde{p} (1 + \varepsilon_1) - p + 2k_P \left(\frac{\tilde{\upell}_1}{\tilde{\upell}_2} \varepsilon_1 + \varepsilon_2 + \mathcal{O}(\varepsilon^2)\right)
    \label{eq:rect-stress}
\end{align}
and analogously for $\overline{\sigma}_{22}$.
Here, we used the force-balance equation \eqref{eq:rect-fb} and the incompressibility constraint $(1 + \varepsilon_1)(1 + \varepsilon_2) = 1$ which is enforced by the pressure $p$. FIn the third line, we made a small strain approximation.
Equation~\eqref{eq:rect-stress} can be brought into the form of a tensorial stress-strain relation
\begin{equation}\label{eq:rect-constitutive}
    \overline{\bm{\sigma}} = \tilde{p} \, \bm{\varepsilon} + 2k_P \, \mathrm{diag}\! \left(
        \tfrac{\tilde{\upell}_1}{\tilde{\upell}_2} {-} 1, \tfrac{\tilde{\upell}_2}{\tilde{\upell}_1} {-} 1\right) \cdot \bm{\varepsilon} + (\tilde{p} - p) \, \mathbb{I}
\end{equation} 
This entails an effective shear modulus $\tilde{p}/2$ with an anisotropic correction due to cell shape anisotropy.
Importantly, perimeter elasticity also enters through the reference pressure $\tilde{p}$.

To calculate $\tilde{p}$ perturbatively for small $k_P$, first consider the unperturbed case $k_P = 0$. The macroscopically stress-free steady state is given by $\tilde{\upell}_{1,2}^{\,0} = \tau_{2,1}/\tilde{p}^{\,0}$. Incompressibility $\upell_1 \upell_2 = \upa_0$ then fixes the reference pressure $\tilde{p}^{\,0} = \sqrt{\tau_1 \tau_2/\upa_0}$ which also sets the effective shear modulus $\tilde{p}^{\,0}/2$.

We now introduce the small dimensionless parameter $\hat{k}_P = k_P/\tilde{p}^{\,0}$, which amounts to measuring force in units of $\sqrt{\tau_1\tau_2}$ and length in units of $\sqrt{\upa_0}$.
Expanding to first order $\upell_\mu = \upell_\mu^{\,0} + \hat{k}_P \upell_\mu^{\,1}$, $p = p^0 + \hat{k}_P p^1$ yields
\begin{align}
    \tilde{\upell}_{1,2} &= \frac{\tau_{2,1} + \hat{k}_P \tilde{p}^{\,0} (\Pi^0 - \Pi_0)}{\tilde{p}^{\,0} + \hat{k}_P \tilde{p}^1} \\
    &= \tilde{\upell}_{1,2}^{\,0} + \hat{k}_P (\Pi^0 - \Pi_0 - \tilde{\upell}^{\,0}_{1,2} \tilde{p}^{\,1}/\tilde{p}^{\,0}), \label{eq:rect-length}
\end{align}
where $\Pi^0 = 2\tilde{\upell}^0_1 + 2\tilde{\upell}^0_2$.
The incompressibility constraint, expanded to first order, yields
\begin{align}
    \tilde{p} &\approx
    \tilde{p}^{\,0} + k_P \frac{\Pi^0(\Pi^0 - \Pi_0)}{4 \upa_0}
\end{align}
The effective shear modulus $\tilde{p}/2$ is reduced if the target perimeter $\Pi_0$ exceeds the unperturbed rest perimeter $\Pi^0$. Conversely, $\Pi_0 < \Pi^0$ leads to stiffening. 
Substituting the solution for $\tilde{p}^{\,1}$ into Eq.~\eqref{eq:rect-length} gives
\begin{equation}
    \tilde{\upell}_{1,2} = \tilde{\upell}_{1,2}^{\,0} + \frac{\hat{k}_P}{2} (\Pi^0 - \Pi_0) \left( 1 - \frac{\tau_{2,1}}{\tau_{1,2}} \right)
\end{equation}
A target perimeter that exceeds the rest perimeter exaggerates shape anisotropy, while a smaller target perimeter ($\Pi_0 < \Pi^0$) makes the rest cell shape more isotropic.

Taken together, the above calculations show that the mechanical ground state and effective shear modulus are set, to leading order, by dominant junctional tensions. Weak perimeter elasticity contributes corrections of order $\hat{k}_P = k_P/\tilde{p}^0$, but does not enter at leading order.

\begin{figure}
    \centering
    \includegraphics[scale=1.2]{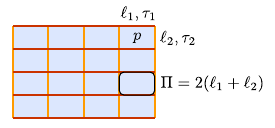}
    \caption{
    Rectangular foam with edge lengths $\upell_{1,2}$, perimeter $\Pi$, and potentially anisotropic tensions $\uptau_{1,2}$.
    }
    \label{fig:rect_foam}
\end{figure}

\section{Batchelor stress on a periodic lattice}
\label{app:batchelor_periodic}

Consider a periodic lattice, shown in Fig.~\ref{fig:lattice-setup-main}, composed of identical tension triangles with edges $\tvec_\mu$, $\mu = 12,23,31$, and cell edge vectors $\rvec_\mu = \upell_\mu \tvec^\perp_\mu/\uptau_\mu$.
(Note that we use the index $\mu$ to distinguish these triangle edge vectors from the triangulation vertex positions, which are labeled with indices $i,j,k$.)
The displacement vectors between cell centroids are given by $\mathbf{c}_{12} = \mathbf{c}_2 - \mathbf{c}_1 = -\rvec_{23} + \rvec_{31}$, and cyclic permutations thereof. The isogonal deformation tensor $\Fiso$ is defined via $\Fiso \cdot \tvec_\mu = \mathbf{c}_\mu$. We can write this in matrix form
\begin{equation}
    \Fiso \begin{pmatrix}
        | & | \\
        \tvec_{12} & \tvec_{23} \\
        | & |
    \end{pmatrix}
    =
    \begin{pmatrix}
        | & | \\
        \mathbf{c}_{12} & \mathbf{c}_{23} \\
        | & |
    \end{pmatrix}.
\end{equation}
Solving for $\Fiso$ yields
\begin{subequations}
\begin{align}
    \Fiso
    &=
    \begin{pmatrix}
        | & | \\
        \mathbf{c}_{12} & \mathbf{c}_{23} \\
        | & |
    \end{pmatrix}
    \begin{pmatrix}
        | & | \\
        \tvec_{12} & \tvec_{23} \\
        | & |
    \end{pmatrix}^{\!-1}
    \\ &=
    \frac{1}{2\tilde{\upa}} 
    \begin{pmatrix}
        | & | \\
        \mathbf{c}_{12} & \mathbf{c}_{23} \\
        | & |
    \end{pmatrix}
    \begin{pmatrix}
        \text{---}\! & \!\tvec^\perp_{23}\! & \!\text{---} \\
        \text{---}\! & \!-\tvec^\perp_{12}\! & \!\text{---} 
    \end{pmatrix}
    \\ &=
    \frac{1}{2\tilde{\upa}} (\mathbf{c}_{12} \otimes \tvec^\perp_{23} - \mathbf{c}_{23} \otimes \tvec^\perp_{12})
\end{align}
\end{subequations}
where $2 \tilde{\upa} = \tvec_{12} \wedge \tvec_{23}$ is twice the area of the tension triangle.
Now we can substitute the expressions for $\mathbf{c}_\mu$ in terms of $\rvec_{12}$ and use that $\tvec_{12} + \tvec_{23} + \tvec_{31} = 0$ to arrive at the expression Eq.~\eqref{eq:isogonal-deformation-lattice}:
\begin{equation} 
    \Fiso = \frac{1}{2\tilde{\upa}} \sum_\mu \rvec_\mu \otimes \tvec^\perp_\mu
    = \frac{1}{2\tilde{\upa}} \sum_\mu \upell_\mu \uptau_\mu \, \hat{\rvec}_\mu \otimes \hat{\rvec}_\mu 
\end{equation}
From this, we recognize that $\Fiso$ is always symmetric, as we expect for isogonal deformations. 
Moreover, in the above equation, we already recognize the form of the Batchelor formula Eq.~\eqref{eq:batchelor_general}. However, the area is that of the tension triangle rather than that of the cell tiling. Using that $\det \Fiso = \upa/(2\tilde{\upa})$, where $\upa$ is the cell area $\upa = \mathbf{c}_{12} \wedge \mathbf{c}_{23}$, we find the Batchelor stress as
\begin{equation}
    \overline{\bm{\sigma}} = \frac{\Fiso}{\det\Fiso} -p_0\mathbb{I}=  \frac{1}{\upa} \sum_\mu \upell_\mu \uptau_\mu \, \hat{\rvec}_\mu \otimes \hat{\rvec}_\mu  -p_0\mathbb{I}.
\end{equation}

\section{The Minkowski lift}\label{app:minkowski}

This appendix introduces a somewhat abstract, but very useful representation of M\"{o}bius diagrams.
We define the \emph{Minkowski lift} of a point $\mathbf{r}\in \mathbb{R}^2$ by
\begin{align}
    \label{eq:Minkowski_lift}
    \rvec \mapsto L(\rvec) = \begin{pmatrix} 1 & r_1 & r_2 & |\rvec|^2/2\end{pmatrix} \in \mathbb{R}^{4}
\end{align}
Quadratic equations in $\mathbf{r}$ become linear equations in $L(\mathbf{r})$. Thus, in the lift, circular arcs,  M\"{o}bius transformations, and so on, can be studied via linear algebra.
Such lifts are a standard tool for Voronoi-type diagrams~\cite{Aurenhammer1987, Boissonnat.etal2006}, and form the basis of practical algorithms for computing the diagram cells from the seed points and weights.

The Minkowski lift owes its name to the following ``Minkowski'' metric on $\mathbb{R}^{4}$:
\begin{align}
    \eta(x, x') =  -(x_0x_3' +x_3 x_0') + (x_1x_1' +x_2x_2') 
\end{align}
One verifies that $\eta$ has signature $(-1,1,1,1)$.
The Minkowski metric has three important properties. First, it allows representing the cell potentials $\Phi_i$ of an MD as linear functions~\cite{Boissonnat.etal2006}:
\begin{align}
\label{eq:Minkowski_potentials}
\begin{aligned}
    \Phi_i(\mathbf{r}) &= - \eta(L(\mathbf{r}), v_i),  \\
    v_i &= \upp_i \, \begin{pmatrix} 1 & \uptau_{i,1} & \uptau_{i,2} & \tfrac{1}{2}|\tvec_i|^2-\uptheta_i \end{pmatrix} \in \mathbb{R}^{4}
\end{aligned}
\end{align}
Second, it is easy to verify that it encodes the distance between lifted points
\begin{align}
    \label{eq:Minkowski_Euclidean_distance}
    \eta(L(\mathbf{r}),L(\mathbf{r}')) = -\frac{|\mathbf{r}-\mathbf{r}'|^2}{2}
\end{align}
Finally, $\eta$ also encodes the tensions:
\begin{align}
    \label{eq:Minkowski_tensions}
    \eta\big(v_i- v_j, v_i- v_j\big) = |\nabla\Phi_i - \nabla\Phi_j|^2 = \uptau_{ij}^2
\end{align}
For $\upp_i = 1$, this reduces to the statement $\uptau_{ij}^2 = |\tvec_i-\tvec_j|^2$.

In the lift, the envelope function becomes
\begin{align}
    \Psi(\mathbf{r}) &= - \max_i \eta(L(\mathbf{r}), v_i)
\end{align}
This makes $\Psi$ into the so-called support function of the convex hull of the points $v_i\in \mathbb{R}^4$. Convex-hull techniques are a standard method for generating Voronoi and power diagrams~\cite{Aurenhammer1987}.

\subsection{M\"{o}bius transformations in the Minkowski lift}

After lifting to $\mathbb{R}^4$, the cell potentials become linear equations. Similarly, M\"{o}bius transformations (MTs) become linear transformations, namely, the Lorentz transformations (LTs) that leave $\eta$ invariant.
The isomorphism between the Lorentz and M\"{o}bius groups is well-known in special relativity~\cite{Weiss.Hillman1998}.

By Eq.~\eqref{eq:Minkowski_Euclidean_distance}, $\eta(L(\mathbf{r}), L(\mathbf{r}))=0$: the lifted points are Minkowski null vectors.
More precisely, due to the normalization $L(\rvec)_0 = 1$, lifted points are in one-to-one correspondence with null rays:
\begin{align}
    L(\mathbf{r}) \leftrightarrow \{x \in \mathbb{R}^4 \;\text{s.th.} \; x \parallel L(\mathbf{r}) \}
\end{align}
The ray $(0,0,0,1)$ corresponds to the point $\mathbf{r} = \infty$ ``at infinity''. Physically, the null rays form what is called the ``celestial sphere''.

Given an MT $M$ with conformal factor $\lambda$, we now define the corresponding LT $\bm{\Lambda}$ (not to be confused with the discrete conformal factor $\Lambda_i$ from Sec.~\ref{sec:circle_packs}). On the lift, we can define the following:
\begin{align}
    \bm{\Lambda}: L\big(\mathbf{r}\big) \mapsto  L\big(M(\mathbf{r})\big) / \lambda(\mathbf{r})
\end{align}
To show that $\bm{\Lambda}$ preserves the Minkowski norm, recall the identity Eq.~\eqref{eq:inversion_identity} for MTs:
\begin{align}
    |M(\mathbf{x}) - M(\mathbf{y})|^2 = \lambda(\mathbf{x})\lambda(\mathbf{y})\,|\mathbf{x}-\mathbf{y}|^2
\end{align}
Combining this with Eq.~\eqref{eq:Minkowski_Euclidean_distance} leads to 
\begin{align}
\begin{aligned}
    &\eta\big(\bm{\Lambda}\cdot L(\mathbf{r}), \bm{\Lambda}\cdot L(\mathbf{r}')\big) \\ &=  \eta\big(\lambda(\mathbf{r})^{-1}L(M(\mathbf{r})), \lambda(\mathbf{r'})^{-1}L(M(\mathbf{r'}))\big) \\
    &= -\frac{|M(\mathbf{r})-M(\mathbf{r}')|^2}{2\lambda(\mathbf{r})\lambda(\mathbf{r}')}  = -\frac{1}{2}|\mathbf{r}-\mathbf{r}'|^2 \\
    &= \eta\big(L(\mathbf{r}), L(\mathbf{r}')\big)
\end{aligned}
\end{align}
Finally, $\bm{\Lambda}$ can be linearly extended to all other points (this works since it preserves the Minkowski metric).

\subsection{M\"{o}bius invariance of MDs}\label{sec:MT_covariance}

With these tools, it is easy to show how MDs behave under MTs. Let $v_i$ be the Minkowski representation of an undeformed MD (cf. Eq.~\eqref{eq:Minkowski_potentials}), and apply an MT $\mathbf{r}' = M(\mathbf{r})$. In the lift, $M$ corresponds to a Lorentz transformation $\bm{\Lambda}$. We claim that the deformed tiling is an MD with representation $v_i' = \bm{\Lambda} v_i$. 
To show this, note
\begin{align}
    \Phi_i'\left(\mathbf{r}'\right) &= -\eta \left(v_i', L(\mathbf{r}') \right) =  -\eta \left(\bm{\Lambda} v_i, \lambda(\mathbf{r}) \bm{\Lambda} \cdot L(\mathbf{r}) \right) \\
    &= -\lambda(\mathbf{r}) \eta \left(v_i,   L(\mathbf{r}) \right)   = \lambda(\mathbf{r}) \Phi_i(\mathbf{r})
\end{align}
since $\bm{\Lambda}$ leaves $\eta$ invariant. This implies:
\begin{align}
\begin{aligned}
    \Phi_i(\rvec) < \Phi_j(\rvec) \, &\Leftrightarrow \, \lambda(\mathbf{r}) \Phi_i(\rvec) <  \lambda(\mathbf{r})\Phi_j(\rvec) \\ 
    &\Leftrightarrow \,  \Phi_i'\left(\mathbf{r}'\right) < \Phi_j'\left(\mathbf{r}'\right)
\end{aligned}
\end{align}
Therefore, the cells of the MD $v_i'$ are precisely the images of the original cells.
Furthermore, the envelope function transforms as
\begin{align}
    \begin{aligned}
    \Psi'\big(\rvec' = M(\rvec)\big) &= \min_i \Phi_i'(\rvec') = \min_i \lambda(\rvec)\Phi_i(\rvec) \\ &= \lambda(\rvec) \Psi(\rvec)
    \end{aligned}
\end{align}
Finally, the deformed tiling has the same tensions: by Eq.~\eqref{eq:Minkowski_tensions}, the tensions are encoded in the Minkowski distance between the $v_i$, which is invariant under the LT $\bm{\Lambda}$.

\section{All force-balanced tilings arise from MDs}{\label{app:balanced_completeness}}

The proof is based on a discrete version of the Poincar\'{e} lemma. In continuous vector analysis, $\mathrm{curl}\,\boldsymbol{v}=0 \Rightarrow \boldsymbol{v} = \nabla V$ -- a curl-free vector field is (locally) the gradient of a potential. 
In the discrete setting, an antisymmetric per-edge quantity $\mathrm{v}_{ij} = - \mathrm{v}_{ji}$ with zero curl around faces is the difference of a vertex field~\cite{Desbrun.etal2005}:
\begin{align}
    \label{eq:discrete_Poincare}
    \mathrm{v}_{ij}+\mathrm{v}_{jk}+\mathrm{v}_{ki}=0 \;\; \Rightarrow 
    \mathrm{v}_{ij} = \mathrm{V}_i-\mathrm{V}_j
\end{align}

\subsection{The constant-pressure case}\label{app:isogonal_exhaustive}

Given a constant-pressure tiling, the (rotated) force vectors $\uptau_{ij} \hat{\mathbf{n}}_{ij}$ are curl-free due to the Young--Dupr\'{e} condition Eq.~\eqref{eq:tension_balance} for force balance. Thus, they can be integrated to obtain a set of tension vertices $\tvec_i$ so that $\uptau_{ij} \hat{\mathbf{n}}_{ij}  = \tvec_j -\tvec_i$. This is the Maxwell--Cremona construction encountered in Sec.~\ref{sec:lattice}.

Next, we must reconstruct the additive MD weights $\uptheta_i$~\cite{DeGoes.etal2014}. 
Note that along cell interfaces:
\begin{align}
   |\rvec-\tvec_i|^2 -|\rvec-\tvec_j|^2=2\uptheta_{ij} =\mathrm{const} \text{ on int. }ij
\end{align}
Since the cell interface is orthogonal to $\tvec_i-\tvec_j$.
The discrete curl of $\uptheta_{ij}$ is zero since the three interfaces $ij, jk,ki$ intersect at vertex $\rvec_{ijk}$:
\begin{align}
\begin{aligned}
    2\left(\uptheta_{ij} + \uptheta_{jk} + \uptheta_{ki}\right) &=  (|\rvec-\tvec_i|^2 -|\rvec-\tvec_j|^2) \\
    &\quad+ (|\rvec-\tvec_j|^2 -|\rvec-\tvec_k|^2)  \\
    &\quad+ (|\rvec-\tvec_k|^2 -|\rvec-\tvec_i|^2) = 0 
\end{aligned}
\end{align}
Therefore, $\uptheta{ij}$ is integrable, $\uptheta_{ij} =\uptheta_i-\uptheta_j$.
We conclude that
\begin{align}
   \tfrac{1}{2}|\rvec-\tvec_i|^2 - \uptheta_i =\tfrac{1}{2}|\rvec-\tvec_j|^2 - \uptheta_j \text{{ on interface} }ij
\end{align}
Thus, the tiling is a power tiling, and the vertex values $\uptheta_i$ define the isogonal potential.

\subsection{The general case}

When cell interfaces are curved, the normals $\hat{\nvec}_{ij}$ are no longer constant along interfaces and do not form a well-defined per-edge field.
Instead, one works in the Minkowski lift of Sec.~\ref{app:minkowski}, where the quadratic equations for the MD cells become linear equations, and the discrete Poincar\'{e} lemma can be applied.

In force balance, cell interfaces $ij$ are circular arcs. In the lift Eq.~\eqref{eq:Minkowski_lift}, these circular arcs become linear equations, defined by a row vector $\mathbf{v}_{ij}$ per edge:
\begin{align}
    \eta\big(\mathbf{v}_{ij}, L(\rvec)\big)  = 0  \;  \Leftrightarrow \; \mathbf{r} \in C_i\cap C_j
\end{align}
We can rescale $\mathbf{v}_{ij}$ so that $\eta\big(\mathbf{v}_{ij}, \mathbf{v}_{ij}\big)  = \uptau_{ij}^2$. 
In terms of the circular arc radius $\mathrm{R}_{ij}$ and center $\boldsymbol{\uprho}_{ij}$, we have:
\begin{align}
    \mathbf{v}_{ij} = \frac{\uptau_{ij}}{\mathrm{R}_{ij}} \begin{pmatrix} 1 & \uprho_{ij, 1} & \uprho_{ij, 2} & \frac{|\boldsymbol{\uprho}_{ij}|^2-\mathrm{R}_{ij}^2}{2} 
    \end{pmatrix}
\end{align}
Since $\mathrm{R}_{ij}$ is signed, $\mathbf{v}_{ij} =-\mathbf{v}_{ji}$.

We now argue that $\mathbf{v}_{ij}$ is curl-free $\mathbf{v}_{ij} + \mathbf{v}_{jk} + \mathbf{v}_{ki} = 0$, where $ijk$ is a vertex with position $\rvec_{ijk}$.
The 0-component is curl-free due to the Young--Laplace law, summed around a vertex, Eq.~\eqref{eq:Young_Laplace_vertex}. 
The middle component is curl-free due to the Young--Dupr\'{e} law Eq.~\eqref{eq:tension_balance} at vertex $\mathbf{r}_{ijk}$ (since $(\boldsymbol{\uprho}_{ij} -\mathbf{r}_{ijk})/\mathrm{R}_{ij} = \hat{\nvec}_{ij}$ is the local normal). The $3^\mathrm{rd}$ component is curl-free because the three circular arcs $ij,jk,ki$ must intersect at $\mathbf{r}_{ijk}$.
Thus, $\mathbf{v}_{ij}$ can be integrated, $\mathbf{v}_{ij} = v_j-v_i$. The resulting MD vertices and weights can be read off from Eq.~\eqref{eq:Minkowski_potentials}. 
Since $\eta\big(\mathbf{v}_{ij}, L(\rvec)\big) =\eta\big(v_j-v_i, L(\rvec)\big) = \Phi_i(\rvec) -\Phi_j(\rvec)$, the MD reproduces the original cell tiling's interfaces.

\section{Geometric constraints on force-balanced tilings}{\label{app:Noll_constraints}}

This appendix rederives two key results from Ref.~\cite{Noll.etal2020}. Namely, balanced circular-arc tilings obey certain geometric constraints \emph{independently} of the values of pressures and tensions. 
To see this, we use angle complementarity $\tilde{\upgamma}^k_{ij} = \pi - \upgamma_{ij}^k$, and the sine law for a tension triangle $\uptau_{ij} \propto \sin\tilde\upgamma_{ij}^k $ (the triangle circumradius is the proportionality constant). 
The Young--Laplace law Eq.~\eqref{eq:Young_Laplace_vertex} (no ``curl'' of pressure around vertices) can thus be formulated entirely in terms of the tiling geometry:
\begin{align}
    \label{eq:Noll_contraint_1}
    \upkappa_{ij} \sin \upgamma_{ij}^k + \upkappa_{jk} \sin \upgamma_{jk}^i  +  \upkappa_{ki} \sin \upgamma_{ki}^j= 0 
\end{align}
A second constraint arises from a telescoping product of tensions around a vertex. Denoting $j_1, j_2, \dots$ the neighbors of cell $i$ in clockwise order:
\begin{align}
    \label{eq:Noll_contraint_2}
    1 = \prod_{k} \frac{\uptau_{i j_k}}{\uptau_{i j_{k+1}}} = \prod_{k}  \frac{\sin \upgamma_{ij _k }^{j_{k+1}}}{\sin \upgamma_{ij_{k+1}}^{j_{k}}}
\end{align}
since every $\uptau_{i j_k}$ appears once in the numerator and once in the denominator.

Together, Eqs.~\eqref{eq:Noll_contraint_1}-\eqref{eq:Noll_contraint_2} yield $V+C=3C$ constraints.  A generic circular-arc polygonal tiling has $2V+E=7C$ degrees of freedom, namely the 2D-positions of the vertices and the radii of the circular arcs. A force-balanced tiling thus has $7C-3C=4C$ degrees of freedom, exactly as many as an MD.

\section{Linear interpolation on triangulations}{\label{app:interpolation}}

To translate between discrete and continuous spaces, we define piecewise affine ``hat'' basis functions $\phi_i(\tvecIt)$ in triangulation space. The $\phi_i$ are defined by their values at vertices, $\phi_i(\tvec_j)=\delta_{ij}$, and linear interpolation in between.
Note that $\phi_i(\tvecIt)=0$ for $\tvecIt$ outside the triangles neighboring $i$. 
Crucially, for linear interpolation on a mesh, the mesh faces must be triangular. It is difficult to define a set of basis functions at the dual vertices $\rvec_{ijk}$, and the existence of a convenient interpolation scheme is one of the advantages of working with the triangulation.

We will need expressions for the linear-hat functions and their gradients (see Ref.~\cite{Crane2025}). Using barycentric coordinates $b_i,b_j,b_k$ for triangle $(ijk)$, we have $\phi_i(\tvecIt) = b_i(\tvecIt)$. Barycentric coordinates are defined by 
\begin{align}
    \begin{aligned}
    \boldsymbol{\zeta} &:= \begin{pmatrix}
        b_i & b_j
    \end{pmatrix}, \quad b_k = 1-b_i-b_j \\
    \boldsymbol{b}&=\mathcal{T}_{ijk}^{-1}\cdot(\tvec-\tvec_k), \quad \mathcal{T}_{ijk} := \left(\tvec_i-\tvec_k, \tvec_j-\tvec_k \right).
    \end{aligned}
\end{align}
In each triangle $ijk$, the gradient is constant and can be expressed in terms of the basis functions as
\begin{align} \label{eq:finite_elements}
    \left(\nabla\phi_i\right)_{ijk} = \tfrac{1}{2\tilde{\upa}_{ijk}} (\tvec_j-\tvec_k)^\perp
\end{align}
where $\tilde{\upa}_{ijk} = \tfrac12 \det\mathcal{T}_{ijk}$ is the area of triangle $(ijk)$.
These finite element gradients fulfill the relations
\begin{subequations} \label{eq:finite-element-relations}
\begin{align}
    &\nabla\phi_i \cdot (\tvec_i-\tvec_j) = 1 \\
    &\nabla\phi_i \cdot \nabla\phi_j = \frac{-1}{2\tilde{\upa}_{ijk}} \cot\tilde{\upgamma}^k_{ij} \quad i \neq j \\
    &|\nabla \phi_i|^2 = \frac{1}{2\tilde{\upa}_{ijk}} (\cot\tilde{\upgamma}^j_{ik} + \cot\tilde{\upgamma}^k_{ij})
\end{align}
\end{subequations}

These basis functions, equivalent to linear Lagrange elements in finite-element-method terminology, enable linear interpolation and the discretization of differential operators. Given values of a scalar function on triangular vertices $\uph_i$, one can linearly interpolate $h(\tvecIt) = \sum_{i} \uph_i \phi_i(\tvecIt)$. The gradient is piecewise constant on each triangle $(ijk)$
\begin{equation} \label{eq:discrete_gradient}
    (\nabla \uph)_{ijk} = \sum_{l\in\{i,j,k\}} \kern-0.5em \uph_l \nabla \phi_l
\end{equation}

\subsection{Discrete Laplacian}{\label{app:discrete_Laplacian}}

We now provide two definitions of the discrete Laplace operator, using first the finite-element and second the finite-volume method. Crucially, for 2D triangulations, the two definitions coincide, a fact that we will exploit below to compute the stress tensor on general triangulations.

The finite-element method defines the Laplace operator on a function $h(\tvecIt)$ in the weak form:
\begin{align}
\label{eq:Laplace-from-FEM}
\begin{aligned}
    (\Delta \uph)^\mathrm{FEM}_i :&= \int \phi_i \Delta h \, d^2\tau
     = -\int \nabla\phi_i \cdot \nabla h \, d^2\tau \\
    &= - \sum_{j\sim i} \left(\int \nabla\phi_i \cdot \nabla \phi_j \, d^2\tau \right) \uph_j  
\end{aligned}
\end{align}
where we integrated by parts, used $h(\tvecIt) = \sum_j\phi_j(\tvecIt) \, \uph_j$, and exploited that $\phi_i$ vanishes on the integration boundary.
We can evaluate the matrix elements of the discrete Laplacian using the relations in Eq.~\eqref{eq:finite-element-relations}. 
The result is referred to as the cotangent Laplacian:
\begin{align}
    \label{eq:cotan_laplacian}
    (\Delta \uph)_i = \frac12 \sum_{j\sim i} (\cot \tilde{\upgamma}_{ij}^k + \cot\tilde{\upgamma}_{ij}^l) (\uph_j-\uph_i)
\end{align}
Here, $\tilde{\upgamma}_{ij}^k$, $\tilde{\upgamma}_{ij}^l$ are the two angles opposite edge $(ij)$.
[Note that $(\Delta \uph)_i$ represents the Laplacian \emph{integrated} over cell $i$; to approximate the continuum Laplace operator, one must additionally divide by the cell area, $\Delta h (\tvec_i) \approx (\Delta \uph)_i  / \upa_i$.]

The finite-volume method defines the Laplace operator using the divergence formula, integrated over the Voronoi cell $i$ corresponding to vertex $i$:
\begin{align} \label{eq:Laplace-from-div-SI}
\begin{aligned}
    (\Delta \uph)^\mathrm{FV}_i :&=
    \int_i (\Delta h) da = \int_i \div (\nabla h) da \\
    &= \int_{\partial i} \nabla h \cdot \hat{\nvec} \, d\upell \approx \sum_{j \sim i}  \frac{\uph_j-\uph_i}{\uptau_{ij}} \, \upell^V_{ij}.
\end{aligned}
\end{align}
Here, we exploited that the normal $\hat{\nvec}$ of the Voronoi cell is parallel to the triangulation edge vector $\tvec_i-\tvec_j$.
The Voronoi edge length is $\upell_{ij}^V = \tfrac{1}{2}(\cot \tilde{\upgamma}_{ij}^k + \cot\tilde{\upgamma}_{ij}^l) \uptau_{ij}$. We conclude that the finite-element and finite-volume Laplacians Eqs.~\eqref{eq:Laplace-from-FEM} and \eqref{eq:Laplace-from-div} are equivalent. In terms of the matrix elements, we have:
\begin{align}
    \label{eq:matrix_elements_FEM_FV}
    \frac{\upell^V_{ij}}{\uptau_{ij}} = -\int \nabla\phi_i \cdot \nabla \phi_j \, d^2\tau 
\end{align}

\subsection{Isogonally weighted Laplacian}{\label{app:isogonal_laplacian}}

Eq.~\eqref{eq:Laplace-from-div} is not restricted to a Voronoi tiling. It also makes sense for non-Voronoi, isogonal tilings, replacing the Voronoi edge lengths $\upell_{ij}^\mathrm{V}$ by the isogonally deformed ones $\upell_{ij}$:
\begin{align}
   (\Delta \uph)_i =  \sum_{j \sim i} \frac{\upell_{ij}}{\uptau_{ij}} (\uph_j-\uph_i)
\end{align}
This weighted Laplace operator depends on the isogonal mode, and reproduces the cotangent Laplacian for $\theta_i=0$~\cite{DeGoes.etal2014}.
In the continuum limit, Eq.~\eqref{eq:Laplace-from-div} corresponds to an anisotropic Laplacian:
\begin{align}
\begin{aligned}
    \Delta h &\approx \frac{1}{\det \Fiso} \nabla_{\!\tvec} \cdot \big[(\det \Fiso) \, \Fiso^{-1} \cdot \nabla_{\!\tvec} h\big]
\end{aligned}
\label{eq:isogonal_Laplacian_continuum}
\end{align}
It can be rationalized as follows. For $\Fiso = \mathbb{I}$, the weighted Laplacian reduces to the conventional Laplacian. We then apply an isogonal deformation 
to the divergence-formula based discretization $(\Delta h)_i = a^{-1}\int\Delta h = - a^{-1}\oint \nabla h \cdot \hat{\nvec} d\ell$.
The interface normal $\hat{\nvec} d\ell$ transforms by $(\det \Fiso)\Fiso^{-1}$ (the so-called adjugate of $\Fiso$), whereas the area $a$ changes by $\det \Fiso$.
We verified Eq.~\eqref{eq:isogonal_Laplacian_continuum} numerically for a periodic lattice.

Note that Eq.~\eqref{eq:isogonal_Laplacian_continuum} does not correspond to the Laplace-Beltrami operator with metric $F$ (the generalization of the Laplace operator to curvilinear coordinates). The Laplace-Beltrami operator requires $\sqrt{\det\Fiso}$ instead of $\det \Fiso$.

\section{Geometry of Voronoi and power tilings}
\label{app:Voronoi}

The areas, edge lengths, and ``intersection fractions'' (intersection points between cell and triangulation edges) of Voronoi and power diagrams can be computed ``intrinsically'' purely in terms of the triangulation edge lengths $\uptau_{ij}$ and angles $\tilde{\gamma}_{ij}^k$. 
These formulas can be found in Refs.~\cite{Meyer.etal2003,DeGoes.etal2014}.

In the Voronoi tiling, the length $\upell_{ij}$ of an edge $ij$ in quadrilateral $ijkl$ has a simple form:
\begin{align}
    \upell_{ij}^\mathrm{V} = |\rvec_{ijk}^\mathrm{V}-\rvec^\mathrm{V}_{ijl}| = \frac{\uptau_{ij}}{2} (\cot\tilde{\upgamma}_{ij}^k +\cot\tilde{\upgamma}_{ij}^l)
\end{align}
Isogonal displacements move vertices $\rvec_{ijk}$, Eq.~\eqref{eq:isogonal_action}, changing the edge length to:
\begin{align}
     \upell_{ij} = \upell_{ij}^\mathrm{V} + \hat{\tvec}_{ij}^\perp \cdot \left[ (\nabla \uptheta)_{ijk} - (\nabla \uptheta)_{ijl} \right]
     \label{eq:isogonal_length}
\end{align}
where $(\nabla \uptheta)_{ijk}$ is the discrete gradient operator defined in Eq.~\eqref{eq:discrete_gradient}. Explicitly, one has:
\begin{align}
\label{eq:dual_length}
\begin{aligned}
    \upell_{ij} &= \frac{\uptau_{ij}}{2}\big(\cot\tilde{\upgamma}^k_{ij}+\cot\tilde{\upgamma}^l_{ij}\big) \\
    &+ \frac{1}{\uptau_{ij}} \big(   
    \uptheta_{jk} \cot\tilde{\upgamma}^{i}_{jk} 
    +\uptheta_{ik} \cot\tilde{\upgamma}^{j}_{ik}
     \\
    &+\uptheta_{jl} \cot\tilde{\upgamma}^{i}_{jl}
    +\uptheta_{il} \cot\tilde{\upgamma}^{j}_{il}
    \big) 
\end{aligned}
\end{align}
where $\uptheta_{jk}=\uptheta_j-\uptheta_k$.
As a consequence, the finite-volume Laplacian $\Delta$ (i.e., its matrix elements) is also defined purely intrinsically.

The area of a Voronoi cell can be computed as 
\begin{align}
    \label{eq:Voronoi_area}
    \upa_i^\mathrm{V} &= \frac{1}{4} \sum_{j\sim i} \upell_{ij}^\mathrm{V} \cdot \uptau_{ij}
\end{align}
To see this, decompose the cell into triangles with base $\upell_{ij}$, tip $\tvec_i$, and thus height $\tau_{ij}/2$. The areas of power tiling cells are
\begin{align}
    \label{eq:PT_area}
    \upa_i &= \sum_{j\sim i} \left( \frac{1}{4} \upell_{ij} \cdot \uptau_{ij} +\frac{1}{2} \frac{\upell_{ij}}{\uptau_{ij}} (\uptheta_i - \uptheta_j) \right)
\end{align}
Both result also holds for non-planar power tilings. 

Lastly, we can compute the \emph{intersection point} between corresponding cell and triangulation edges.
In a Voronoi tiling, cell interfaces intersect triangulation edges at their midpoint.
In a power diagram, the intersection point is moved to:
\begin{align}
    \label{eq:intersection_fraction}
    \uph_{ij} = \uptau_{ij}/2+(\uptheta_i-\uptheta_j)/\uptau_{ij}
\end{align}
This quantity is defined purely intrinsically. Eq.~\eqref{eq:intersection_fraction} follows from the condition for cell interfaces in power diagrams, Eq.~\eqref{eq:power_cell}.

\section{Exact result for the macroscopic stress}{\label{app:stress}}

This Appendix presents two exact results for the macroscopic stress tensor, for Voronoi diagrams and power diagrams with quadratic weights.
We thus consider tilings with constant pressure and straight cell interfaces.
The decomposition Eq.~\eqref{eq:Sigma_definition} of the stress tensor specializes to:
\begin{align}
    \label{eq:stress_decomposition}
    \overline{\boldsymbol{\sigma}} &= -\mathbf{S} + \mathrm{Tr}[\mathbf{S}]\,\mathbb{I} - p_0\mathbb{I}, \;\; \overline{\mathbf{S}} = \frac{1}{A} \sum_{ij\in A}\frac{\upell_{ij}}{\uptau_{ij}} \tvec_{ij}\otimes\tvec_{ij}
\end{align}
Indeed, an individual interface contributes to $\overline{\mathbf{S}}$ as
\begin{align}
    \begin{aligned}
    \int \uptau_{ij} \hat{\nvec}_{ij} \otimes \hat{\nvec}_{ij} \delta_{ij}(\rvec) \,d^2 r
    &= \upell_{ij}\uptau_{ij} \hat{\nvec}_{ij} \otimes \hat{\nvec}_{ij} \\
    &= \frac{\upell_{ij}}{\uptau_{ij}} \tvec_{ij} \otimes \tvec_{ij}
    \end{aligned}
\end{align}

\subsection{Batchelor stress on generic Voronoi tilings}{\label{app:voronoi_proof_new}}

We now show that the Voronoi stress, averaged over any tension triangle $(\tvec_i,\tvec_j,\tvec_k)$, vanishes. In the average, each tensile cell interface is split between two adjacent triangles at the triangulation edge.

In the first, potentially anisotropic, term $\mathbf{S}$, we recognize the matrix elements $\upell_{ij}/\uptau_{ij}$ of the finite-volume Laplacian, Eq.~\eqref{eq:Laplace-from-div}. 
For a Voronoi tiling, $\upell_{ij} = \upell_{ij}^\mathrm{V}$, the finite-volume Laplacian agrees with the finite-element Laplacian.
By Eq.~\eqref{eq:matrix_elements_FEM_FV}, the term becomes:
\begin{align}
    \label{eq:FV_term}
   \overline{\mathbf{S}}^\mathrm{V} = - \frac{1}{A}\sum_{ij\in A} \left(\int \nabla\phi_i {\,\cdot\,} \nabla \phi_j \, d^2r \right) \tvec_{ij}\otimes\tvec_{ij}
\end{align}

We now consider the area formed by a single tension triangle, so $\nabla\phi_i$ is constant, and the integral and $1/A$ cancel.
Following Ref.~\cite{Desbrun.etal2013}, we form a ``partition of unity'' from the $\phi_i$ functions. 
Differentiating $\sum_i \phi_i(\tvecIt) = 1$ and $\sum_i \phi_i(\tvecIt) \tvec_i = \tvecIt$ yields
\begin{subequations}
\begin{align}
    \label{eq:partition_of_unity_a}
    \sum_i \nabla \phi_i &= 0 \\
    \label{eq:partition_of_unity_b}
    \sum_i  \nabla \phi_i \otimes \tvec_i &= \mathbb{I}
\end{align}
\end{subequations}
To use these identities, we expand Eq.~\eqref{eq:FV_term}:
\begin{align}
\begin{aligned}
     \overline{\mathbf{S}}^\mathrm{V}= 
     &-\sum_i \left(\sum_{j\neq i} \nabla {\phi_j}\right) \cdot \nabla \phi_i \tvec_i\otimes \tvec_i  \\
     &+\sum_{i\neq j} (\nabla \phi_i \cdot \nabla \phi_j ) \tvec_i\otimes \tvec_j
\end{aligned}
\end{align}
By Eq.~\eqref{eq:partition_of_unity_a}, $\sum_{j \neq i} \nabla\phi_j = -\nabla\phi_i$, so we can combine the two terms
\begin{align}
\begin{aligned}
    \overline{\mathbf{S}}^\mathrm{V} &= \sum_{i,j}  (\nabla \phi_i \cdot \nabla \phi_j )\tvec_i\otimes \tvec_j \\
    &= \left(\sum_i \tvec_{i} \otimes \nabla {\phi_i} \right) \left(\sum_j \nabla {\phi_j}  \otimes \tvec_{j} \right) =\mathbb{I}
\end{aligned}
\end{align}
where the final line used Eq.~\eqref{eq:partition_of_unity_b}.
Therefore, the stress Eq.~\eqref{eq:stress_decomposition} is isotropic and constant (when averaged over each triangle).

Combining both terms in Eq.~\eqref{eq:stress_decomposition},
\begin{align}
  \overline{\boldsymbol{\sigma}}^\mathrm{V} &= (1-p_0)\,\mathbb{I} = 0
\end{align}
Since we chose units where the reference pressure $p_0=1$, the Voronoi stress vanishes.

This proof only applies in 2D; in higher dimensions, the finite-volume Laplacian and the finite-element Laplacian differ~\cite{Alexa.etal2020}, so the argument above does not work.
Nevertheless, the Voronoi stress tensor is still zero on average over sufficiently large regions due to Eq.~\eqref{eq:stress_bound}.

\subsection{Change in Batchelor stress by isogonal mode}{\label{app:isogonal_proof_new}}

We now turn to how the isogonal mode changes the stress tensor. We assume that the isogonal potential is a \emph{quadratic} function of triangulation vertex positions. Locally, any isogonal potential can be approximated by a quadratic function. Since a constant $\uptheta_i$ does not alter the tiling, and a linear $\uptheta_i$ creates a rigid translation, we consider 
\begin{align}
    \uptheta_i = \tfrac{1}{2}\tvec_i^T \cdot \bm{\Theta} \cdot \tvec_i
\end{align}
with a constant, symmetric matrix $\bm{\Theta}$. The deformation tensor is
\begin{align}
    \mathbf{F} = \mathbb{I} - \bm{\Theta}
\end{align}
Recall that the cells in the resulting power tiling are defined by Eq.~\eqref{eq:power_cell}:
\begin{align}
    \label{eq:power_cell_appendix}
    C_i^\mathrm{P} = \{\rvec \, : \,  \tfrac{1}{2} |\rvec-\tvec_i|^2 - \uptheta_i < \tfrac{1}{2} |\rvec-\tvec_j|^2 - \uptheta_j \}
\end{align}
Crucially, these power cells are the image of a set of Voronoi cells in deformed coordinates.
Define the deformed coordinates and Voronoi centroids by
\begin{align}
      \rvec' = \mathbf{F}^{-1/2} \rvec , \quad \tvec_i'  = \mathbf{F}^{1/2} \tvec_i
\end{align}
The resulting Voronoi cells read
\begin{align}
    \label{eq:deformed_Voronoi}
    {C}_i' = \{{\rvec'} \, : \,  \tfrac{1}{2}|{\rvec'}-{\tvec'}_i|^2  < \tfrac{1}{2}|{\rvec'}-{\tvec'}_j|^2, i\neq j \}
\end{align}
A direct calculation shows that
\begin{multline}
    \left(\tfrac{1}{2}|\rvec-\tvec_i|^2 - \uptheta_i\right) - \left(\tfrac{1}{2}|\rvec-\tvec_j|^2 - \uptheta_j\right)  \\
    = \tfrac{1}{2}|{\rvec'}-{\tvec'}_i|^2  - \tfrac{1}{2}|{\rvec'}-{\tvec'}_j|^2 
\end{multline}
Therefore, $\rvec\in C_i^\mathrm{P} \Leftrightarrow \tilde{\rvec}\in \tilde{C}_i$, or $C_i^\mathrm{P} = F^{1/2} {C'}_i$. 

In the preceding section, we showed that the Voronoi stress tensor is isotropic, which applies to the tiling $\tilde{C}_i$.
To compute the isogonal stress, we thus compute how the different pieces in Eq.~\eqref{eq:stress_decomposition} transform. 
First, the total area transforms like $A =(\det \mathbf{F}^{1/2}){A'}$. Second, $\tvec_{ij} = \mathbf{F}^{-1/2} {\tvec'}_{ij}$. Third, we represent an interface $ij$ by its vector length $\mathbf{n}_{ij} = \upell_{ij} \hat{\mathbf{n}}_{ij}$ (in 3D, $\mathbf{n}_{ij}$ is the vector area of a plane segment). Vector lengths/areas transform according to:
\begin{align}
    \mathbf{n}_{ij} = (\det \mathbf{F}^{1/2}) \mathbf{F}^{-1/2} \,   {\mathbf{n}'}_{ij}
\end{align}
Since $\tvec_{ij}=\mathbf{F}^{-1/2}\tilde\tvec_{ij}$, it follows
\begin{align}
    \frac{\upell_{ij}}{\uptau_{ij}}  = \frac{(\det \mathbf{F}^{1/2}) | \mathbf{F}^{-1/2 }  \tilde{\mathbf{n}}_{ij}|}{|\uptau_{ij}|} = (\det \mathbf{F}^{1/2}) \frac{{\upell_{ij}'}}{{\uptau'}_{ij}} 
\end{align}

We can now transform the non-trivial part of the stress tensor in Eq.~\eqref{eq:stress_decomposition}. The factors of $\det \mathbf{F}^{1/2}$ cancel:
\begin{align}
\begin{aligned}
    \overline{\mathbf{S}} &= \frac{1}{A}\sum_{ij\in A}\frac{\upell_{ij}}{\uptau_{ij}} \tvec_{ij}\otimes\tvec_{ij} \\
    &= \mathbf{F}^{-1/2} \left( \frac{1}{\tilde A} \sum_{ij\in \tilde A}
    \frac{{\upell_{ij}'}}{{\uptau}_{ij}'}  {\tvec{ij}'} \otimes  {\tvec_{ij}'}\right) \mathbf{F}^{-1/2} \\
    &= \mathbf{F}^{-1/2} \cdot {\overline{\mathbf{S}}'} \cdot  \mathbf{F}^{-1/2} =  \mathbf{F}^{-1}
\end{aligned}
\end{align}
since for Voronoi tiling, ${\overline{\mathbf{S}}'} = \mathbb{I}$. By Eq.~\eqref{eq:stress_decomposition}, the tensile stress reads
\begin{align}
    \overline{\boldsymbol{\sigma}} &= -\mathbf{F}^{-1} +\mathrm{Tr}[\mathbf{F}^{-1}]\,\mathbb{I} -p\mathbb{I}.
\end{align}
In 2D, we can use the matrix inverse identity
\begin{align}
    \label{eq:2_2_inverse}
    \mathbf{M}^{-1} = \frac{\mathrm{Tr}[\mathbf{M}]\, \mathbb{I} - \mathbf{M}}{\det \mathbf{M}} = \frac{\epsilon\cdot \mathbf{M} \cdot \epsilon^T}{\det \mathbf{M}}
\end{align}
to conclude that
\begin{align}
    \overline{\boldsymbol{\sigma}} &= \frac{\mathbf{F}}{\det \mathbf{F}}  - p\mathbb{I}.
\end{align}
Finally, let us linearize for small isogonal strain $\mathbf{F}^{-1} \approx \mathbb{I} + \bm{\Theta}$, and use $\Theta_{ab} = \partial_a\partial_b \theta$. One finds that the isogonal potential acts as a stress function:
\begin{align}
    \overline{\boldsymbol{\sigma}} = -\nabla\nabla \uptheta + (\Delta\uptheta) \mathbb{I}
\end{align}

\section{Mechanical duality between tension networks and granular materials}\label{app:duality}

\subsection{Discrete Legendre duality}

The mechanical dual introduced in Sec.~\ref{sec:legendre} is in equilibrium. Hence, the dual stress tensor $\sigma^*$ should be balanced and derived from an Airy function. 
Indeed, Ref.~\cite{DeGoes.etal2013} argued that the dual's Airy function is
\begin{align}
    \label{eq:dual_airy}
\begin{aligned}
    \mathrm{g}(\tvec) &= \sum_i \left(\tfrac{1}{2}|\tvec_i|^2 - \uptheta_i \right) \phi_i(\tvec) \\
    \boldsymbol{\sigma}^* &= \nabla^\perp\nabla^\perp \mathrm{g}
\end{aligned}
\end{align}
The function $\mathrm{g}(\tvec)$ extends the vertex values $\mathrm{g}(\tvec_i)$, Eq.~\eqref{eq:envelope_legendre}, by linear interpolation in each triangle, Eq.~\eqref{eq:interpolation}.
An argument analogous to Sec.~\ref{sec:envelope_airy} shows that  Eq.~\eqref{eq:dual_airy} is the dual Airy function: $\mathrm{g}(\tvec)$ is linear in each triangle (indeed, $\nabla \mathrm{g} = \rvec_{ijk}$ is the power cell vertex). Hence, the Hessian $\nabla\nabla \mathrm{g}(\tvec)$ is singular on triangular edges $ij$. The singularity produces precisely the local force dipoles of the dual stress tensor Eq.~\eqref{eq:batchelor_sigma}.

On the other hand, per Eq.~\eqref{eq:envelope_legendre}, the Legendre transform $\mathrm{g}^*$ is precisely the Airy function of the primal stress (up to the uniform pressure contribution).
Thus, primal and dual Airy functions are Legendre duals, so the primal and dual stresses are inverses of one another (since the stresses are singular distributions, this inverse does not hold pointwise, but after averaging). 
In particular, the dual stress is uniform and isotropic for the Voronoi configuration.

One can also see explicitly that the primal $\bm{\sigma}$ and dual stress $\bm{\sigma}^*$ must be inverses. We compare the corresponding Batchelor formulas at a cell $i$:
\begin{subequations}
\begin{align}
    \overline{\bm{\sigma}}^* &= \frac{1}{A^*} \sum_{ij\in A^*} \upell_{ij} \uptau_{ij} \; \hat{\tvec}_{ij}  \otimes  \hat{\tvec}_{ij} \\
    \overline{\bm{\sigma}} &= \frac{1}{A} \sum_{ij \in A} \uptau_{ij} \upell_{ij} \; \hat{\tvec}_{ij}^\perp  \otimes  \hat{\tvec}_{ij}^\perp
    \label{eq:batchelor_sigma}
\end{align}
\end{subequations}
where $A$ and $A^*$ are corresponding areas in the cell tiling and the triangular network. Because dual and primal stress tensors are rotated by $\pi/2$ with respect to one another, $\overline{\bm{\sigma}}^* \cdot \overline{\bm{\sigma}} \propto \mathbb{I}$. 
We verify this result numerically for an example in Fig.~\ref{fig:inverse-stress-check}.

\subsection{Discretization of the dual stress tensor}{\label{app:stress_discretization}}

In this appendix, we provide another perspective on how to relate discretized and continuum descriptions of stress on triangulations.
We work with the ``mechanical Legendre dual'' of the tension network (a triangular ``truss'' network with force $\upell_{ij}$ on each link).
(Like in Sec.~\ref{sec:legendre}, pressure terms in energy and stress will be omitted in this appendix since the pressure $\upp_i$ has no counterpart in the mechanical dual).
We noted in Sec.~\ref{sec:legendre} that the dual stress is the inverse of the physical stress. The advantage of working with the dual is that one can use linear interpolation on the triangular mesh.
We show how to discretize a continuum dual stress $\bm{\sigma}^*$ to edge values $\upsigma_{ij}^*$, which can be interpreted as forces on each edge.
As we will now show, this scheme is physically consistent: discretizing the continuum equation $\mathrm{div}\;\bm{\sigma}^*=0$ reproduces the ``microscopic'' force-balance equations (Eq.~\eqref{eq:divergence_discrete}).

Following Ref.~\cite{DeGoes.etal2013}, a continuous stress tensor field $\bm{ \sigma}^*(\tvecIt)$ can be discretized onto edges $ij$ as
\begin{equation} \label{eq:stress_projection}
    \upsigma^*_{ij} = -\int \nabla\phi_i(\tvecIt) \cdot \bm{\sigma}^*(\tvecIt) \cdot  \nabla\phi_j(\tvecIt) \, d^2\tau
\end{equation}
Continuum force balance requires $\mathrm{div}\:\bm{\sigma}^*=0$. Importantly, one obtains this discretization:
\begin{equation} \label{eq:divergence_discrete}
    (\mathrm{div}\: \bm{\sigma}^*)_i = -\sum_{j} \upsigma^*_{ij} (\tvec_j - \tvec_i)
\end{equation}
Indeed, use Eq. \eqref{eq:partition_of_unity_b} in the form $\mathbb{I} = \sum_{j} (\tvec_j-\tvec_i) \otimes \nabla \phi_j^T$ integration by parts, and the definition of $\upsigma_{ij}^*$:
\begin{align}
   (\div \bm{\upsigma}^*)_i &=  \int \phi_i (\nabla\cdot\bm{\sigma}^*) d^2\tau = - \int \bm{\sigma}^* \cdot \nabla\phi_i \, d^2\tau 
   \nonumber \\
    &= \sum_{j} (\tvec_i-\tvec_j) \int \nabla\phi_j^T \cdot  \bm{\sigma}^* \cdot \nabla\phi_i d^2\tau \nonumber \\ &=
    \sum_{j } \upsigma_{ij}^* (\tvec_i-\tvec_j)
\end{align}
Now compare Eq.~\eqref{eq:divergence_discrete} to the discrete force balance equation for the mechanical dual
\begin{align}
    \sum_{j} \frac{\upell_{ij}}{\uptau_{ij}} (\tvec_j - \tvec_i) = 0
    \label{eq:force_balance_primal}
\end{align}
Hence, for a balanced stress, the discretized values $\upsigma^*_{ij}$ correspond to:
\begin{align}
    \upsigma_{ij}^*=\upell_{ij} / \uptau_{ij}
    \label{eq:discrete_stress_length_ratio}
\end{align}
Eq.~\eqref{eq:discrete_stress_length_ratio} is a form of the Maxwell--Cremona correspondence: the edge stresses $\upsigma_{ij}^*$ are represented geometrically as the lengths $\upell_{ij}$ of an orthogonal tiling. 

Vice versa, to reconstruct a continuum stress tensor $\tilde{\sigma}$ from discrete stresses $\tilde{\upsigma}_{ij}$, we use the (dual) Batchelor formula:
\begin{align}
    \bm{\sigma}^* = \frac{1}{\tilde A}\sum_{ij\in \tilde A} \upell_{ij} \uptau_{ij} \; \hat{\tvec}_{ij} \otimes  \hat{\tvec}_{ij} = \frac{1}{\tilde A}\sum_{ij\in \tilde A} \upsigma^*_{ij} \; \tvec_{ij} \otimes \tvec_{ij}
    \label{eq:primal_Batchelor}
\end{align}
A direct calculation, using the finite-element relations Eq.~\eqref{eq:finite-element-relations}, shows that Eqs.~\eqref{eq:stress_projection} and \eqref{eq:primal_Batchelor} are compatible: discretizing the interpolated stress gives back the same edge values $\upsigma^*_{ij}$:
\begin{align}\label{eq:rediscretizing}
    &\int \nabla\phi_i \cdot \left(\frac{1}{\tilde{A}}\upsigma^*_{ij} \; \tvec_{ij} \otimes \tvec_{ij} \right) \cdot \nabla\phi_j d^2\tau  = \upsigma^*_{ij}
\end{align}

Like the physical stress, the dual stress is constant and isotropic in the Voronoi configuration. To confirm this, we compute the discrete edge stresses $\upsigma^*_{ij}$ of a constant, isotropic stress using Eq.~\eqref{eq:stress_projection}: 
\begin{equation} \label{eq:identity_projection}
     \mathbb{I}_{ij}=\frac{\cot\tilde{\upgamma}^k_{ij}+\cot\tilde{\upgamma}^l_{ij}}{2}
\end{equation} 
Comparing with Eq.~\eqref{eq:cotan_length}, we recognize $\upell_{ij}^\mathrm{V}/\uptau_{ij}$, the Voronoi edge length.

\subsection{Emergent elasticity in granular media}\label{app:elasticity_granular}

There has also been recent interest in the emergent elastic behavior in granular matter~\cite{Nampoothiri.etal2022}. Similar to ATNs, granular materials like hard disk packings do not have a stress-strain relation at the (microscopic) grain level.
Appealing to an analogy with (tensor) electromagnetism~\cite{Gromov.Radzihovsky2024}, Ref.~\cite{Nampoothiri.etal2022} postulates that the stress tensor $\sigma_{ab}$ and the momentum density obey a generalized version of Maxwell's equations. Ref.~\cite{Nampoothiri.etal2022} emphasizes that this is a phenomenological hypothesis, since the equivalent of Ampere's law has no clear mechanical origin.
In the static limit, this assumption entails that $\epsilon_{ab} \partial_b \epsilon_{cd} \partial_d \sigma_{ac} =0$; i.e., the stress is integrable and derives from a vector field $\varphi_a$.
Ref.~\cite{Nampoothiri.etal2022} further assumes ``linear dielectric'' behavior, i.e., a linear stress-strain relationship between $\sigma_{ab}$ and $\varphi_a$.
These two hypotheses amount to treating a granular system as an effectively elastic medium, which is in good agreement with numerical simulations (once the elastic moduli are fitted to the data).

By contrast, the present work and the companion paper~\cite{Claussen.etal2026} derive the emergent elastic behavior from the microscopic force balance constraints. The effective Airy potential $\theta$ has a direct microscopic interpretation (the isogonal mode), and one can compute the effective elastic moduli in terms of the microscopic model. 
In general, the stress-free configuration is not realizable because of the incompatibility between tension ``density'' and cell density. This is generically the case when the tension triangulation has Gaussian curvature. The resulting residual stress is not integrable and thus does not obey Ampere's law postulated in Ref.~\cite{Nampoothiri.etal2022}.
Indeed, in Eq.~\eqref{eq:biharmonic}, we showed that, in general, $\epsilon_{ab} \partial_b \epsilon_{cd} \partial_d \sigma_{ac}=\Delta^2 \theta \neq 0$.

\section{Discrete conformal maps and MTs}\label{app:discrete_conformal}

\subsection{Discrete conformal maps of power diagrams}

This appendix provides additional detail on the behavior of power diagrams under MTs. As noted in Sec.~\ref{sec:MT_covariance}, power cells are defined by the power distance to power circles $S_i$ of radius $\sqrt{2\uptheta_i}$ around $\tvec_i$~\cite{Aurenhammer1987,DeGoes.etal2014}. 
The cell vertex $\rvec_{ijk}$ is the center of the \emph{face circle}, the unique circle that intersects all power circles orthogonally. If $\uptheta_i=0$, the face circle reduces to the circumcircle of the Voronoi construction.

Applying an MT to the set of power circles $S_i$ yields a new set of power circles $M(S_i)$ (Fig.~\ref{fig:Moebius-triangle}, bottom right).
For example, $M(\rvec)=\rvec/|\rvec|^2$, changes circle centers and radii to:
\begin{align}
    \tvec_i' =  \frac{\tvec_i}{|\tvec_i|^2-2\uptheta_i}, \quad \sqrt{2\uptheta_i'} = \frac{\sqrt{2\uptheta}_i}{|\tvec_i|^2-2\uptheta_i}
\end{align}
Note that the centers of the mapped circles are not the images of the original centers, i.e.\ $\tvec_i' \neq M(\tvec_i)$. Similarly, the scale factors no longer equal the conformal factors evaluated at vertices, $\upp_i^{-1} = \uplambda_i \neq \lambda_M(\tvec_i)$, but rather the ratio of circle radii.
We define the scale factors by the ratios of undeformed and deformed circles:
\begin{align}
    \upp_i = \sqrt{\uptheta_i / \uptheta_i'}
\end{align}
In the limit $\uptheta_i\rightarrow 0$, one recovers $\upp_i = |\tvec_i|^2 = 1/\lambda_M(\tvec_i)$, the definition of an undecorated conformal map.

Eq.~\eqref{eq:inversion_identity} showed that the Euclidean distance $|\rvec-\tvec_i|$ is effectively rescaled by an MT. It is not hard to check that a similar identity holds for the power distance $\Phi^\mathrm{P}(\rvec, S_i)$:
\begin{align}
    \label{eq:inversion_identity_power}
    \Phi^\mathrm{P}\big(M(\rvec), M(S_i)\big) = \pm \lambda(\rvec) \lambda_i \, \Phi^\mathrm{P}\big(\rvec, S_i\big)
\end{align}
where $\lambda(\rvec)$ is the conformal factor, and $\lambda_i = \text{radius}\big(M(S_i)\big) / \text{radius}\big(S_i\big)$.
We now apply an MT $M$ to a power diagram, whose cells are defined by
\begin{align}
    C_i = \big\{\rvec : \;\;  \Phi^\mathrm{P}\big(\rvec, S_i\big) < \Phi^\mathrm{P}\big(\rvec, S_j\big) \big\}
\end{align}
Using Eq.~\eqref{eq:inversion_identity_power}, the deformed cells $C_i' = \{\rvec'=M(\rvec):\; \rvec\in\ C_i \}$ correspond to the potentials
\begin{align}
    \Phi_i'(\rvec') = \frac{ \Phi^\mathrm{P}\big(\rvec', M(S_i)\big)}{\lambda_{i}} = \upp_i\left(\tfrac{1}{2}|\rvec'-\tvec_i'|^2-\uptheta_i' \right)
\end{align}
This is a general MD with parameters
\begin{subequations}
\begin{align}
    \tvec_i' &= \text{center}\big(M(S_i)\big) \\
    \upp_i &= \text{radius}\big(S_i\big) / \text{radius}\big(M(S_i)\big) \\
    \uptheta_i' &= \tfrac{1}{2}\,\text{radius}^2\big(M(S_i)\big) =  \upp_i^{-2} \uptheta_i 
\end{align}
\end{subequations}
This confirms Eq.~\eqref{eq:power_transform} presented in the main text.

\subsection{Triangle-wise MTs}{\label{app:angles_MWVT}}

In the main text, we noted that a discrete conformal map can be thought of as a set of triangle-wise MTs $\{M_{ijk}\}$, alternatively to its definition as a set of scale factors $\uplambda_i = \upp_i^{-1}$. Fig.~\ref{fig:discrete_conformal} shows an example of a discrete conformal map that is not globally Möbius.
In each quadrilateral $ijkl$ composed of adjacent triangles $ijk$ and $ijl$, the local MTs must fit together so that the scale factors are consistent~\cite{Bobenko.Lutz2024}:
\begin{align}
    \lambda_{M_{ijk}}(\tvec_i) &= \lambda_{M_{ijl}}(\tvec_i) = \upp_i^{-1}, \nonumber \\ 
    \lambda_{M_{ijk}}(\tvec_j) &= \lambda_{M_{ijl}}(\tvec_j) = \upp_j^{-1}.
\end{align}
This definition is equivalent to Eq.~\eqref{eq:discrete_conformal_definition}.

The action of a discrete conformal map on the triangulation (rescaling) and the action on the dual cell tiling (MT) are compatible.
We showed this explicitly for a global MT above, but it also holds for general discrete conformal maps, defined by local MTs $M_{ijk}$. The MTs of adjacent triangles $(ijk)$, $(ijl)$ applied to the infinite extension of their shared Voronoi edge $(ij)$, yield the same mapped edge. 
This is because the mapped edge is defined by $\upp_i\,|\rvec'-\tvec_i'|^2= \upp_j\,|\rvec'-\tvec_j'|^2$, and, by construction
\begin{align}
\begin{aligned}
    |M_{ijk}'(\tvec_i)| &= |M_{ijl}'(\tvec_i)| = \upp_i^{-1}, \\
    |M_{ijk}'(\tvec_j)| &= |M_{ijl}'(\tvec_j)| = \upp_j^{-1}, \\
    M_{ijk}(\tvec_i) &= M_{ijl}(\tvec_i) = \tvec'_i, \\
    M_{ijk}(\tvec_j) &= M_{ijl}(\tvec_j) = \tvec'_j
\end{aligned}
\end{align}
However, in general $M_{ijk}(\rvec_{ijk}) \neq M_{ijl}(\rvec_{ijk})$. Put differently, the local MTs do not act continuously on individual points, but they do act continuously (even smoothly) on the circular arcs of a cell tiling~\cite{Bobenko.Lutz2024}. This allows piecing adjacent MTs together.

This observation leads to an alternative proof that the Yound-Dupr\'{e} law is fulfilled in MDs. We need to show that the vertex angles in an MD 
$(\tvec_i', \upp_i)$ are complementary to the underlying tension triangulation with edge lengths $\uptau_{ij}= \sqrt{\upp_i\upp_j} \,|\tvec'_i-\tvec'_j|$.
To do so, we find an MT that locally maps an ordinary power diagram onto a given MD.
Consider a single tension triangle $(ijk)$ and an MT $M_{ijk}$ that has scale factors $\lambda_{M_{ijk}}(\tvec_l) = \upp_l^{-1}$ for $l \in \{i,j,k\}$. Note that this fixes three of the six real coefficients of the MT and leaves rigid rotation and translation degrees of freedom, which can then be used to fit adjacent triangles together.

As we saw in Eq.~\eqref{eq:Moebius_multiplicative_Voronoi}, $M_{ijk}(\tvec_i) = \tvec'_i$ acts as a discrete conformal map on the $\tvec_i$.
Applied to the Voronoi edges, $M_{ijk}$ yields the circular arcs of the multiplicative diagram with seed points $\tvec'_i$ and scale factors $\upp_i$. 
Force balance at the mapped vertex $M_{ijk}(\rvec_{ijk})$ trivially follows because MTs preserve intersection angles.
As an aside, observe that under $M_{ijk}$, the Voronoi edges $(ij)$, $(jk)$, $(ik)$, extended to infinite lines, map to circles which intersect twice: Once at $M_{ijk}(\rvec_{ijk})$ and once at $\rvec_{ijk}^*$, the image of the point at infinity (Fig.~\ref{fig:Moebius-triangle}). 
This implies that the centers of these circles must be collinear~\cite{Moukarzel1997,Noll.etal2020}.

\section{Continuum limit of isogonal modes}{\label{app:nonflat_isogonal_continuum}}

The tension vectors $\tvec_{i(\rvec)}$ can be thought of as forming a ``coordinate system'' for the cell tiling. To measure physical distances (for example, between cell-centroids) in these coordinates, we need to use a metric. The isogonal deformation $\Fiso$ defines precisely this \emph{physical} metric, via $\mathbf{g}^\mathrm{I} = \Fiso\cdot(\Fiso)^{T}$. This is because $\Fiso$ maps tension vertices to cell centroids.

The metric $\mathbf{g}^\mathrm{I}$ defines a Laplace-Beltrami operator, the generalization of the Laplace operator to curvilinear coordinates. This operator reads
\begin{align}
    \begin{aligned}
    \Delta_{\mathbf{g}^\mathrm{I}} h &= \frac{1}{\sqrt{\det \mathbf{g}^\mathrm{I}}} \,\nabla_{\!\tvec}  \cdot \bigl[\sqrt{\det \mathbf{g}^\mathrm{I}} \; (\mathbf{g}^\mathrm{I})^{-1} \cdot \nabla_{\!\tvec}  h \bigr] \\
    &=  \frac{1}{\det \Fiso} \,\nabla_{\!\tvec}  \cdot \bigl[\left(\det\Fiso\right) \; (\Fiso\cdot\Fiso^T)^{-1} \cdot \nabla_{\!\tvec}  h \bigr]
    \end{aligned}
\end{align}
In App.~\ref{app:isogonal_laplacian}, we argue that in the continuum limit $\Delta_{\mathbf{g}^\mathrm{I}} h = (\det \Fiso)^{-1}\nabla_{\!\tvecIt} \cdot ( \det \Fiso \; (\Fiso)^{-1} \nabla_{\!\tvecIt} h)$, where $\Fiso$ is the isogonal deformation tensor. Combined with the continuum Young--Laplace law $\nabla_{\!\tvecIt} p = \Fiso^{-1} \nabla_{\!\tvecIt} \lambda^{-1}$ [Eq.~\eqref{eq:pressure_transform}], $\Delta_{\mathbf{g}^\mathrm{I}} \log p = K $ implies the following continuum equation for the conformal factor
\begin{multline}
   (\det \Fiso)^{-1}\nabla_{\!\tvecIt} \cdot \big[ \det \Fiso \; (\Fiso)^{-2} \nabla_{\!\tvecIt} \log \lambda\big]\\
   = \Delta_{\mathbf{g}^\mathrm{I}} \log \lambda = -\lambda^{-2} K
\end{multline}
This is simply the usual Liouville Eq.~\eqref{eq:liouville} for the conformal factor, now expressed in cell-space coordinates.

We therefore arrive at the following interpretation. An isogonal deformation $\Fiso$ stretches and compresses cells, defining the physical metric of the cell tiling, $g^\mathrm{I} = \Fiso \cdot(\Fiso)^{T}$ (see Fig.~\ref{fig:curved-tri-iso-conf}). 
On a curved tension manifold, the isogonal mode acts in the local tangent plane of the non-planar triangulation without changing the curvature $K$.

Explicitly, the isogonal displacement in the presence of conformal deformation is given by
\begin{align}
    r_a \mapsto r_a^\mathrm{I} &= r_a  + (\mathbf{g}^{-1}(\tvecIt) )_{ab} \partial_b  \theta  \nonumber \\ &= r_a + \lambda^{2} \partial_ a \theta
    \label{eq:isogonal_metric_gradient}
\end{align}
The Riemannian metric reflects that the isogonal displacement takes place on the tension manifold. This matches the form for the isogonal displacement for a non-flat tension metric identified in the companion paper~\cite{Claussen.etal2026}. For a flat conformal deformation $\rvecIt \mapsto f(r_1+ir_2)$, Eq.~\eqref{eq:isogonal_metric_gradient} can be derived by means of a Taylor expansion:
\begin{align}
    f(z+\nabla\theta) &\approx  f(z) +( \partial_z f)\nabla\theta 
    \\ &= f(z) + \lambda^2 \nabla_{f(z)} \theta 
    \nonumber
\end{align}
This makes it clear that the continuum interpretation is valid when the isogonal displacement is small compared to the scale $1/\nabla \lambda \sim 1/\sqrt{K}$ of the tension manifold curvature.

\section{Macroscopic stress compared to the continuum theory}\label{sec:macroscopci_stress_SVD}

Here, we connect Eq.~\eqref{eq:stress_conformal_qualitative} for the stress transformation by a conformal map to a result of the continuum theory from Ref.~\cite{Claussen.etal2026}.
Let $\lambda \mathbf{R}$, a scale-rotation, be the Jacobian of the conformal map, and $\Fiso$ be the isogonal deformation tensor. Eq.~\eqref{eq:stress_conformal_qualitative} means:
\begin{align}
    \bm{\sigma} = \frac{p_0}{\lambda}  \frac{\mathbf{R} \cdot\Fiso \cdot \mathbf{R}^T}{\det \Fiso}
    \label{eq:stres_conformal}
\end{align}
The stress has a simple expression in terms of the total deformation tensor $\Fiso^\mathrm{tot} = \lambda \mathbf{R} \cdot \Fiso$, which maps the Voronoi tiling to the final MD.
Comparing this factorization with the  with the singular value decomposition $\Fiso^\mathrm{tot}=\mathbf{R}(\phi_{\mathrm{I}})\cdot \bm{\Sigma}_{\mathrm{tot}} \cdot \mathbf{R}^T$, yields, after a few algebraic steps:
\begin{align}
    \bm{\sigma} = p_0  \frac{\mathbf{R}(\phi_{\mathrm{I}}) \cdot \bm{\Sigma}_{\mathrm{tot}} \cdot \mathbf{R}(\phi_{\mathrm{I}})^T}{\det \bm{\Sigma}_{\mathrm{tot}}}
    \label{eq:stress_polar_decomp}
\end{align}
This SVD form of the stress precisely matches the continuum theory (Sec. B.2 in Ref.~\cite{Claussen.etal2026}), now obtained by coarse-graining the microscopic stress tensor.

\section{Triangle shape via local tension configuration parameters}
\label{app:LTC}

The local configuration of tension is encoded in the shape of tension triangles. In Refs.~\cite{Brauns.etal2024,Claussen.etal2024}, we introduced local tension configuration (LTC) parameters based on the singular value decomposition of the linear map $T$ from an equilateral reference triangle to the target triangle.
On a triangle, $\mathbf{T}$ can be defined by linear interpolation. Given 2D vertices $\bm{\upzeta}_{1,2,3}$ of the equilateral reference triangle and $\bm{\uptau}_{1,2,3}$ of the target tension triangle, we can define $\mathrm{T}(\bm{\zeta}) = \sum_{i=1,2,3} \phi_i(\bm{\zeta}) \,\tvec_i$ using the barycentric interpolation of App.~\ref{app:interpolation}. $\mathbf{T}$ is therefore a linear map.

The singular value decomposition of a $2{\times}2$ matrix $\mathbf{T} = \mathbf{R}(\phiT) \cdot \bm{\Sigma} \cdot \mathbf{R}(\psi)^\mathrm{T}$ specifies two angles and a diagonal matrix of stretch factors
\begin{align}
    \bm{\Sigma} := \sqrt{S} \; \mathrm{diag} (m, 1/m), \; m = \left(\frac{1+|\muT|}{1-|\muT|}\right)^{\!\tfrac14}
\end{align}
The scale factor $S$ is proportional to the area of the triangle (i.e.\ the local scale of tension). In the following, we will focus on scale-invariant features of the triangle shape, characterized by the tuple $(\muT, \psi) \in \mathbb{C} \otimes \mathbb{R}$, which we refer to as LTC parameters.
The angle $\psi$ accounts for the orientation of shear relative to the orientation of the equilateral reference triangle.
It therefore tunes the triangle's shape between acute and obtuse. 
The reference triangle edges after rotation by $\psi$ read
\begin{equation}
    \hat{\bm{\psi}}_\mu := \begin{pmatrix}
        \cos\bigl(-\psi + \frac{2(\mu - 1) \pi}{3}\bigr) \\
        \sin\bigl(-\psi + \frac{2(\mu - 1) \pi}{3}\bigr)
    \end{pmatrix}
\end{equation}
Due to invariance of shape under permutation of the edge labels $\mu = \{1,2,3\}$, $\psi$ can be restricted to the fundamental domain $[-\pi/6, \pi/6]$.
Applying the stretch and second rotation yields the triangle edge vectors 
\begin{equation} \label{eq:T-from-LTC}
    \tvec_i = \mathbf{R}(\phiT) \cdot \bm{\Sigma}_T \cdot \hat{\bm{\psi}}_\mu.
\end{equation}
The angle $\psi$, to which we refer as ``LTC phase'', determines the orientation of the triangle's principal axes in real space.
This ``extrinsic'' shape information is contained in the tensor
\begin{equation}
    \mathbb{T} := \sum_i \tvec_i \otimes \tvec_i = \frac{3}{4} \mathbf{R}(\phiT) \cdot \bm{\Sigma}_T^2 \cdot \mathbf{R}(\phiT)^{\mathrm{T}},
\end{equation}
which is independent of the LTC phase $\psi$.
The Beltrami coefficient of the quadratic form $\mathbb{T}$ is given by $\muT = |\muT| e^{2 i \phiT}$ with $|\muT| = (2\tr[\tilde{\mathbb{T}}^2])^{1/2}/\tr[\mathbb{T}]$, where $\tilde{\mathbb{T}}$ denotes the traceless part of $\mathbb{T}$.
It compactly encodes information about the magnitude and orientation of tension anisotropy.

\section{T1 threshold}
\label{app:T1_threshold}

We calculate the T1 threshold for a kite formed from two identical tension triangles (i.e., we work in the regular lattice of Sec.~\ref{sec:lattice} with representative edges $\mu=12,23,31$). 
To parametrize the tension triangle shape, we use the local tension configuration (LTC) parameters; see App.~\ref{app:LTC}. Because the overall scale of $\upell_i$ is irrelevant, the conformal mode does not play a role in the T1-threshold.

To find the T1 threshold as a function of the isogonal deformation (i.e., macroscopic stress) and LTC parameters, we need to find the edge lengths $\upell_\mu$.
From Eq.~\eqref{eq:isogonal-deformation-lattice}, one sees that $\Fiso$ is linear in the $\upell_\mu$ such that finding the edge lengths is achieved by solving a linear set of equations. 
Since the overall scale factor $\det  \Fiso$ is irrelevant, we can arbitrarily fix it with the simple linear constraint
\begin{equation}
    \upell_{12} + \upell_{23} + \upell_{31} = 1
\end{equation}
The linear system of equations to be solved for $\upell_\mu$ then reads
\begin{align}
\begin{aligned}
    \mathbf{F} \cdot \tvec_\mu &= \mathbf{c}_\mu =\upell_{\mu+1} \hat{\tvec}_{\mu+1}^\perp - \upell_{\mu+2} \hat{\tvec}_{\mu+2}^\perp \\
    \sum_\mu \upell_\mu &= 1.
\end{aligned}
\end{align}
where $\mathbf{c}_\mu$ is the displacement vector between cell centroids (cf.\ App.~\ref{app:batchelor_periodic}) and $\mu+1,\mu+2$ indicate cyclic shifts.
Taking the dot product with $\tvec_{\mu+2}$ and shifiting indices, the solution reads
\begin{equation} \label{eq:length-from-F-solution}
    \frac{\upell_\mu}{\uptau_\mu} \propto   -\tvec_{\mu+1}\cdot\Fiso\cdot \tvec_{\mu+2}
\end{equation}
The proportionality factor is the same for all three $\upell_\mu$. 
The T1 threshold $\upell_\mu = 0$ thus becomes
\begin{align}
    \tvec_{\mu+1}\cdot\Fiso\cdot \tvec_{\mu+2} \stackrel{!}{=}0
\end{align}
We now use the map $\mathbf{T}$ to map the tension vectors $\tvec_\mu$ to the edges $\boldsymbol{\upzeta}_\mu$ of the equilateral reference triangle, leading to
\begin{align}
    \label{eq:T1_threshold_T_F}
    \boldsymbol{\upzeta}_{\mu+1}\cdot\left(\mathbf{T}^\mathrm{T} \cdot\Fiso\cdot \mathbf{T}\right)\cdot\boldsymbol{\upzeta}_{\mu+2} \stackrel{!}{=}0
\end{align}
From this, the T1-threshold can be computed in terms of the LTC parameters (which arise from the SVD of $\mathbf{T}$), and the SVD of $\mathbf{F}$:
\begin{align}
    \mathbf{F} = \mathbf{R}(\phiI) \cdot \bm{\Sigma}_\mathrm{I} \cdot \mathbf{R}(\phiI)^\mathrm{T}
\end{align}
where $\Sigma_I = \mathrm{diag}(s^\mathrm{I}_1, s^\mathrm{I}_2)$ is the diagonal matrix of isogonal principal strains. 
For the T1 threshold, the scale $\det \Sigma_I$ is irrelevant -- only the orientation and magnitude of anisotropy of deformation matter.
This information is captured by the isogonal Beltrami coefficient
\begin{align}
    \muI = \frac{s_1^\mathrm{I} - s_2^\mathrm{I}}{s_1^\mathrm{I} + s_2^\mathrm{I}} \; e^{2 i \phiI}
\end{align}
For explicit calculations, we parametrize $s_1^\mathrm{I} = 1 + |\muI|$, $s_2^\mathrm{I} = 1 - |\muI|$.

A T1 happens when $\upell_\mu(\muT, \muI, \psi) = 0$ for one of the edges $\mu$, implicitly defining a hypersurface $\mathcal{C}_\mathrm{T1}$ in the configuration space $(\muT, \muI, \psi)$ which bounds the admissible configurations.
Due to rotational symmetry, only the relative orientation $\phiTI := \phiI - \phiT$ of shear deformation and tension anisotropy is relevant.
Some slices of this $\mathcal{C}_\mathrm{T1}$ in the polar $(|\muI|, \phiTI)$ plane are shown in Fig.~\ref{fig:T1-threshold_q-sweep} for various values of $|\muT|$ and $\psi$.
Parameterizing $\mathcal{C}_\mathrm{T1}$ as a graph $|\muI^\mathrm{T1}|$ over $(|\muT|, \phiTI, \psi)$ defines a yield strain -- and hence yield stress.
Note however that for large anisotropy $|\muT| > 1/2$, there is both a lower and an upper bound on $\muI$: a minimum amount of isogonal strain is required to avoid a T1 due to tension anisotropy
[see Fig.~\ref{fig:T1-threshold_q-sweep}(d)].

For isotropic tension ($|\muT| = 0$), Eq.~\eqref{eq:T1_threshold_T_F} yields the explicit expression
\begin{multline}
    |\muI^\mathrm{T1}|\bigl(|\muT| = 0, \phiTI, \psi\bigr) = \\
    \big[2 \cos 2(-\tfrac{\pi}{6} + (\tfrac{\pi}{6} + \phiTI - \psi \; \mathrm{mod} \; \tfrac{\pi}{3}))\big]^{-1}.
\end{multline}
The corresponding graph in the polar $(|\muI|,\phiTI)$ plane is shown in Fig.~\ref{fig:T1-thresh}(b).
This case corresponds to an ordinary fluid foam where all surface tensions are identical. 
Passive T1s are driven by external (boundary) forces acting on the tissue, which drive $|\muI|$ toward $|\muI^\mathrm{T1}|$.

Through controlling tensions, cells can manipulate $|\muT|$ and thus drive \emph{active} T1s.
In particular, for $|\muT| = 1/2, \psi = 0$, the yield strain in the sector $|\phiTI| < \pi/4$ vanishes; Fig.~\ref{fig:T1-thresh_psi-marginalized}(b). This is therefore the critical tension anisotropy for which active T1s happen in the absence of isogonal deformation, i.e.\ in the macroscopically stress-free state. 
Active T1s happen most readily when the principal axis of tension anisotropy is aligned with one of the cell interfaces ($\psi = 0$).

\subsection{Marginal T1 threshold}\label{app:marginal_T1}

In a disordered tissue, tension triangles with all shape phases $\psi$ occur.
Thus, given $\muI, \muT$, we can define the minimal edge length that occurs across all $\psi$
\begin{multline}
    \upell_\mathrm{min}(|\muI|, |\muT|, \phiTI) :=  \\
    \min_{\mu \in \{12,23,31\}} \min_{\psi \in [0, \pi/6]} \upell_\mu(|\muI|, |\muT|, \phiTI, \psi)
\end{multline}
From this, we can define the ``marginal'' T1 threshold as the locus $\upell_\mathrm{min}(|\muI|, |\muT|, \phiTI) = 0$.

Eq.~\eqref{eq:T1_threshold_T_F}, leads to an explicit formula for the marginal T1 threshold. Using the SVD $\mathbf{T} = \mathbf{R}(\phiT)\cdot\bm{\Sigma}_T\cdot \mathbf{R}(-\psi)$, the T1 threshold becomes
\begin{align}
    \label{eq:T1_marginal}
    \min_{\psi, \mu} \left(\mathbf{R}(\psi)\boldsymbol{\upzeta}_{\mu+1}\right)^T\cdot \mathbf{M}\cdot   \left(\mathbf{R}(\psi)^T\boldsymbol{\upzeta}_{\mu+2}\right) \stackrel{!}{=} 0, \\
    \text{with} \quad \mathbf{M} := \bm{\Sigma}_T \cdot \mathbf{R}(\phiTI) \cdot \bm{\Sigma}_\mathrm{I} \cdot \mathbf{R}(\phiTI)^T \cdot \bm{\Sigma}_T
\end{align}
The matrix $\mathbf{M}$ is $\psi$-independent, so that the condition Eq.~\eqref{eq:T1_marginal} only depends on the anisotropy of $\mathbf{M}$. 
Since $\boldsymbol{\zeta}_{\mu+1}, \boldsymbol{\zeta}_{\mu+2}$ make a $2\pi/3$-angle, one finds that the eigenvalue anisotropy of $\mathbf{M}$ must be at $1/2$ for Eq.~\eqref{eq:T1_marginal} to vanish.
Note that $\mathbf{M}$ has the same eigenvalues as $\Fiso \cdot \left( \mathbf{T}\mathbf{T}^T\right)$ and thus captures the ``sum'' of tension and isogonal-strain anisotropy. We can therefore express $\mathbf{M}$'s anisotropy magnitude in terms of the Beltrami coefficients, leading to
\begin{equation} \label{eq:T1-marginalized-Beltrami}
    \left(1-|\muI|^2\right)\left(1-|\muT|^2\right)
    = \frac{3}{4}\left(1+|\muI|\,|\muT|\cos2\phiTI\right)^2
\end{equation}
This locus is rendered in Fig.~\ref{fig:T1-thresh_psi-marginalized}(a) in the $(|\muI|, |\muT|)$-plane for various values of $\phiTI$. It is manifest from Eq.~\eqref{eq:T1-marginalized-Beltrami} that the T1 threshold is symmetric under the exchange of $|\muI|$ and $|\muT|$. In other words, the anisotropy magnitudes in tension space and real space play equivalent roles in driving T1s, i.e.\ there is a correspondence (or duality) between active and passive T1s. 
Fig.~\ref{fig:T1-thresh_psi-marginalized}(b) shows T1 threshold contours in the $(|\muI|, \phiTI)$-polar plot for different values of $|\muT|$.

\begin{figure*}
    \centering
    \includegraphics{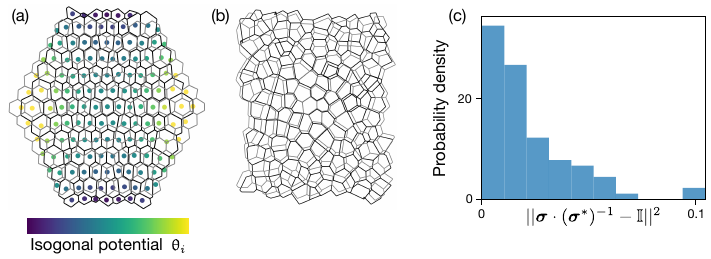}
    \caption{
    (a)~Parabolic isogonal mode (color code) results in pure shear.
    (b)~Disordered cell tiling. A disordered tension triangulation was generated via hard-disk packing~\cite{Claussen.etal2024}. Gray shows the Voronoi dual of the triangulation, deformed by the addition of an isogonal deformation into the black tiling. The isogonal mode comprises a parabolic (shear) and a white noise component.
    (c)~Evaluation of the inverse-stress formula on the disordered tiling (b). Per-cell dual and primal stresses were calculated using the Batchelor formula Eq.~\eqref{eq:batchelor_sigma}.
    }
    \label{fig:inverse-stress-check}
\end{figure*}

\begin{figure*}[p]
    \centering
    \includegraphics[width=0.5\textwidth]{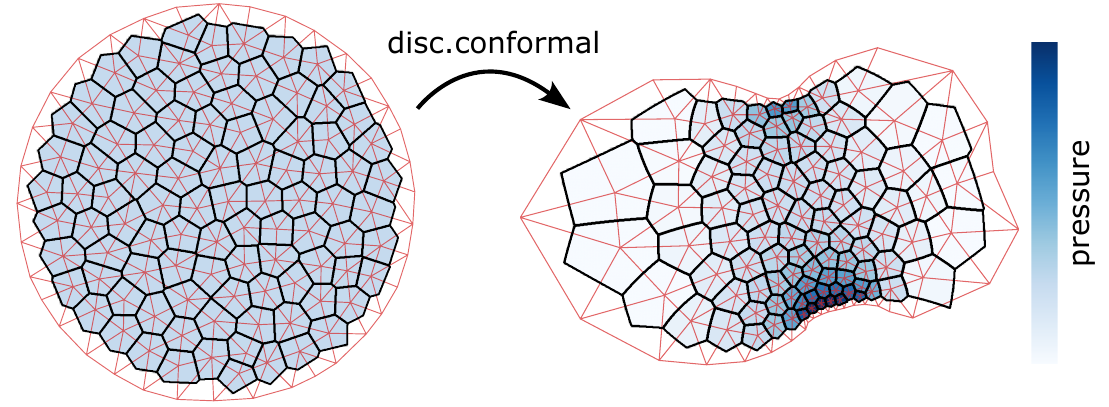}
    \caption{Example of a non-trivial discrete conformal map from the disk to an irregular shape. Note that the map is not globally a Möbius transformation. The dual cell tiling is the corresponding MVD.
    }
    \label{fig:discrete_conformal}
\end{figure*}

\begin{figure*}[p]
    \centering
    \includegraphics[scale=1.3]{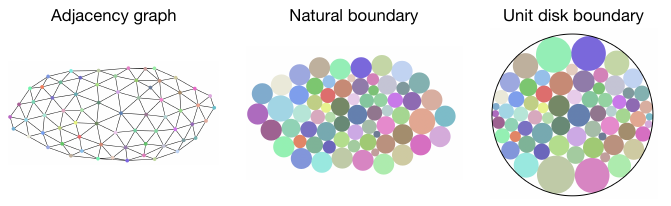}
    \caption{Boundary conditions for Thurston circle packs.}
    \label{fig:Thurston-in-disk}
\end{figure*}

\begin{figure*}[p]
    \centering
    \includegraphics{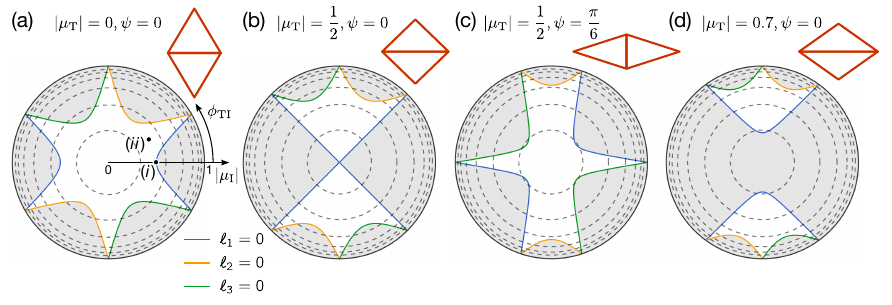}
    \caption{
    (a)~T1 threshold for isotropic tension $|\muT| = 0$ (same as panel (b) in Fig.~\ref{fig:T1-thresh}).
    (b)~Critical case $|\muT| = 1/2, \psi=0$, where the Voronoi edge vanishes $\upell_1^\mathrm{V} = 0$, i.e.\ a T1 takes place in the stress-free reference configuration ($|\mu_I| = 0$). This is the prototypical case of an active T1.
    (c)~For the same tension anisotropy $|\muT| = 1/2$ but a ``cable-like'' tension configuration $\psi = \pi/6$, the T1 threshold remains at finite $\mu$.
    (d)~For higher tension anisotropy, active T1s happen even when the isogonal deformation is oriented opposite to the axis of tension anisotropy. }
    \label{fig:T1-threshold_q-sweep}
\end{figure*}

\begin{figure*}[p]
    \centering
    \includegraphics{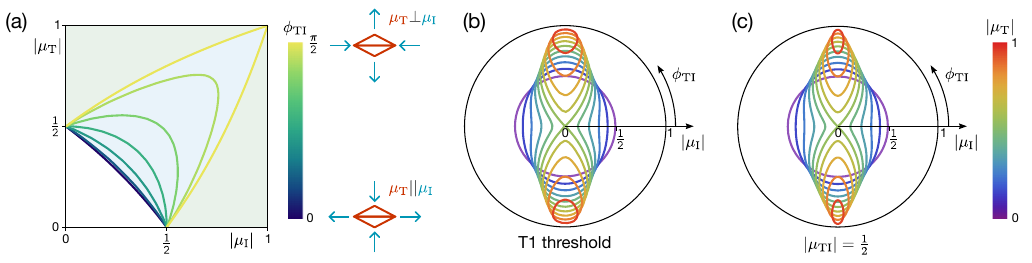}
    \caption{(a)~Marginalized T1 threshold as a function of the tension and isogonal Beltrami coefficients, $\muT$ and $\muI$, shown in the plane of their magnitudes.
    The threshold depends on the relative phase of the two Beltrami coefficients, $2\phi_\mathrm{TI} = \arg \muT -\arg \muI$, which measures the angle between the principal axis of tension anisotropy and the principal axis of isogonal strain.
    The shaded area indicates the region where at least one of the $\upell_i$ is negative for some value of $\psi$, indicating that a T1 must have happened.
    Note the symmetry under exchange of $\muT$ and $\muI$.
    (b)~Marginalized T1 thresholds as a polar plot $(|\muI|,\phi_\mathrm{TI})$ for a range of $|\muT|$. Exchanging $\muI \leftrightarrow \mu_\mathrm{T}$ yields the same plot.
    (c)~Contours along which the ``Beltrami sum'' of $\muT$ and $\muI$, has magnitude 1/2. This provides a good approximation for the true marginalized T1 threshold shown in (b). 
    }
    \label{fig:T1-thresh_psi-marginalized}
\end{figure*}

\begin{table*}[h!]
    \renewcommand{\arraystretch}{1.1}
    \centering
    \begin{tabular*}{0.45\linewidth}{l@{\extracolsep{\fill}}l}
    \toprule
    Symbol \hspace{3em} & Description \\
    \midrule
    $i,j,\dots$ & Cell indices \\
    $V, E, C$ &  No. of tiling vertices, edges, and cells \\
    $a,b,\dots$ & Coordinate indices \\
    $\left(\mathbf{v}\otimes\mathbf{w}\right)$ & Tensor product \\
    $\epsilon_{ab}, \, (\cdot)^\perp$ & Clockwise rotation by $\pi/2$ \\
    $\mathbf{R}(\phi)$ & Rotation matrix by $\phi$ \\
    $\hat{\nvec}$ & Unit normal vector \\
    $C_i$ & Tiling cell \\
    $\upa_{i}$ & Cell area \\
    $\Pi_{i}$ & Cell perimeter \\
    $\gamma_{ij}^k$ & Tricellular vertex angle \\
    $\rvec_{ijk}$ & Tricellular vertex position \\
    $\upell_{ij}$ & Cell edge lengths \\
    $\upkappa_{ij}$ &  Line curvature \\
    $\mathrm{R}_{ij}$ &  Curvature radius \\
    $\upp_{i}$ & Intracellular pressures \\
    $P(\upa_i)$ & Pressure equation of state \\
    $p_0$ & Reference pressures (set to $p_0=1$) \\
    $\uptau_{ij}$ & Interfacial tension \\
    $\tvec_{i}$ & 2D tension vertex \\
    $\tilde{\gamma}_{ij}^k$ & Tension triangulation angle \\
    $\updelta_i$ & Angle defect \\
    $\tilde\upa_{ijk}$ & Triangle area \\
    $\tilde{\cdot}$ & Triangulation quantity (except in App.~\ref{app:perturbation}) \\
    $(\cdot)'$ & (Conformally) deformed quantity \\
    $\phi_i(\tvecIt)$ & Piecewise-linear interpolation \\
    $(\nabla \uph)_{ijk}$ & Discrete gradient of a function \\
    $(\Delta \uph)_{i}$ & Discrete Laplacian of a function \\
    \bottomrule
    \end{tabular*}
    \hspace{0.05\linewidth}
    \begin{tabular*}{0.45\linewidth}{l@{\extracolsep{\fill}}l}
    \toprule
    Symbol \hspace{3em} & Description \\
    \midrule
    $\nabla\nabla h$ & Hessian of a function \\
    $\Phi_i$ & Cell potential in MDs \\
    $\Psi$ & Envelope function \\
    $(\cdot)^\mathrm{V}$ & Voronoi diagram \\
    $(\cdot)^\mathrm{P}$ & Power diagram \\
    $(\cdot)^\mathrm{C}$ & Multiplicative Voronoi diagram \\
    $\uptheta_i, \theta$ & Discrete, continuous isogonal potential  \\
    $\uplambda_i=1/\upp_i, \, \lambda$ & Discrete, continuous conformal factor \\
    $\bm{\sigma}$ & Stress tensor \\
    $\mathbf{S}$ & Stress tensor decomposition tensor \\
    $\overline{\bm{\sigma}}$ & Averaged stress tensor \\
    $\Fiso$ & Isogonal deformation tensor \\
    $(\cdot)^*$ & Legendre dual \\
    $\bm{\sigma}^*$ & Dual stress tensor \\
    $\bm{f}$ & Conformal map \\
    $M$ & M\"{o}bius transformation \\
    $\mathcal{D}, \mathcal{P}$ & Strain and cross-ratio operators \\
    $\boldsymbol{\upzeta}_i, \Lambda_i$ &  Circle packing centers and radii \\
    $\mathbf{T}$ & Map to reference triangle \\
    $\bm{\Sigma}$ & Diagonal singular value matrix \\
    $\psi$ & LTC phase \\
    $\phi_\mathrm{T}, \phi_\mathrm{I}$ & Tension and strain anisotropy \\
    $\mu_\mathrm{T}, \mu_\mathrm{I}$ & Beltrami coefficients \\
    $\mathcal{C}_\mathrm{T1}$ &  T1-threshold \\
    $\bm{g}$ & Tension metric \\
    $\bm{a}$ & Adjacency metric \\
    $\xivecIt$ & Lagrangian cell-label coordinates\\
    $K$ &  Gaussian curvature \\
    \bottomrule
    \end{tabular*}
    \caption{Overview of notation used in the main text.}
    \label{tab:notation}
\end{table*}

\end{document}